\documentclass[11pt,leqno]{article}
\usepackage[utf8]{inputenc}
\usepackage{authblk}
\usepackage{geometry}
\usepackage{natbib}
\usepackage{graphicx}
\usepackage{pdflscape}
\usepackage{appendix}
\usepackage{threeparttable}
\usepackage{booktabs}
\usepackage{amssymb}
\usepackage{amsbsy}
\usepackage{bm}
\usepackage{amsmath}
\usepackage{amsthm}
\usepackage{graphicx}
\usepackage{mathtools}
\usepackage{rotating}
\usepackage{setspace}
\usepackage{diagbox}
\usepackage{footmisc}

\usepackage[table]{xcolor}    

\usepackage{colortbl}

\usepackage{tikz}
\usepackage{enumitem}
\newlist{steps}{enumerate}{1}
\setlist[steps, 1]{label = Step \arabic*:}

\usepackage{dcolumn}
\newcolumntype{d}[1]{D{.}{.}{#1}}

\usepackage[colorlinks=true,urlcolor=nblue,linkcolor=nblue,citecolor=nblue]{hyperref}
\usepackage{doi} 

\usepackage{caption}
\usepackage{subcaption}
\definecolor{nblue}{HTML}{000660}

\title{\textbf{Predictive Density Combination Using a Tree-Based Synthesis Function}\thanks{The views expressed herein are solely those of the authors and do not necessarily reflect the views of the Bank of Canada, the Federal Reserve Bank of Cleveland, or the Federal Reserve System. We thank Mike West and conference and seminar participants at Notre Dame and \"{O}rebro, including Christiane Baumeister, Drew Creal, Luca Rossini, and Mattias Villani for helpful comments. Niko Hauzenberger gratefully acknowledges financial support from the Austrian Science Fund (FWF, ZK-35) and the Austrian Central Bank (Anniversary Fund, project no. 18763). Florian Huber gratefully acknowledges financial support from the Austrian Science Fund (FWF, ZK-35).}}
\author{}
\date{}

\begin{document}

\maketitle
\thispagestyle{empty}
\vspace*{-6.5em} 
\begin{center}
\end{center}

\normalsize

\vspace*{1em}
\begin{minipage}{.49\textwidth}
  \large\centering Tony Chernis\\[0.25em]
  \small Bank of Canada, Canada
\end{minipage}
\begin{minipage}{.49\textwidth}
  \large\centering Niko Hauzenberger\\[0.25em]
  \small University of Strathclyde, United Kingdom\\[0.25em]
  \small University of Salzburg, Austria
\end{minipage}

\vspace*{1.5em}

\begin{minipage}{.49\textwidth}
  \large\centering Florian Huber\\[0.25em]
  \small University of Salzburg, Austria
\end{minipage}
\begin{minipage}{.49\textwidth}
  \large\centering Gary Koop\\[0.25em]
  \small University of Strathclyde, United Kingdom
\end{minipage}

\vspace*{1.5em}

\begin{minipage}{.9\textwidth}
  \large\centering James Mitchell\\[0.25em]
  \small Federal Reserve Bank of Cleveland, United States
\end{minipage}

\vspace*{2em}
\begin{center}
\begin{minipage}{0.925\textwidth}
\begin{abstract} 
\setcounter{page}{0}
\doublespacing
\noindent Bayesian predictive synthesis (BPS) provides a method for combining multiple predictive distributions based on agent/expert opinion analysis theory and encompasses a range of existing density forecast pooling methods. The key ingredient in BPS is a ``synthesis'' function. This is typically specified parametrically as a dynamic linear regression. In this paper, we develop a nonparametric treatment of the synthesis function using regression trees. We show the advantages of our tree-based approach in two macroeconomic forecasting applications. The first uses density forecasts for GDP growth from the euro area's Survey of Professional Forecasters. The second combines density forecasts of US inflation produced by many regression models involving different predictors. Both applications demonstrate the benefits -- in terms of improved forecast accuracy and interpretability -- of modeling the synthesis function nonparametrically.
\end{abstract}
\end{minipage}
\end{center}
\begin{center}
\begin{minipage}{0.825\textwidth}
\bigskip
\begin{tabular}{p{0.2\hsize}p{0.65\hsize}} 
\textbf{Keywords:} &  Forecast density combination, Bayesian nonparametrics, Bayesian predictive synthesis \\
\textbf{JEL Codes:} & C11, C32, C53 \\
\end{tabular}
\end{minipage}
\end{center}

\normalsize \newpage \doublespacing

\section{Introduction}
It is commonplace when forecasting macroeconomic variables, such as output growth or inflation, to consider a large number of competing predictive densities. These density forecasts might come from different reduced-form or structural models, and/or be subjective and come from surveys. How to combine these densities is an open question being addressed by a growing literature.\footnote{See, among many others, \cite{mitchell_evaluating_2005, wallis_combining_2005, hall_combining_2007, geweke_optimal_2011, kk2012, billio2013time, aastveit2014nowcasting, conflitti_optimal_2015, chernis_nowcasting_2022, knotek_real-time_2022, aastveit_quantifying_2022, capek2023macroeconomic, diebold2022aggregation}.} The literature concludes that combined density forecasts tend to be more accurate and more robust than single-model approaches that ignore model uncertainty; for a review see \cite{Aastveit_Review}. One issue is that traditional forecast combination techniques are often linear and do not exploit information besides the forecasts and the target variable. Contrast this with a policymaker who combines forecasts nonlinearly and uses external information, such as on the current state of the economy or financial conditions, to help determine how much weight to attach to the different forecasts. We propose a novel technique that mimics this practice. Our approach combines density forecasts nonparametrically while allowing the combination weights to be determined by information that may be external to the forecasting models.

Key to our approach is Bayesian predictive synthesis (BPS). It has emerged, as extended into a time-series context by \cite{mcalinn2019dynamic}, as a general method of density forecast combination with a strong theoretical basis. BPS draws on an earlier Bayesian literature on agent or expert opinion analysis \citep{west_modelling_1992}, and provides a formal and theoretically justified method for pooling densities. It can be shown to nest many previous approaches \citep[see, for example, Section 2.2 of][]{mcalinn2019dynamic} and has been used successfully in various applications in economics, such as \cite{mcalinn2020multivariate}, \cite{chernis2023BPS}, and \cite{aastveit_quantifying_2022}. In this paper we develop density forecast combination strategies within the BPS framework. 

In existing implementations of BPS, the so-called ``synthesis function,'' which determines the weight attached to each density, needs to be specified parametrically. Common choices, as made in the aforementioned papers, are to assume that the synthesis function takes the form of a dynamic linear regression, with parameters that are allowed to change over time typically as random walk processes. This specification of the synthesis function thus allows the weights on competing density forecasts to evolve over time as linear Gaussian random walks.  But such an assumption may or may not be valid. Misspecification occurs if the weights depend on other factors or if they follow a different law of motion than a random walk.

These considerations motivate the present paper. BPS has theoretically rigorous foundations, but the manner in which it has been implemented in practice risks misspecification due to the adoption of particular and untested parametric assumptions. We therefore propose to use flexible nonparametric techniques to specify the synthesis function. Specifically, we use regression trees. In conventional (single-model) forecasting applications, tree-based models of the conditional mean have proven highly successful \citep[see, for example,][]{clark2021tail, huber2022inference, huber2020nowcasting}. A small number of other papers have used nonparametric techniques to combine predictive densities \citep[for example,][]{jin2022infinite, bassetti2018bayesian, bassetti2023inference}. However, unlike our proposed method, these other papers neither use regression trees nor fit explicitly within the formal BPS framework. 

While regression trees have become a popular way to estimate nonparametric regressions, here we propose to use them differently. 
Similarly to \cite{coulombe2020macroeconomy}, \cite{deshpande2020vcbart}, and \cite{hhkm2023tvpbart}, who provide a nonparametric treatment to the parameters rather than the
variables in single-equation and VAR models, we model the coefficients in the synthesis function with regression trees (RT). Accordingly we label our version of BPS, BPS-RT. The synthesis function remains linear in the parameters, which, as we will demonstrate, aids in interpretation. Use of regression-tree methods requires the choice of covariates, which we call ``weight modifiers.'' These weight modifiers help determine the weights attached to the competing density forecasts. Conventional BPS does not make use of weight modifiers, given that the weights are typically assumed to follow random walks. Thus, in popular implementations of BPS any relevant information in the form of additional covariates is neglected.\footnote{Notable recent exceptions are \cite{Villani}, who following \cite{LI2023}, let the weights in linear density forecast combinations depend on (potentially time-varying) exogenous variables. As \cite{Villani} explain, such linear pools are one specific instance of BPS. Letting the combination weights in linear pools change over time according to these ``pooling variables,'' as in the more general (nonlinear) BPS framework that we consider, can offer more flexibility than assuming that the combination weights follow an assumed autoregressive process; cf. \cite{DELNEGRO2016}.} But decision makers when combining competing density forecasts may wish to condition their forecasts on such ``outside'' information. For example, they may wish to let the weights on the different forecasting models vary with the  state of the economy or vary as a function of the features of each forecast density. Our tree-based specification for the synthesis function is able to condition on both ``global'' (that is, information not associated with a particular forecaster) and ``local'' (that is, information associated with a given forecaster) variables when determining the weights. In our tree-based synthesis function, the weights on each density forecast are dynamically determined via a sequence of decision rules. BPS-RT allows the decision maker to combine predictive densities in a highly flexible way and to distill optimally all relevant information contained in the predictive densities and weight modifiers. The fact that the synthesis function remains conditionally linear in the parameters helps the decision maker interpret the combined density and better understand the role each individual density is playing in the combination. We will show how BPS-RT can be used to understand the role of model incompleteness, agent clustering, and the time-varying importance of the different weight modifiers.

The next section of the paper introduces and motivates BPS in theory and then discusses how it has been implemented in the existing literature. It then proposes our generalization, BPS-RT, and explores its properties. 
Section 3 demonstrates the utility of BPS-RT by undertaking two forecasting applications. The first application takes the individual forecaster density forecasts from the European Central Bank Survey of Professional Forecasters (ECB SPF) and combines them. The second application forecasts US inflation using a commonly used large set of indicators. The predictive densities that are synthesized are produced by regression models using the different indicators.  We find that BPS-RT produces well-calibrated and accurate forecasts. Notably, we find that single-tree models perform best, in contrast to standard recommendations when using regression trees. This suggests that a relatively parsimonious weight scheme with few changes in weights is supported by the data. The superior performance of BPS-RT stems from its better ability to explain periods of volatility, such as the global financial crisis that affected euro area GDP growth and the post-COVID inflation period in the US. Zooming in on the best performing BPS-RT specification in the US inflation application, we show how the combination forecasts from BPS-RT can be interpreted. BPS-RT can be used to understand the role of model incompleteness, agent (forecast) clustering, and the time-varying importance of the different weight modifiers. We find little model set incompleteness during the post-COVID inflation period, suggesting that BPS-RT's success comes from its ability to successfully forecast inflation using the underlying models with changes in the combination weights driven by a time trend. This contrasts with the earlier period of lower inflation, when business cycle indicators are shown to be more important. Section 4 concludes. Online Appendix A provides details on Bayesian inference of BPS-RT and Appendix B provides additional empirical results, as referenced in the main paper.

\section{Bayesian Predictive Synthesis with Regression Trees}\label{sec:BPS}
In Section \ref{sec2.1}, we provide some background on BPS, distinguishing between BPS in theory and its use in practice in extant empirical applications. Then in Section \ref{Sec:BPS-RT} we explain how regression trees can be used to provide a more flexible way of operationalizing BPS.

\subsection{Bayesian Predictive Synthesis}\label{sec2.1}

\subsubsection{BPS in Theory}
BPS is a foundational theoretically coherent Bayesian method for combining predictive densities.\footnote{For a general description of BPS, see  \cite{mcalinn2019dynamic}; specific implementation details related to our applications are discussed below and in Appendix A.} The theory of BPS provides a pooled predictive distribution for the variable being forecast (say, GDP growth) given a set of individual density forecasts. 
Operationally, this pooled predictive distribution is produced using Markov chain Monte Carlo (MCMC) methods involving two steps. In the first step, draws are taken from the individual predictive densities for GDP growth. These draws are then, in effect, treated in a second step as explanatory variables in a time-series model where the dependent variable is the outcomes for GDP growth. This time-series model amounts to the synthesis function. Standard choices for this function are typically based on linearity, either simply a constant linear relationship or a dynamic relationship where the linear coefficients evolve over time according to a random walk. As pointed out by \cite{aastveit_quantifying_2022}, this means that BPS can be thought of as a multivariate regression relating the target variable (GDP growth) to the forecasts for GDP growth, which are treated as generated regressors. We make use of this generated regressor interpretation below. 

More formally, at time $t$ a decision maker $\mathcal{D}$ is confronted with $h$-step-ahead forecast densities for variable $y_{\tau+h}$ produced by $J$ different agents, experts, or models, where $\tau$ ranges from 1 to $t$. At each forecast origin, $\tau$, we label these predictive densities $\{\pi_{j \tau}(y_{\tau+h})\}_{j=1}^J$. These densities, available from time $1$ through $t$, form the information set $\mathcal{H}_{t}$ of $\mathcal{D}$ and can, in principle, be of any distributional form. $\mathcal{D}$ then forms an incomplete joint prior $p(y_{t+h}, \mathcal{H}_t) = p(y_{t+h}) \times \mathbb{E}\left(\prod_j \pi_{jt}(x_{jt+h|t}) \right)$ with $\bm x_{t+h|t} = (x_{1t+h|t}, \dots x_{Jt+h|t})'$ denoting latent agent states (that is, draws from the agent-specific predictive densities). These agents' forecasts  target $t+h$ but, under the prior, are made using information through time $t$. The prior is incomplete, in the sense that $\mathcal{D}$ only forms an expectation of the product of the agent densities. Agent opinion analysis theory \citep{west_modelling_1992-1, west_modelling_1992}, extended to a time-series context by \cite{mcalinn2019dynamic}, shows that the posterior conditional density for $y_{t+h}$ under this incomplete prior takes the form:
\begin{equation} \label{eq:bps}
    p(y_{t+h}|\bm \Psi_{t+h},\mathcal{H}_t) = \int \alpha(y_{t+h}|\bm x_{t+h|t},\bm \Psi_{t+h})\prod_{j=1}^J \pi_{jt}(x_{jt+h|t}) dx_{jt+h|t},
\end{equation}
where $\alpha(y_{t+h}|\bm x_{t+h|t},\bm \Psi_{t+h})$ denotes the synthesis function that reflects how $\mathcal{D}$ combines her prior information with the set of expert-based forecasts; $\bm \Psi_{t+h}$ denotes a matrix of parameters and latent states that control the properties of the synthesis function, $\alpha(.)$.

\subsubsection{BPS in Practice}
Theory offers no guide as to the specific choice of the synthesis function, $\alpha(y_{t+h}|\bm x_{t+h|t},\bm \Psi_{t+h})$. But a common choice in empirical applications, used, for example, in \cite{mcalinn2019dynamic}, \cite{mcalinn2020multivariate}, and \cite{aastveit_quantifying_2022}, is to assume a dynamic linear regression model treating the draws from the $J$ competing densities as generated regressors, $\bm x_{t+h|t}$. 
Our synthesis functions will have a dynamic regression form, but we will use a non-centered parameterization \citep[see][]{SFS_W}:
\begin{equation}\label{eq:synthesis}
\alpha(y_{t+h}| \bm x_{t+h|t}, \bm \Psi_{t+h}) = \mathcal{N}\left(y_{t+h} | c_{t+h} +   \sum_{j=1}^J  (\gamma_{j} + \beta_{jt+h}) x_{jt+h|t}, \sigma_{t+h}^2\right),
\end{equation}
 where $c_{t+h}$ is a time-varying intercept assumed to follow a random walk, $\bm \gamma = (\gamma_{1}, \dots, \gamma_{J})'$  are time-invariant weights, and  $\bm \beta_{t+h} = (\beta_{1t+h}, \dots, \beta_{Jt+h})'$ denotes the time-varying combination weights. As discussed above, a common choice in the literature is to assume that the weights, $ \beta_{jt+h}$, evolve as a random walk (RW) with innovation covariance matrix $\bm V$, leading to a version of BPS that we label ``BPS-RW.'' When implementing BPS-RW in our empirical applications below, we make standard choices for the prior and MCMC method. In particular, they are similar to those used in \cite{hauzenberger2022fast}. The only difference is that we use the hyperparameter-free horseshoe prior instead of the normal-Gamma prior, so as to have a prior that is comparable to the one used with our regression-tree model. Accordingly, we do not provide additional details here on drawing the time-varying weights for BPS-RW; see \cite{hauzenberger2022fast} for details.

 In all of our implementations of BPS, including BPS-RW, we consider two versions: one that assumes stochastic volatility (SV) and another that is homoskedastic.  In the SV case, the error variance, $\sigma_{t+h}^2$,  changes over time. We assume that the log-volatilities $\varsigma_{t+h} := \log \sigma_{t+h}^2$ evolve according to an AR(1) model with autoregressive coefficient $\rho_\varsigma$, unconditional mean $\mu_\varsigma$, initial value $\varsigma_0$, and error variance $\sigma^2_\varsigma$. The prior choices for these parameters are given in Appendix \ref{app:prior}. Homoskedastic cases are obtained setting $\sigma^2_\varsigma$ to zero.  Below, for notational ease, we do not explicitly note those parameters relating to SV in the conditioning arguments. 
 
All of our implementations of BPS also include a time-varying intercept, $c_{t+h}$, which is assumed to follow a random walk. As discussed below, $c_{t+h}$ is included to allow for model incompleteness. 
Further econometric details are provided in Appendix \ref{app:A}. 

With these notational conventions established, $\bm \Psi_{t+h} = 
\left(\bm \gamma, \{c_\tau, \bm \beta_{\tau}, \sigma_\tau \}^{t+h}_{\tau=0}, \bm \theta \right)$, where $\bm \theta$ will be method-specific parameters that define the law of motion of latent states or appear in the hierarchical priors (such as $\bm V$ in the case of BPS-RW).  
  
The synthesis function, $\alpha(y_{t+h}|\bm x_{t+h|t}, \bm \Psi_{t+h})$, is quite flexible, given that the weights it attaches to each of the $J$ densities are dynamic and because it allows for time-varying error variances. We can also see that while Eq. ({\ref{eq:synthesis}}) implies a Gaussian density conditional on $\bm \beta_{t+h}$, $\bm x_{t+h|t}$, and $\sigma_{t+h}^2$, when carrying out predictive inference we marginalize out the unknowns  of the model, leading to a predictive density that can be highly non-Gaussian; see Eq. (\ref{eq: pred_dens}) below.

In contrast with other approaches to combining models and density forecasts, such as Bayesian model averaging (BMA), the weights on each density are restricted neither to lie between zero and one nor to sum to unity. In the case of BPS-RW, the degree of change in the weights will depend on the  magnitude of the state innovation variances for these parameters: small values imply slow, smooth adjustment of the weights over time, large values allow for bigger sharper changes.

Two additional aspects of this parameterization of the synthesis function are worth noting before we introduce our regression-tree approach, which provides a more flexible nonparametric representation of the synthesis function.

First, as a special case, we define a static version of BPS that assumes time-invariant weights $\bm \beta_\tau = \bm 0_J$ for all $\tau$ but  leaves $\bm \gamma$ unrestricted. We label this instance of BPS, which assumes the  combination weights to be constant over time, ``BPS-CONST.''

Second, the presence of both an intercept and an error in the synthesis function means that these versions of BPS allow for model set ``incompleteness'' \citep{Geweke2010}. That is, they allow the ``true'' (but unknown) model not to be in $\mathcal{D}$'s model space; see, for example, \cite{billio2013time} and \cite{AastveitJBES_2018}. A conventional model combination scheme such as BMA sets both intercept and error to zero. The fact that the intercept, $c_{t+h}$, and error variance, $\sigma^2_{t+h}$, are both time varying provides additional flexibility when modeling the degree of model set incompleteness. Note that these specific assumptions are equivalent to embedding a popular benchmark for forecasting (especially of inflation) -- the unobserved components SV (UCSV) model of \cite{SW_UCSV} - within our set of now $J+1$ density forecasts. This is also related to an alternative treatment of model set incompleteness in BPS that adds a fictitious baseline predictive density to the set of densities being synthesized \citep[see, for instance, the discussion in Section 2.2.3 of][]{TallmanWest2023}.\footnote{\cite{diebold2022aggregation} also add a fictitious forecaster in their ECB SPF application that, like ours below, combines forecaster-level density forecasts.} In our case, this baseline predictive density comes from a UCSV model. But importantly, as when estimating a mixture density, the parameters of the UCSV density are estimated simultaneously with the weights in the synthesis function. 


To carry out predictive inference, we need to compute the predictive distribution. We do so by simulation. Let $y_{T+h}$ denote a future realization of our target variable at time $T+h$ and let $\mathcal{H}_{T}$ denote the set of agent densities that are available at time $T$ but target $T+h$. The predictive density, in our case, is obtained as follows:
\begin{equation}
    p(y_{T+h}|\mathcal{I}_{T}) = \int \int p(y_{T+h}|\bm \Psi_{T+h},  \mathcal{H}_{T}) ~d\bm \Psi ~ d \mathcal{H}, \label{eq: pred_dens}
\end{equation}
where $\mathcal{I}_{T}$ indicates the information set up to time $T$ and $\bm \Psi_{T+h}$ are the latent states (projected forward to time $T+h$). We can simulate from (\ref{eq: pred_dens}) by simulating from the joint posterior of the agents and states, projecting the states forward to time $T+h$, and then using the synthesis function in (\ref{eq:synthesis}) to produce a combined forecast draw. By doing so, we integrate out the unknowns of the model and the resulting predictive density can be highly non-Gaussian and feature heavy tails, multi-modalities, and/or skewness.

\subsection{Bayesian Predictive Synthesis with Regression Trees (BPS-RT)}\label{Sec:BPS-RT}

In this paper, our proposal is to relax the restrictions in BPS-RW by considering more flexible forms of time variation in $\bm \beta_{t+h}$. Specifically, we use techniques from machine learning to model the dynamic evolution of the weights, $\bm \beta_{t+h}$, in a nonparametric manner as a function of additional weight modifiers. This treatment can be contrasted with the alternative of treating the function, $\alpha$, itself nonparametrically. We follow \cite{chipman2010bart} and use Bayesian additive regression trees (BART) to estimate the regression trees. BART consists of a set of priors for the tree structure and the terminal nodes (the leaf parameters) and a likelihood for data in the terminal nodes.

BPS-RT differs from existing implementations of BPS through both the hierarchical priors used on elements in $\bm \gamma$ and $\bm \beta_{t+h}$ and by incorporating additional covariates into $\mathcal{D}$'s information set. These are stored in a $K_\gamma$ vector $\bm z_j^\gamma$ and a $K_\beta$ vector $\bm z^\beta_{j t+h|t}$, both containing additional ``data'' known to $\mathcal{D}$ through period $t$.

We postulate a nonlinear relationship between the weights and the weight modifiers through functions $\mu^{\gamma}_j(\bm z_j^\gamma)$ and $\mu^{\beta}_{j}(\bm z_{jt+h|t}^\beta)$ that determine the state transition equation that can be interpreted as a prior. In particular, we assume:
\begin{equation}
\begin{aligned}
\gamma_j  \sim \mathcal{N}(\mu^{\gamma}_j(\bm z_j^\gamma), \tau^{\gamma}_j) \quad \text{and} \quad \beta_{jt+h} \sim \mathcal{N}(\mu^{\beta}_{j}(\bm z_{jt+h|t}^\beta), \tau^{\beta}_{j}), 
\end{aligned}  \label{eq: bps_rt}
\end{equation}
where  $\tau^{\gamma}_j$ and $\tau^\beta_j$ denote prior scaling parameters. For convenience, we define $\mu^{\gamma}_j := \mu^{\gamma}_j(\bm z_j^\gamma)$  and $\mu^\beta_{jt+h} := \mu^{\beta}_{j}(\bm z_{jt+h|t}^\beta)$.   The best way to illustrate the effect the scaling parameters have on the actual estimates of the weights is to consider a re-parameterization of the synthesis function. 
Integrating out $\gamma_j$ and $\beta_{jt+h}$ by plugging Eq. (\ref{eq: bps_rt}) into Eq. (\ref{eq:synthesis}) yields:
\begin{equation}
    y_{t+h} = c_{t+h} + \sum_{j=1}^J \left[\underbrace{\left(\mu^{\gamma}_j(\bm z_j^\gamma) + \sqrt{\tau^{\gamma}_j}\nu^\gamma_{j}\right)}_{\gamma_j}x_{jt+h|t} + \underbrace{\left(\mu^{\beta}_j(\bm z_{jt+h|t}^\beta) + \sqrt{\tau^{\beta}_{j}}\nu^\beta_{jt+h}\right)}_{\beta_{jt+h}} x_{jt+h|t} \right ] + \sigma_{t+h} u_{t+h}, \label{eq: synth_integrated}
\end{equation}
with $\nu^\gamma_j, \nu^\beta_{jt+h} \sim \mathcal{N}(0, 1)$ denoting process innovations. 
The innovations, $\nu^\gamma_j$ and $\nu^\beta_{jt+h}$, and the corresponding scaling terms control the degree of dispersion of the actual weights from those expected under the prior mean.  If the scalings are close to zero, the posterior of $\gamma_j$ and $\beta_{jt+h}$ is pulled toward the prior mean and the resulting estimates will be close to  $\mu^{\gamma}_j(\bm z_j^\gamma)$ and $\mu^{\beta}_{j}(\bm z_{jt+h|t}^\beta)$ and so strongly depend on $\bm z_j^\gamma$ and $\bm z_{jt+h|t}^\beta$. If this is not the case, the resulting scaling parameters would be larger so that  substantial deviations from the prior means are more likely. Another feature of this representation is worth emphasizing. As opposed to a model that directly approximates the synthesis function nonparametrically, the specification in  (\ref{eq: synth_integrated}) introduces interaction terms of the form $\mu_j^\gamma(\bm z_j^\gamma) \times x_{jt+h|t}$. This specific form might reduce the risk of overfitting by introducing more structure on the space of functions that we approximate.

We approximate the prior mean functions through a sum-of-trees model with $S$ trees:
\begin{equation}\label{eq:bart}
\mu^{\gamma}_j( \bm z^{\gamma}_{j})   \approx \sum_{s=1}^{S} g(\bm z^{\gamma}_{j}|\mathcal{T}^{\gamma}_s, \bm \phi^{\gamma}_{s})
\quad \text{and} \quad \mu_{jt+h}^\beta = \mu^{\beta}_{j}(\bm z^{\beta}_{jt+h|t}) \approx \sum_{s=1}^{S} g(\bm z^{\beta}_{jt+h|t}|\mathcal{T}^{\beta}_s, \bm \phi^{\beta}_{s}),
\end{equation}
where $g$ denotes a tree function that is parameterized by so-called tree structures, $\mathcal{T}^{n}_s$, and terminal node parameters, $\bm \phi^n_s$, for $n \in \{\beta, \gamma\}$. The basic idea behind a single tree is that the tree structures describe sequences of disjoint sets. These sets partition the input space (determined by exogenous covariates, $\bm z^\gamma_{j}$ and $\bm z^\beta_{jt+h|t}$, respectively). Each of these sets is associated with a particular terminal node parameter. In our case, the terminal node parameters serve as prior expectations for the $\gamma_j$s and for the $\beta_{jt+h}$s. The input space is associated with vectors of variables, $\bm z^{\gamma}_{j}$ and $\bm z^{\beta}_{jt+h|t}$,  which we refer to as weight modifiers. Note that $\bm z^{\beta}_{jt+h|t}$ could include quantities (such as moments from the agent-specific predictive densities) that explicitly target $t+h$ but are available at time $t$.

There are two main justifications for our BPS-RT modeling approach. First, there is no reason to restrict attention, as in BPS-RW, to random walk specifications for the evolution of $\beta_{t+h}$. BPS-RW implies, at a given point in time, a linear relationship between $y_{t+h}$ and $\bm x_{t+h|t}$. This assumption might be warranted in tranquil periods. But, in unusual times, nonlinearities could be present, and exploiting these might lead to more accurate forecasts. Our regression-tree approach allows for flexibility in the way such nonlinearities are modeled and lets the ``data speak.'' Second, and this holds across all existing instances of BPS not just BPS-RW, an implicit assumption made is that the information set available to $\mathcal{D}$ comprises exclusively the agent-based forecast densities.\footnote{As mentioned in footnote 2, an exception is \cite{Villani}, who when combining density forecasts using the linear opinion pool also let the weights depend on  exogenous variables. Our BPS-RT model generalizes to consider BPS combinations beyond the linear special case and to allow for nonlinearities in how the weight modifiers affect the weights.} But, in principle, additional unmodeled information is available to $\mathcal{D}$ and might help inform evolution of the weights. In our BPS-RT approach,  the weight modifiers, $\bm z^{\gamma}_{j}$ and $\bm z^{\beta}_{jt+h|t}$, comprise this extra information. 

These weight modifiers might include characteristics of the agents' forecasts not directly reflected in their predictive distributions or other common (to agents) factors, such as general  information about the macroeconomic environment. For example, $\bm z^{\gamma}_{j}$ might contain summary metrics of overall past forecast performance (such as the average historical forecast performance) for each agent. Or, as noted above, $\bm z^{\beta}_{jt+h|t}$ might contain more granular and time-varying information, such as  time-varying characteristics of the agent-specific predictive densities (say their higher moments and/or time-varying measures of past forecasting performance). We provide specific context and motivate our choice of weight modifiers in the  empirical applications in Section \ref{ssec:spec} below.


To return to the regression tree, note that it is defined by disjoint sets that are determined by splitting rules of the form $z^{\beta}_{k,jt+h|t} \le d_k$ or $z^{\beta}_{k,jt+h|t} > d_k$, where $z^{\beta}_{k,jt+h|t}$ is the $k\textsuperscript{th}$ weight modifier for the $j\textsuperscript{th}$ agent/model  and $d_{k}$ is a threshold parameter associated with the $k$\textsuperscript{th} effect modifier, which is estimated from the data. It is important to note, however, that any splitting rule associated with the $k\textsuperscript{th}$ effect modifier is common across agents and periods (that is, it is specific neither to agent $j$ nor to period $t$). Hence, the thresholds $d_k$ and thus the tree structures do not have $t$ or $j$ subscripts and are common across agents/models and time. Since these splitting rules effectively govern the prior mean, $\mu^{\beta}_{jt+h}$, this structure in a sense captures the notion of a pooling prior and reflects the situation that $\mathcal{D}$ decides on the weights associated to each of the different agents based on using additional factors $\bm z^{\gamma}_{j}$ and $\bm z^{\beta}_{jt+h|t}$ according to a set of common decision/splitting rules. The same structure also holds for the $\gamma_j$s, with the difference that the splitting rules controlling $\mu^{\gamma}_{j}$ pool exclusively over the cross-section and not over time (since the $\gamma_j$s are time invariant). 

To see this pooling feature more clearly, consider a BPS-RT model that assumes $\bm \beta_{t+h} = \bm 0_J$ and features only a time-variant part $\bm \gamma$, for which the prior mean $\bm \mu^{\gamma} = (\mu^{\gamma}_1, \dots, \mu^{\gamma}_J)'$ is defined by a single tree ($S=1$) and by a single effect modifier  in $z_j^\gamma$ (that is, $z_j^\gamma$ is a scalar with $K_\gamma = 1$). In this case, the prior on $\gamma_j$ can be written as:
\begin{equation}
    \gamma_j = g(z_j^\gamma|\mathcal{T}_s^\gamma, \bm \phi_s^\gamma) + \sqrt{\tau_j^\gamma} v_j, \quad v_j \sim \mathcal{N}(0, 1).
\end{equation}

If we now compute the difference between $\gamma_j$ and $\gamma_m$ for distinct agents, $j \neq m$, and assume that $z_j^\gamma$ and $z_m^\gamma$ are similar, in the sense that both imply the same decomposition of the input space and are thus located in the same terminal node of the tree, we end up with:
\begin{equation}
    (\gamma_j - \gamma_m) \sim \mathcal{N}(0 , \tau^\gamma_j + \tau^\gamma_m).
\end{equation}

This equation implies that if the tree suggests that the characteristics between agents are so similar that they are grouped together in the same  terminal node, the same prior mean applies and the difference between prior means will be zero.  The presence of the prior scaling parameters $\tau_j^\gamma$ and $\tau_m^\gamma$ will then allow for data-based testing of whether that restriction should be strongly enforced or not. Since both prior means would coincide, setting both $\tau_j^\gamma$ and $\tau_m^\gamma$ to values  close to zero would induce a clustering of  $\gamma_j$ and $\gamma_m$ around $g(z_j^\gamma|\mathcal{T}_s^\gamma, \bm \phi_s^\gamma) =  g(z_m^\gamma|\mathcal{T}_s^\gamma, \bm \phi_s^\gamma)$. Hence, the choice of the prior specified on the scaling parameter $\tau^\gamma_j$ is crucial in determining the clustering behavior of BPS-RT.

Another feature of our prior is that $\mathcal{D}$ adjusts her weights on the agents' densities depending on the (common) macroeconomic environment as captured by the weight modifiers, which might include, as discussed, indicators of the state of the business cycle, measures of economic uncertainty, or deterministic trends. For example, in turbulent times larger weights on component densities that are far from Gaussian and feature, say, heavy tails might lead to better combined density forecasts. Our approach can control for this, if supported by the data.

Note that we estimate the tree structures and the terminal parameters alongside all other unknown parameters and therefore also specify priors for them. We follow here the recommendations of \cite{chipman2010bart} and discuss the remaining model and prior specification issues in detail in Appendix \ref{app:A}. This technical appendix also describes the MCMC methods used to estimate BPS-RT. In summary, these MCMC methods are straightforward. They require a method for predictive simulation from each individual model (to draw from each agent's forecast density) and a method for drawing from the regression-tree model conditional on the individual-agent draws. For BPS-RT the algorithm is taken directly from  \cite{chipman2010bart}.

\subsection{Illustrating BPS-RT}
We now explain how BPS-RT works and allocates combination weights using an illustrative toy example. Assume that, unknown to $\mathcal{D}$, the ``true'' data for $y_t$ are generated by the following threshold model:
\begin{equation}
    y_t =  \begin{cases}
         \rho_1 y_{t-1} + c \rho_2 y_{t-2} + \sigma_0 \nu_t, ~\text{ for } t =1, \dots, 200\\
         c \rho_1 y_{t-1} +  \rho_2 y_{t-2} + \sigma_0 \nu_t,  ~\text{ for } t = 201, \dots, 350
    \end{cases}
\end{equation}
where  $\rho_1 = 0.8, \rho_2 = -0.8$  and $\sigma_0 = 1.2$, $y_0 = y_1 = 0$, $c=1/100$, and $\nu_t \sim \mathcal{N}(0, 1)$.

Then, $J=2$ agents predict $y_t$ as follows (these forecasts are one-step-ahead, $h = 1$):
\begin{align}
    x_{1t} \sim y_{1t} = \mathcal{N}(\rho_1 y_{t-1}, (1-\rho_1^2) \sigma^2_0), \\
    x_{2t} \sim y_{2t} = \mathcal{N}(\rho_2 y_{t-2},  (1-\rho_2^2) \sigma^2_0).
\end{align}

Both agents use forecasting methods with a different type of misspecification. The first agent's forecast is almost correctly specified for the first part of the sample, but the second agent's is substantially misspecified. In the second part of the sample this switches. We would hope that BPS-RT, when combining these two misspecified densities, would put more weight on the first agent when $t \le 200$, then increase the weight on agent 2 when $t > 200$. 

Notice that the structure of the data-generating process (DGP) implies that BPS-RW is severely misspecified, since BPS-RW implies that the combination weights on the two agents evolve smoothly over time. Our more flexible choice of synthesis function, (\ref{eq:synthesis}), conditional on choosing appropriate effect modifiers, as we shall show, is capable of accommodating the break at $t=200$. 

We consider three variables as weight modifiers. The first is a simple deterministic time trend, $z^{\beta}_{1,jt+1|t} = t+1$, that is common to both agents. The remaining two effect modifiers are agent-specific and measure historical forecasting performance.  To capture historical point forecasting performance, we consider each agent's squared forecast error (SFE) as recursively computed at time $t-1$: $z^{\beta}_{2,jt+1|t} =(y_{t} - \mathbb{E}(x_{jt|t-1}))^2$ for $j=1,2$. Then to measure past density forecasting performance, we consider each agent's continuous ranked probability score (CRPS).\footnote{If $F$ is the c.d.f. of the forecast and $y$ the subsequent realization, then $\text{CRPS}(F,y) = \int (F(x) -\mathbf{1}_{x \geq y} )^2 dx$. }   

Our synthesis function is given by Eq. (\ref{eq:synthesis}). To facilitate illustration of BPS-RT, we make some simplifying assumptions. We set the time-invariant weights $\bm \gamma=\bm 0$ and, for the prior on $\bm \beta_{t+1}$, set the scaling parameters equal to zero so that the weights and prior means coincide, and we focus on the single-tree case ($S=1$). For expositional ease, we drop corresponding sub- and super-scripts when there is no loss in meaning. Under these simplifying assumptions, the synthesis function, similarly to (\ref{eq: synth_integrated}), reduces to:
\begin{equation*}
    \alpha(y_{t+1}~|~\bm x_{t+1|t}, \bm z_{t+1|t}, \bm \Psi_{t+1}) = \mathcal{N}\left(y_{t+1}~|c_{t+1}+g(\bm z_1|\mathcal{T}, \phi) x_{1t+1|t} + g(\bm z_2|\mathcal{T}, \phi)  x_{2t+1|t}, \sigma_{t+1}^2\right).
\end{equation*}

This equation shows that with the scaling parameters set equal to zero, we end up with a BART model that assumes the weights depend nonlinearily on $\bm z_{t+1|t}$. 

\begin{figure}[!htbp]
\centering
\caption{Illustration of BPS-RT.}\label{fig:illustr}
\begin{minipage}{0.49\textwidth}
\centering
(a) Decision tree of $\mathcal{D}$ 
\end{minipage}
\begin{minipage}{0.49\textwidth}
\centering
(b) Combination weights over time
\end{minipage}
\begin{minipage}{0.49\textwidth}
\centering
\includegraphics[scale=0.4]{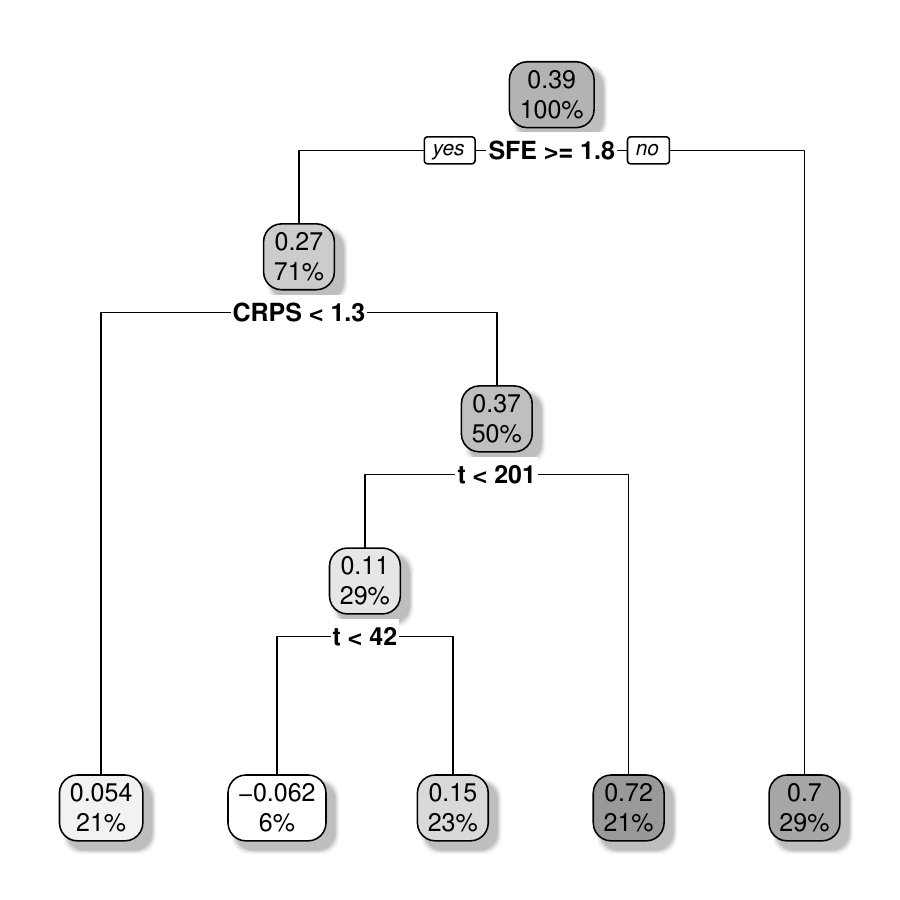}
\end{minipage}
\begin{minipage}{0.49\textwidth}
\centering
\vspace*{20pt}
\includegraphics[scale=0.4]{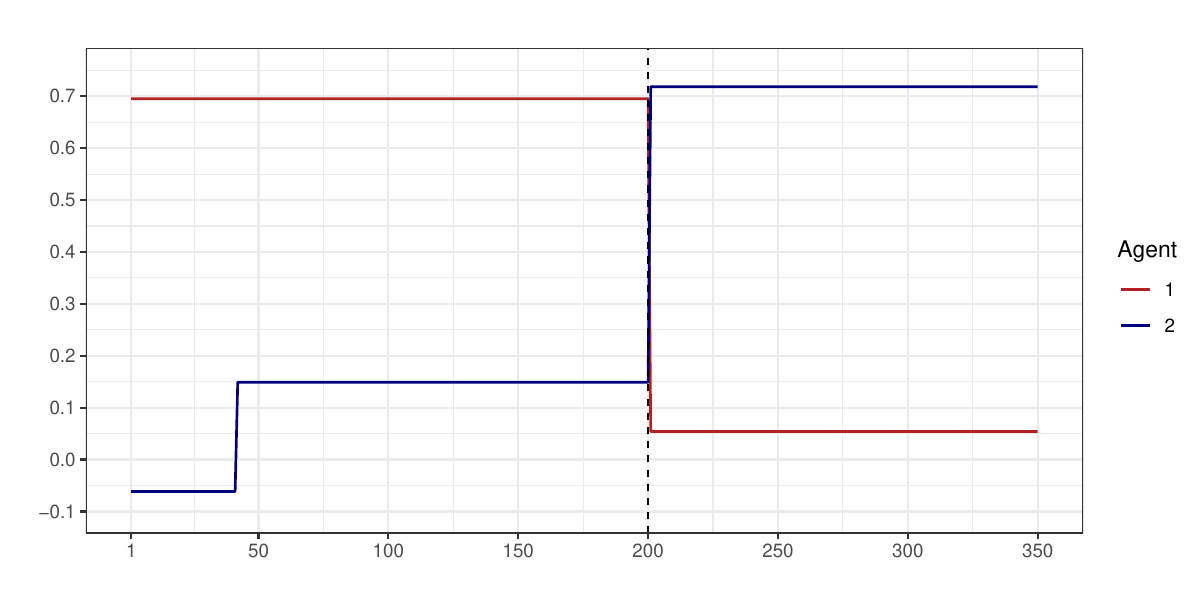}
\end{minipage}
\scriptsize \textbf{Notes:} As the weight modifiers, we use a simple linear time trend, SFE, and CRPS. Each oval box in panel (a) indicates the terminal node parameter of a particular branch and the share (in percent) of observations belonging to this branch. 
\end{figure}

Figure \ref{fig:illustr} depicts in panel (a) the estimated tree  and in panel (b) the temporal evolution of the estimated weights. We emphasize that these weights are in-sample estimates, that is, conditional on data through $T=350$. 

The tree in panel (a) can be understood as follows. Let us start at the bottom of the tree. We see five terminal nodes. Hence, we observe five groups/clusters that define the prior mean  both over time and across agents.
Put differently, there are five ``breaks'' over time and across agents in the prior mean. 

How we pool is defined by the splitting rules. These are understood by turning to the top of the tree. At the root (level 0), the SFE is used as a splitting variable. The threshold parameter is $1.8$ and, hence, if the SFE in $t-1$ is larger than or equal to $1.8$, we move down the left branch of the tree. At the first level, the lagged CRPS shows up as the next threshold variable. If the CRPS is smaller than $1.3$, we end up in a terminal node and set the weight associated with an agent that has an SFE greater than or equal to $1.8$ and a CRPS smaller than $1.3$ equal to $\mathbb{E}(\beta_{jt})=0.054$. These conditions are fulfilled $21$ percent of the time. By contrast, if the CRPS is greater than or equal to $1.3$, we drop down to the second level of the tree. In this segment, time shows up as a splitting variable and if $t \ge 201$, we assign a weight of $0.72$. For $t < 201$ we introduce a further splitting rule that splits the sample once more by testing whether $t < 42$. If this is the case a negative weight of $-0.062$ is applied, whereas if $42 \le t < 201$ the weight is $0.15$.  If the past SFE is smaller than $1.8$, we end up in the right branch of the tree and assign a weight equal to $0.7$.


Hence, the tree suggests that, first and foremost, $\mathcal{D}$ selects agents according to the past performance of their forecasts, since both SFE and CRPS are identified in the estimated tree.  Under our DGP, this implies that weights dynamically update if a given agent issued a poor prediction in the previous period without taking into account the past performance of her forecasts. To understand how these decision rules translate into the actual evolution of model weights, panel (b) shows the weights over time. These indicate that in the first part of the sample, Agent 1 receives substantial weight, while Agent 2 receives relatively little weight. This makes sense, given that the former is only mildly misspecified, whereas the latter features substantial model misspecification.  As expected, given the structural break in the DGP, $\mathcal{D}$ now overweights the second agent, whereas the weight on Agent 1 is now much smaller. 

This simple exercise illustrates how $\mathcal{D}$ incorporates additional information (time and past forecast errors in this case) to combine models. In general, though, the prior scaling parameters in BPS-RT  are greater than zero, and hence, the decision tree gives rise to prior expectations that, in turn, inform the posterior estimates of the weights. Hence, if there is no relationship between the weights and the weight modifiers, the resulting prior variance  would be large and the weights would follow a white noise process.

\section{Two Macroeconomic Forecasting Applications}\label{sec:appl}
We investigate the performance of BPS-RT in two forecasting exercises. 
In the first application, we combine predictive densities of GDP growth for the euro area (EA) produced by individual professional forecasters participating in the ECB Survey of Professional Forecasters (SPF). Beyond its intrinsic interest, this data set is a good testing ground for BPS-RT because it has been used before when comparing alternative density forecast combination methods; see \cite{diebold2022aggregation}, \cite{conflitti_optimal_2015}, and \cite{chernis2023BPS}. Second, we forecast US inflation using a set of autoregressive distributed lag (ADL) regression models. This data set and model set has been used by \cite{stock2003forecasting} and \cite{rossi2014evaluating}, the latter using a similar ADL strategy to create each of the agent's forecast densities. 

These two applications differ not only geographically and in terms of target variables, but also in the number of agents and the nature of the forecast densities the agents provide. The EA GDP growth application features a relatively small number of subjective, most likely judgment-informed, forecasts \citep{european_central_bank_results_2019} that are provided in the form of histograms (with $J = 14$). In contrast, the US inflation application uses a large number of model-based predictive densities, which are continuous and produced with distinct ADL regressions (with $J = 56$). Further details on the design of both applications are provided in the subsequent sub-sections \ref{ssec:easpf} and \ref{ssec:usadl}. Both applications' evaluation samples cover the global financial crisis, the euro area crisis, and the COVID-19 pandemic. Taken together, these two applications enable a comprehensive assessment of BPS-RT. 


\subsection{BPS-RT Specifications}\label{ssec:spec}
We experiment with several different specifications of BPS-RT to draw out how density forecast accuracy varies with the characteristics of the specific synthesis function used. In broad strokes, we look at the importance of time variation, in both weights and volatility, the number of trees, and the choice of weight modifiers. Accommodating temporal instabilities \citep[for example, see][]{RossiJEL} is important in macro-modeling, and so is a natural subject of inquiry, while the number of trees is an important aspect of specifying BART models. Being able to specify weight modifiers is an attractive feature of BPS-RT and allows the combination weights to change based on information exogenous to the individual agents but known to the BPS decision maker. Hence this is also a key area of inquiry. 

We  accordingly investigate the following four specifications of BPS-RT distinguished by their choice of weight modifier(s) and whether that choice introduces cross-sectional (which we label C) or cross-sectional and time variation (which we label TC) in the combination weights seen in (\ref{eq:synthesis}).

\begin{itemize}
    \item \textbf{BPS-RT(C): AVG.-SCORES}: This specification uses as weight modifiers for the cross-sectional varying coefficients ($\bm z^{\gamma}_{j}$) measures of each agent's historical (ex post) forecast accuracy. Specifically, to capture past point and density forecast accuracy, it considers model-specific averages of the mean squared forecast errors (MSEs) and the CRPSs, respectively. These averages are computed recursively to reflect information known only to $\mathcal{D}$ in real time and could help the synthesis function distinguish between ``good'' and ``bad'' forecasters. This specification assumes constant weights over a given estimation window, although the weights are updated recursively through the evaluation period.  
    \item \textbf{BPS-RT(TC): EXO.-IND}: This specification selects as weight modifiers application-specific exogenous indicators. These vary over time but not over the cross-section, implying that they are the same for each agent. These indicators are intended to provide a signal on the state of the economy, prompting BPS-RT to re-weight the individual agents while simultaneously fostering a certain degree of synchronization among them. For example, during periods of high economic uncertainty, financial stress, or elevated inflation expectations, BPS-RT may weight a subset of models more heavily. In the EA SPF application, we use the European economic policy uncertainty (EPU) index of \cite{BakerBloomDavis}, available via \url{https://www.policyuncertainty.com}. In recessions, their uncertainty measure rises; so allowing the combination weights to depend on uncertainty enables them to move with the business cycle. In the US inflation application, we consider measures of inflation expectations and financial conditions. Both variables have been considered in the inflation-at-risk literature \citep{lopez_salido_inflation_at_risk}. Specifically, we consider households' one-year-ahead inflation expectations from the University of Michigan survey and, as a broad measure of financial conditions, the Chicago Fed's national financial conditions index (NFCI). Both measures are available from the Federal Reserve Bank of St. Louis (\url{https://fred.stlouisfed.org}). In our empirical application, where we use a direct forecast design, we lag these exogenous indicators by the forecast horizon $h$, to acknowledge the reality that we do not observe values for them in a future period, $t+h$, but only have information up to $t$. To catch any other time effects, in both applications we also consider a time trend ($t = 1, \dots, T$). 
    \item \textbf{BPS-RT(TC): FEATURES}: In addition to the scores discussed above, this specification considers statistical ``features'' of each agent's predictive density, known to $\mathcal{D}$ in real time. Specifically, we consider the first four moments of each agent's predictive density and the cross-sectional dispersion of the agents. The latter is measured by the standard deviation (at time $t$) across the $J$ agents' (models') mean forecasts. Consideration of these features allows the density combination weights, in effect, to cluster to reflect the marginal properties of the individual forecasts and their disagreement. For example, it may well be that high (ex ante) uncertainty forecasters should be weighted similarly. Besides these features, we consider historical point and density forecast performance, measured by lagged MSEs and lagged CRPSs. For this RT(TC) specification, it is noteworthy that both score measures are used in such a way that the averages of these lagged scores impact the time-invariant part of the weights $\bm \gamma$, while the plain lagged scores impact the weights through the time-varying part $\bm \beta_t$. 
    \item \textbf{BPS-RT(TC): ALL}: This specification includes all the previously discussed weight modifiers. By looking at the weight modifiers individually, and adding features sequentially, we can assess the marginal benefit of each weight modifier. 
\end{itemize}

For each of these four versions of the model, we consider models with SV and homoskedastic errors and we allow the BART specification to either have a single tree ($S = 1$), leading to a Bayesian regression-tree specification \citep[see][]{chipman1998bayesian}, or a large number of trees ($S = 250$), leading to BART. In traditional Bayesian implementations using trees for nonlinear regression, such as \cite{chipman2010bart}, it is generally found that increasing the number of trees, starting at $S=1$, leads to an improvement in forecast performance. But this improvement tends to peter out when the number of trees gets moderately large. The conventional wisdom is that the precise choice of the number of trees is not that important, provided that too small a value is not chosen. This may not be the case in BPS, since the data may prefer to have weights that are reasonably constant over time and change only occasionally. Hence, we choose to focus on single-tree specifications and BART to model the weights in BPS. As we shall see, we find that single-tree methods tend to forecast more accurately. As benchmarks in the forecasting exercises below, we consider both BPS-CONST and BPS-RW (as defined in Section 2.1.2).

\subsection{Forecasting EA Output Growth Using the Survey of Professional Forecasters}\label{ssec:easpf}
 \color{black}
The ECB has been producing the SPF since $1999$. The ECB SPF is the longest running EA survey of macroeconomic forecasts. Each quarter, the survey elicits from a panel of professional forecasters point and probability forecasts of EA inflation and GDP growth at various horizons.\footnote{For a full description of the EA SPF, see \cite{garcia_introduction_2003}.} We consider the two-quarter-ahead forecasts of year-on-year EA GDP growth. On average, there are $50$ responses a quarter from a survey panel of over $100$ professional forecasters. 

There are a couple of features of the forecaster-level density forecasts from the ECB SPF that we have to address in order to combine them. First, survey respondents provide their probability forecasts over given (fixed) ranges. That is, they produce histogram rather than continuous density forecasts. For example, in the $1999$Q$1$ survey, forecasters were instructed to provide their probability forecasts over $10$ bins. The first bin was GDP growth less than 0 percent, with the bins then increasing in intervals of $50$ basis points, until the tenth bin of higher than 4 percent growth. To accommodate the discretized nature of these probability forecasts, rather than fit a continuous density to the histogram (that may or may not have a good fit), we use the histogram forecast data ``as is." We do this by, within our BPS approach, drawing samples for each forecaster directly from the histograms. Details of our algorithm, which involves a Metropolis-Hastings step, are given in Appendix A.2.  Our sampling approach changes over time to capture the fact that the bin definitions have been moved over time. In particular, after shocks such as the global financial crisis and COVID-19, the ECB shifted the bins to allow forecasters to say more about the probabilities in what were, prior to the survey change, the extremes of the distribution. We also have to take a stand on the open intervals at the bottom and top of the histogram. We set the end-points for the histograms equal to the outer bin plus or minus (depending on whether we are at the top or bottom of the histogram) two standard deviations of GDP growth, as estimated using the vintage of GDP data available at the time the forecast was made. 

Second, forecasters enter and exit the panel. This means that the panel is unbalanced. We follow \cite{diebold2022aggregation} in constructing the longest consistent panel possible by dropping forecasters who are regular non-responders and then filling in the occasional missing values for the remaining forecasters. Specifically, we drop forecasters who have not responded for five or more consecutive quarters. This results in a panel of $14$ forecasters. Any missing forecast data for these $14$ forecasters are estimated using a Normal distribution based on the unconditional distribution of GDP growth as estimated in real time.\footnote{We differ from \cite{diebold2022aggregation} in two ways. First, they interpolate missing forecasts based on historical performance. Second, we have a different number of forecasts because we use a different sample and we forecast GDP growth instead of inflation.} 


We then take these 14 forecasters' densities and carry out a recursive out-of-sample evaluation of the alternative BPS specifications over the sample $2005$Q$2$ through $2021$Q$1$. To do this, we first estimate the BPS combinations on a set of training samples that comprise a sequence of expanding windows of GDP and density forecast data. The GDP data used in the training sample are that vintage of GDP data available to the forecasters when they made their forecasts. The first training sample uses forecasts from the five-year period targeting GDP outturns from $1999$Q$3$ through $2004$Q$2$. These forecasts are taken from the surveys administered between $1999$Q$1$ and $2003$Q$4$. Given its publication lags and our desire to approximate the information set available at the time the SPF forecasts are publicly available, the GDP outturns required to estimate the BPS synthesis function over this training sample are taken from the $2004$Q$4$ vintage. This estimated synthesis function then uses the $2004$Q$4$ survey to forecast (out-of-sample) $2005$Q$2$. The training sample and vintage of GDP data are then extended by one quarter, and forecasts are produced for $2005$Q$3$. This process is continued until forecasts are produced for $2021$Q$1$. This set of out-of-sample BPS density forecasts is then evaluated against GDP outturns taken from the June $9$, $2021$ vintage.

\subsection{Forecasting US inflation Using a Set of Indicators from FRED-QD}\label{ssec:usadl}
We follow \cite{rossi2014evaluating} and construct density forecasts of US inflation using a set of autoregressive distributed lag (ADL) models. Each ADL model considers 1 of $27$ indicators taken from the FRED-QD data set \citep{McCrackenNg}, which is commonly used when forecasting macroeconomic aggregates such as inflation in the US. The selected indicators capture movements in assets prices, measures of real economic activity, wages and prices, and money. This rich and diverse set of economic indicators allows the ADL density forecasts of US inflation to display significant heterogeneity. Table \ref{tab:data} in the Appendix provides an overview of the variables used as exogenous predictors and the transformations applied to ensure their stationarity.

We then use each of these ADL models to produce direct forecasts for quarter-on-quarter consumer price (\texttt{CPIAUCSL}) inflation one-quarter-ahead ($h = 1$) and one-year-ahead ($h = 4$). Specifically, for each indicator, $x_{jt}$, for $j=1,...,27$, we estimate the set of ADL models:
\begin{equation}\label{eq:adl}
\pi_{t+h} = \rho_{\pi} \pi_t + \alpha_{\pi} x_{jt} + \varepsilon_{\pi, t+h}, \quad \varepsilon_{\pi, t+h}\sim\mathcal{N}\left(0,\sigma_{\pi, t+h}^2\right), 
\end{equation}
where $\pi_t$ is inflation, $\rho_{\pi}$ is the autoregressive coefficient, and $\alpha_{\pi}$ denotes the coefficient related to the $j\textsuperscript{th}$ exogenous indicator.\footnote{For notational ease, we do not use $j$ subscripts to distinguish parameters in Eq. (\ref{eq:adl}).} We supplement these $j=1,\dots,27$ models with a $28\textsuperscript{th}$ model (the AR(1) model) that sets $x_t = 0$ in Eq. (\ref{eq:adl}). We also allow $\sigma_{\pi, t+h}^2$, the error variance, to be both time-varying and constant. Hence, we estimate $28$ models both with and without SV, delivering, in total, a set of $56$ individual models whose density forecasts we then combine using BPS. All 56 models are estimated using standard Bayesian techniques. Details are provided in Appendix \ref{app:adl}.

We first estimate these models on a training sample from $1970$Q$1$ to $1989$Q$4$. We then iterate forward using a rolling estimation window of $80$ quarters to account for possible structural changes in the US economy. The first ten years of forecasts ($1990$Q$1$ to $1999$Q$4$) are used as a training window to estimate the BPS synthesis functions. The combined forecasts are then assessed on the evaluation sample $2000$Q$1$ to $2022$Q$4$. This evaluation period includes distinct economic periods characterized by different inflation dynamics, including the dotcom crash, the global financial crisis, the COVID-19 period, and the post-pandemic inflationary period.

\subsection{Empirical Results}\label{ssec:results1}
We break the empirical results into three parts presented in the following three sub-sections. First, we evaluate the relative and absolute density forecast accuracy of BPS-RT. Second, we examine why BPS-RT forecasts more accurately than the benchmarks by comparing features of their forecast densities. Third, we demonstrate aspects of interpretability of BPS-RT by examining how BPS-RT can be used to understand the role of model incompleteness, agent clustering, and the time-varying importance of the different effect modifiers.

\subsubsection{Forecast Accuracy}  We evaluate forecast accuracy in several ways. We first evaluate the point (conditional mean) forecasts, extracted from the combined densities, using the root mean squared forecast error (RMSE) loss function. Second, we evaluate the full predictive densities. We emphasize evaluation of the predictive densities rather than the point forecasts. Since the loss functions of forecast users tend not to be quadratic --- as the density forecast literature \citep[see][]{Aastveit_Review} emphasizes --- it is always important to produce and evaluate complete probabilistic forecasts. We measure the relative forecast accuracy of the forecast densities using two popular metrics: CRPS and a tail-weighted CRPS. Both are loss functions that score the density forecast according to the realization that subsequently materializes. CRPS evaluates the ``whole" density, while tail-weighted CRPS focuses on accuracy in the tails \citep{GneitingRanjan}.\footnote{In the empirical appendix we follow \cite{GneitingRanjan} and break CRPS tails into their left and right tails. See Figures \ref{fig:easpfappend} and \ref{fig:usadlappend} in Appendix \ref{app:B}.} We also test the absolute calibration of the combined density forecasts using the \cite{rossi2019alternative} test on the probability integral transforms (PITs); and we assess the temporal stability of forecast performance using the fluctuation test of \cite{giacomini2010forecast}. The results of both these tests are summarized below, with full results presented in Appendix \ref{app:B}. 

Figures \ref{fig:easpfmain} and \ref{fig:usadlmain} report the relative forecast performance of the different models in the EA GDP growth and US inflation applications, respectively, using the RMSE, CRPS, and CRPS-tails loss functions. Each row in these figures reports the relative (to the BPS-RW benchmark) performance of the four BPS-RT specifications as differentiated by whether they use a single tree or 250 trees and whether they have SV or homoskedastic errors. The four columns in the figures refer to which set of weight modifiers is used. 

Looking first at the RMSE panel in Figure \ref{fig:easpfmain} for EA GDP growth, we see little difference between the alternative BPS-RT specifications in terms of their point forecast accuracy. The accuracy of the BPS-RT specifications also tends to be similar to that of BPS-CONST and BPS-RW, with gains/losses in general only around 3 percent. This supports the stylized fact from the forecasting literature that equal-weighted combinations of point forecasts are hard to beat \citep[see][]{TIMMERMANN2006}. Turning to US inflation (Figure \ref{fig:usadlmain}), we do see in the RMSE panel that some of the tree-based methods now improve upon the point forecast accuracy of both benchmarks and in a manner that is statistically significant. Of particular note is the superior performance of the single-tree models, which almost always outperform the more complicated $250$-tree models. We discuss this finding further below.

The CRPS panels in both Figures \ref{fig:easpfmain} and \ref{fig:usadlmain} reveal yet more of a payoff to using BPS-RT, certainly relative to BPS-RW, when we evaluate the whole density. Many of the forecast accuracy gains for BPS-RT are statistically significant. An implication of this finding is that BPS-RW's assumption that the combination weights follow a random walk is not supported by the data. But BPS-CONST, especially when BPS allows for SV, remains competitive for EA GDP growth.  

The CRPS and CRPS tail results echo those under RMSE loss in concluding that single-tree structures, $S=1$, are almost always preferred to $S=250$. The fact that a single-tree model produces more accurate forecasts  contrasts with the conventional wisdom in the wider BART literature; see \cite{chipman2010bart}.  In our case, however, we model the weights, rather than the observed outcomes,  nonparametrically and hence the implied conditional mean relation (see  \autoref{eq: synth_integrated}) introduces more restrictions relative to a standard BART model and hence lessens the risk of overfitting.

While the benefits of allowing for SV are well established in the density forecast literature \citep[see][]{clark2011}, allowing for SV in the BPS combination does not obviously improve the density forecasts from BPS-RT. But recall, and we touch on this again below when showing that these models in fact receive higher combination weights, in the US inflation application half of the components models themselves allow for SV. 


We now focus on comparing forecast accuracy across the first four columns of both Figures \ref{fig:easpfmain} and \ref{fig:usadlmain}. This comparison reveals that the choice of weight modifier does affect forecast accuracy. It is not always the case that using more weight modifiers delivers more accurate forecasts. The benefit of different modifiers varies by application and by which row (which of the four BPS-RT specifications) is consulted. 

Finally, we summarize the results from both the PITs calibration tests and the fluctuation tests. These results are reported in the online appendix for space reasons. The PITs plots (see Figure \ref{fig:pits}) show that the BPS-RT densities are well calibrated and especially so when forecasting EA GDP growth or US inflation one-quarter-ahead.
The fluctuation test of \cite{giacomini2010forecast} reveals that there is temporal variation in the relative performance (under CRPS loss) of the preferred BPS-RT model and BPW-RW. Results (see Figure \ref{fig:fluctest})  indicate that the superior performance of BPS-RT in the EA application is due to better forecasting performance toward the end of the global financial crisis. For US inflation, the better accuracy of BPS-RT is explained by its more accurate density forecasts in the post-lockdown inflationary period. 


\begin{figure}[!htbp]
\centering
\caption{Relative forecast accuracy: EA GDP growth.}\label{fig:easpfmain}
\includegraphics[width=0.9\textwidth,keepaspectratio]{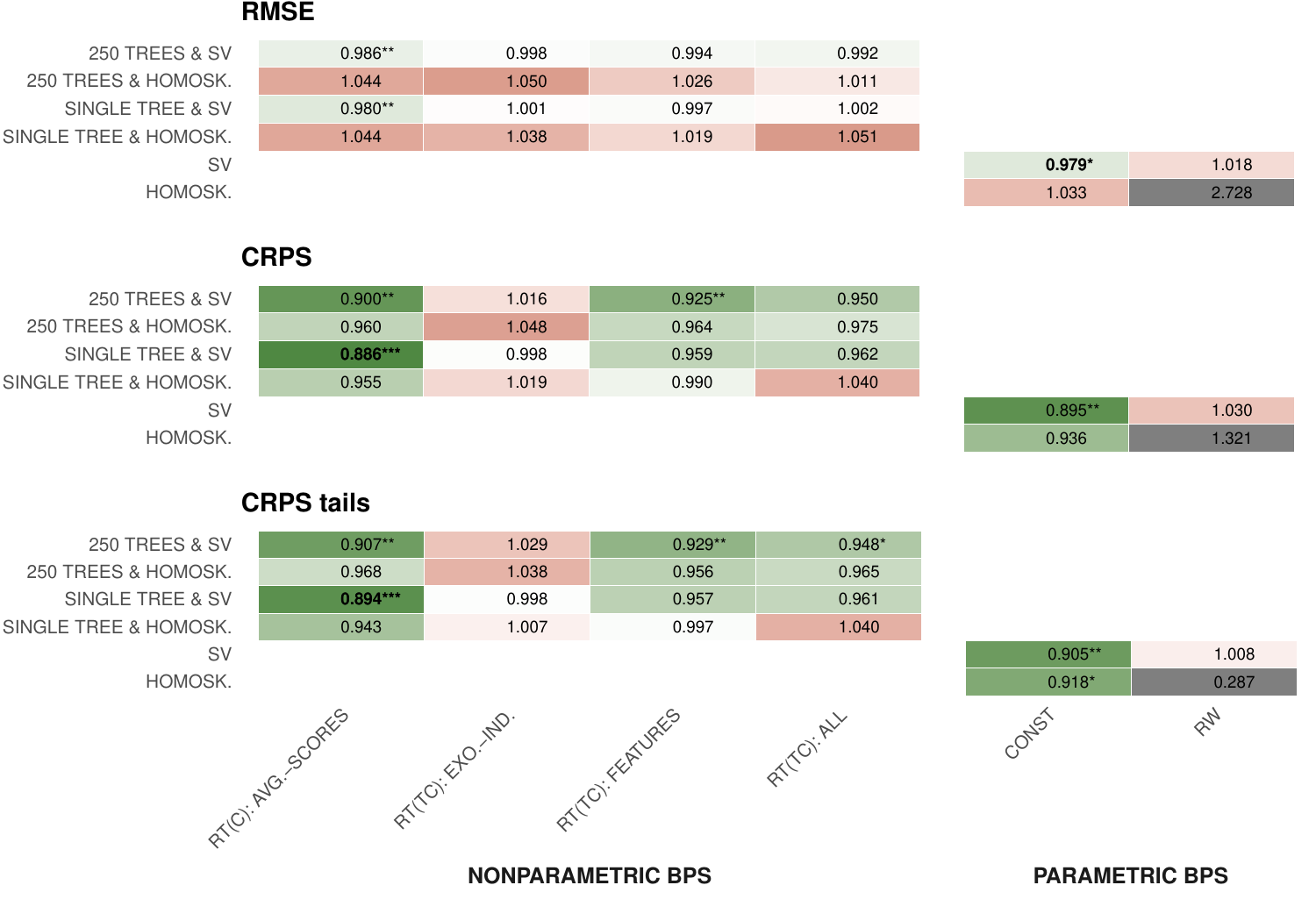}
\caption*{\scriptsize \textbf{Notes:} This figure shows root mean square error (RMSE) ratios, (unweighted) continuous ranked probability score (CRPS) ratios, and a variant of quantile-weighted CRPS ratios that focuses on the tails. The gray shaded entries give the actual scores of our benchmark (BPS-RW with homoskedastic error variances). Green shaded entries refer to models that outperform the benchmark (with the forecast metric ratios below one), while red shaded entries denote models that are outperformed by the benchmark (with the forecast metric ratios greater than one). The best performing model specification by forecast metric is given in bold. Asterisks indicate statistical significance of the \cite{diebold1995dmtest} test, which tests equal forecast performance for each model relative to the benchmark, at the $1$ ($^{***}$), $5$ ($^{**}$), and $10$ ($^{*}$) percent significance levels.}
\end{figure}

\begin{figure}[!http]
\centering
\caption{Relative forecast accuracy: US inflation.}\label{fig:usadlmain}
\begin{minipage}{\textwidth}
\centering
(a) One-quarter-ahead $(h = 1)$
\vspace*{2pt}
\end{minipage}
\begin{minipage}{\textwidth}
\centering
\includegraphics[width=0.8\textwidth,keepaspectratio]{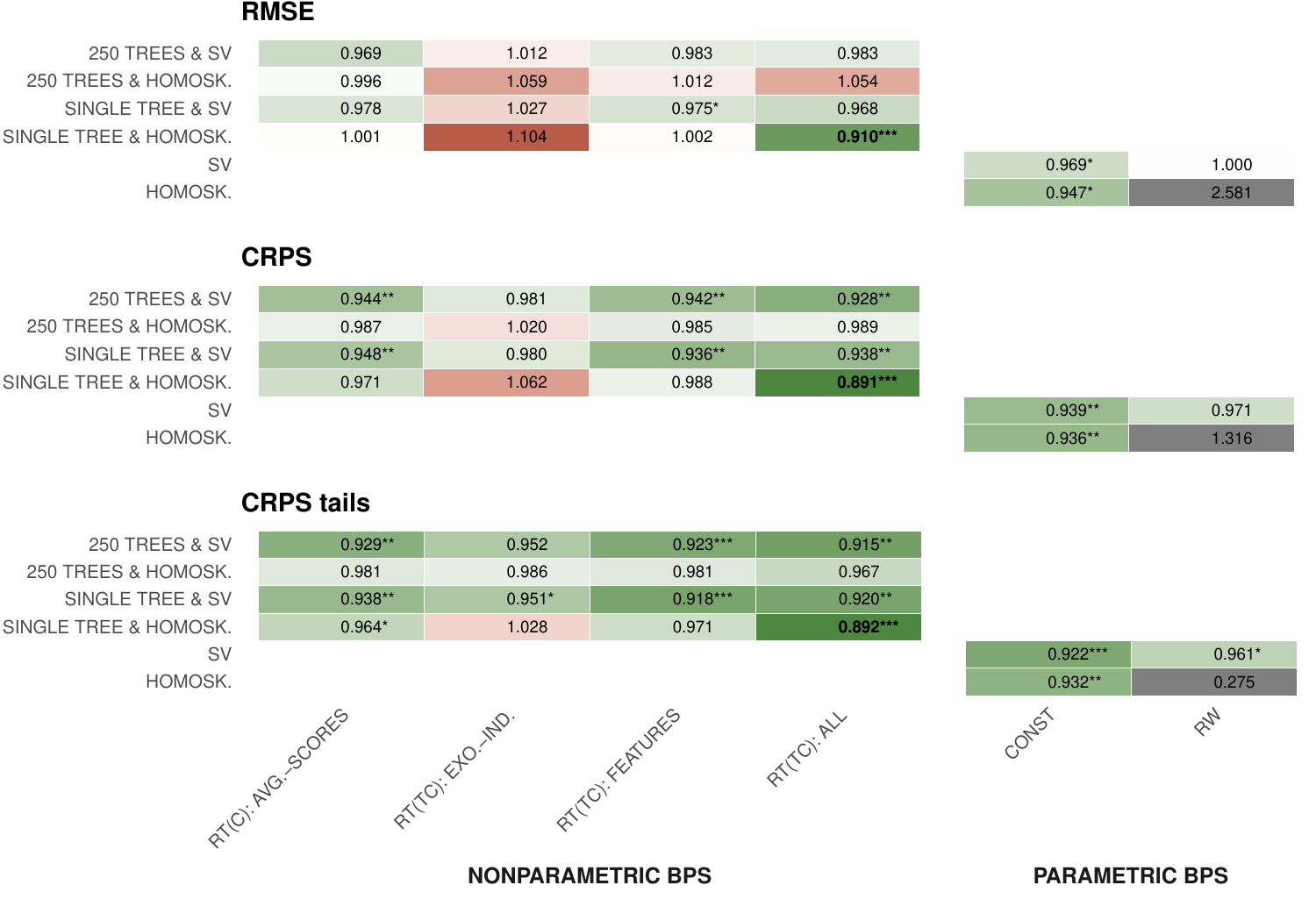}
\end{minipage}
\begin{minipage}{\textwidth}
\centering
\vspace*{2pt}
(b) One-year-ahead $(h = 4)$
\vspace*{2pt}
\end{minipage}
\begin{minipage}{\textwidth}
\centering
\includegraphics[width=0.8\textwidth,keepaspectratio]{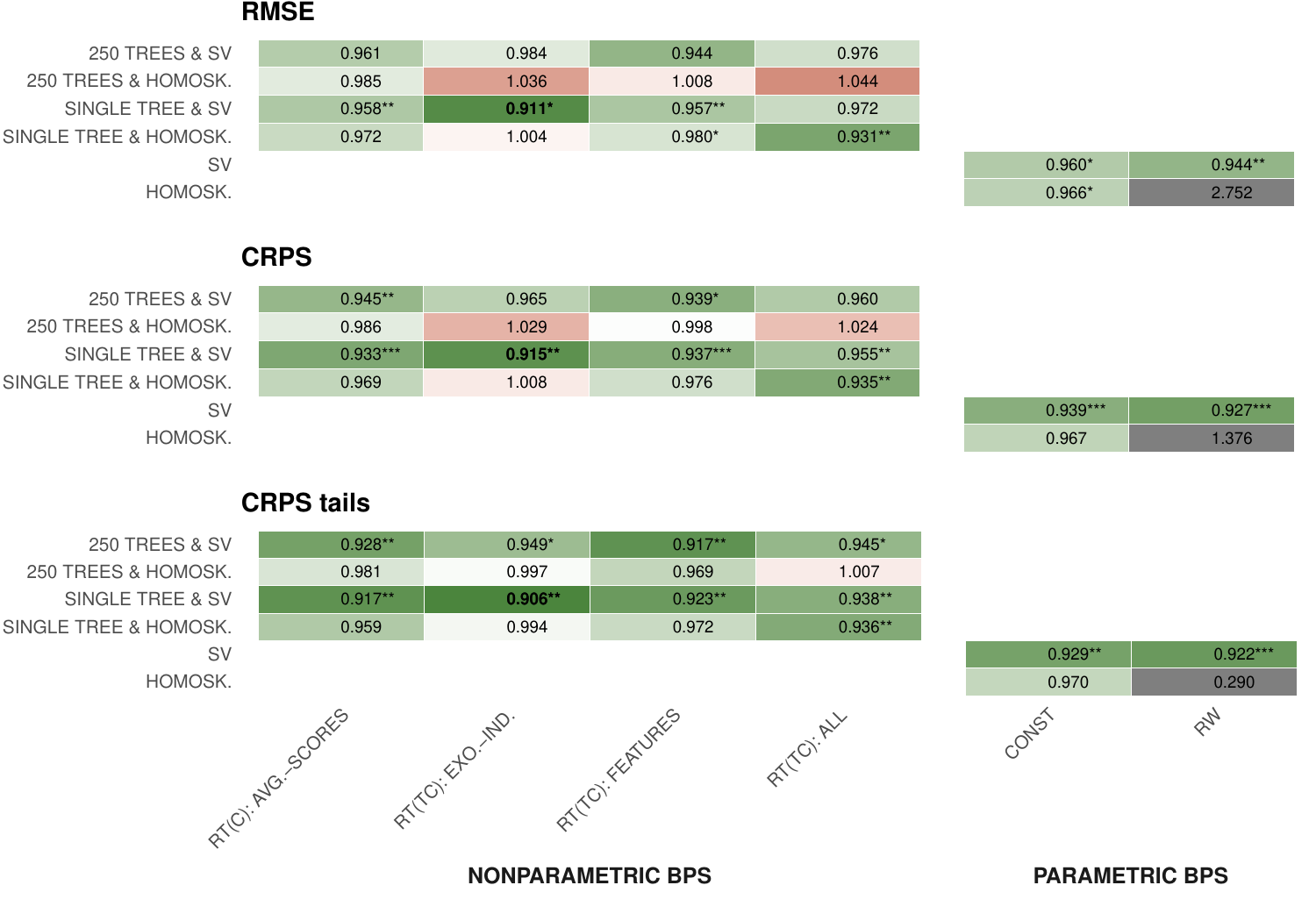}
\end{minipage}
\caption*{\scriptsize \textbf{Notes:} This figure shows root mean square error (RMSE) ratios, (unweighted) continuous ranked probability score (CRPS) ratios, and a variant of quantile-weighted CRPS ratios that focuses on the tails. The gray shaded entries give the actual scores of our benchmark (BPS-RW with homoskedastic error variances). Green shaded entries refer to models that outperform the benchmark (with the forecast metric ratios below one), while red shaded entries denote models that are outperformed by the benchmark (with the forecast metric ratios greater than one). The best performing model specification by forecast metric is given in bold. Asterisks indicate statistical significance of the \cite{diebold1995dmtest} test, which tests equal forecast performance for each model relative to the benchmark, at the $1$ ($^{***}$), $5$ ($^{**}$), and $10$ ($^{*}$) percent significance levels.}
\end{figure}

\clearpage

\subsubsection{Properties of the BPS-RT Density Forecasts}
In this section we examine how and why BPS-RT forecasts more accurately. We focus on the best performing (most accurate) model in each application and compare its forecast densities to those of the benchmark model, BPS-RW.\footnote{As seen from Figures \ref{fig:easpfmain} and \ref{fig:usadlmain}, in the EA GDP growth application, the ``best" BPS-RT specification has a single tree and SV and uses average scores as effect modifiers (i.e., \texttt{RT(C): AVG.-SCORES}). For the US inflation application, the ``best'' BPS-RT specification has a single tree, homoskedastic errors, and the full set of weight modifiers (i.e., \texttt{RT(TC): ALL}).} 

\begin{figure}[!htbp]
\centering
\caption{Difference in probabilities between BPS-RT and BPS-RW}\label{fig:diebold_plot}
\begin{minipage}{\textwidth}
\centering
(a) EA GDP growth  
\vspace*{2pt}
\end{minipage}
\begin{minipage}{\textwidth}
\centering
\includegraphics[width=0.7\textwidth,keepaspectratio]{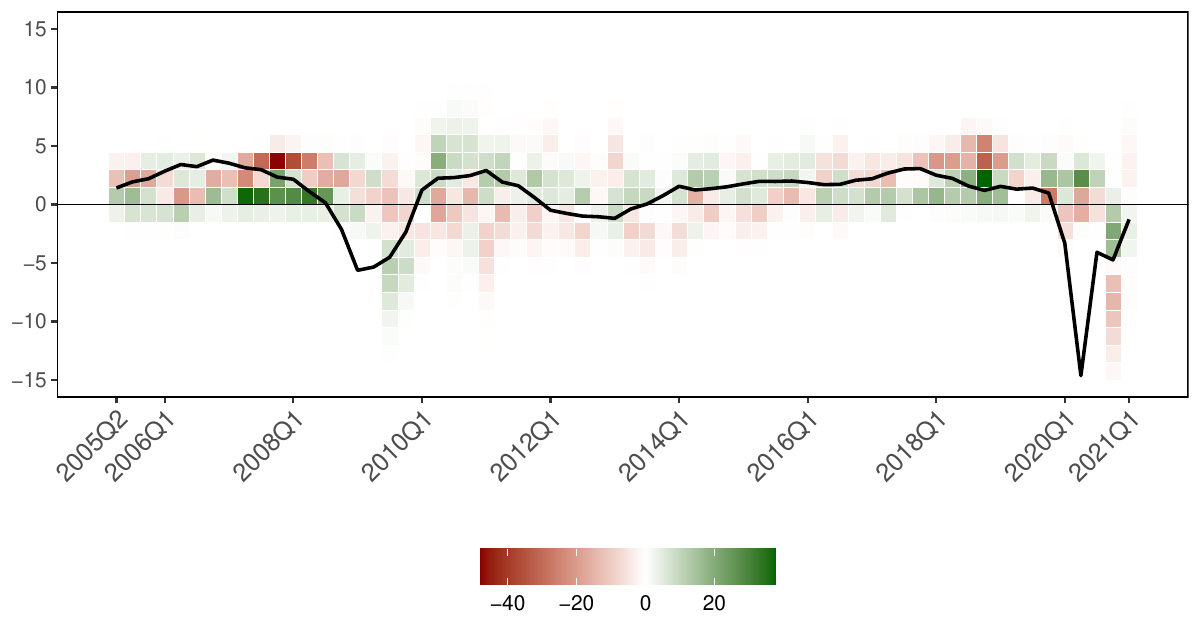}
\end{minipage}
\begin{minipage}{\textwidth}
\centering
\vspace*{5pt}
(b) One-quarter-ahead US inflation ($h = 1$)
\vspace*{2pt}
\end{minipage}
\begin{minipage}{\textwidth}
\centering
\includegraphics[width=0.7\textwidth,keepaspectratio]{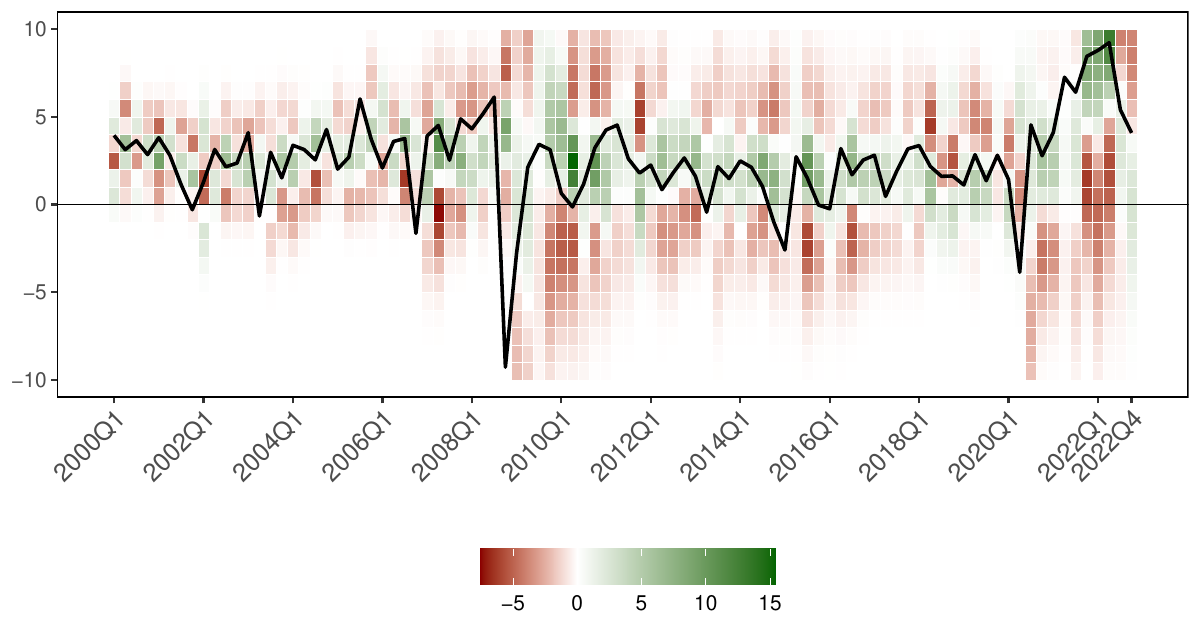}
\end{minipage}
\begin{minipage}{\textwidth}
\centering
\vspace*{5pt}
(c) One-year-ahead US inflation ($h = 4$) 
\vspace*{2pt}
\end{minipage}
\begin{minipage}{\textwidth}
\centering
\includegraphics[width=0.7\textwidth,keepaspectratio]{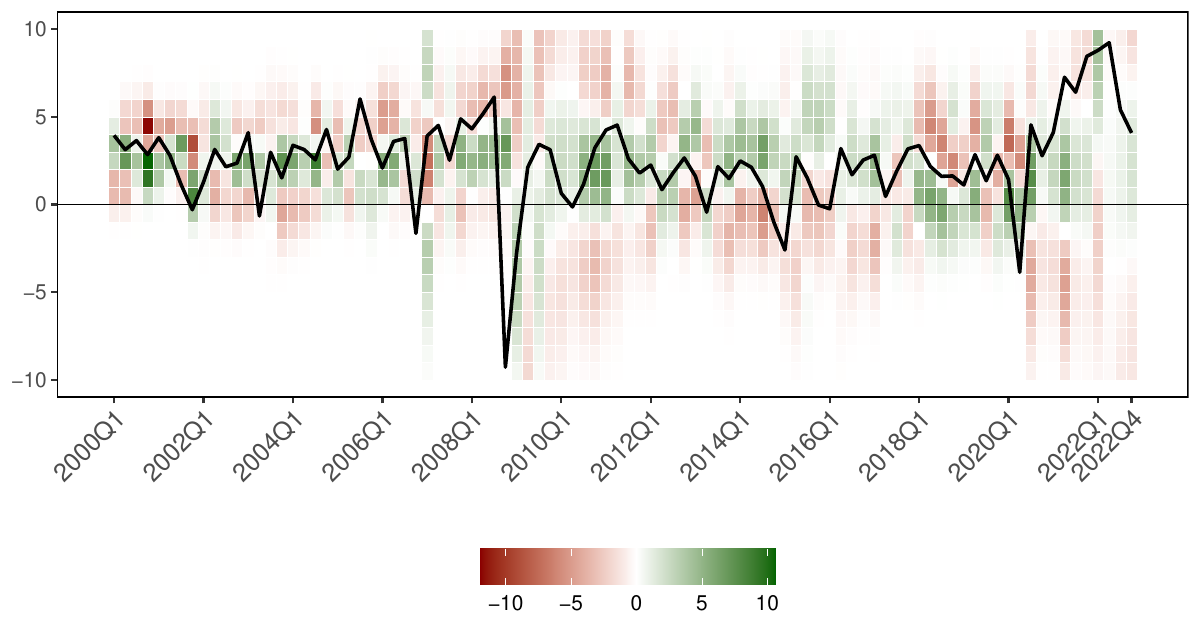}
\end{minipage}
\caption*{\scriptsize \textbf{Notes:} This figure shows the difference in probabilities between the best performing BPS-RT model (in terms of CRPS) and BPS-RW. We define a grid of possible values for EA GDP growth ranging from $-15$ percent to $15$ percent with increments of $1.5$ percent, while we define a grid of possible values for US inflation ranging from $-10$ percent to $10$ percent with increments of $1$ percent. Green (red) shaded cells indicate that the best performing model adds (subtracts) probability relative to the benchmark in the respective region.}
\end{figure}

Figure \ref{fig:diebold_plot} shows a heat map of the difference in probabilities, in intervals of $1.5$ percentage point for EA GDP growth and of $1$ percentage point for US inflation, between BPS-RT and BPS-RW. Green (red) shading indicates that BPS-RT adds (subtracts) probability relative to BPS-RW in that interval. This is the approach pioneered by \cite{diebold2022aggregation} as a way of visualizing the differences between competing density forecasts.\footnote{For an alternative but complementary visualization, Figure \ref{fig:preddens} in Appendix \ref{app:B} shows the temporal evolution of the underlying density forecasts from BPS-RT and the benchmark BPS-RW model over the EA and US evaluation samples.} 

Panel (a) of Figure \ref{fig:diebold_plot} shows that, in general, BPS-RT predictions are less dispersed than BPS-RW with more mass near the subsequent outcomes. Additionally, the BPS-RT density adds probability to low GDP growth outturns prior to the financial crisis and also forecasts higher growth than BPS-RW in both the post-global financial crisis recovery and the rebound from the COVID-19-induced recession.\footnote{As shown in Figure \ref{fig:skew} in the online appendix, in moving the probability mass from the centers to the left tail of the forecast density, BPS-RT captures asymmetries in the forecast densities. While there is some evidence of heightened downside risk asymmetries to GDP growth in the course of the financial crisis, consistent with the growth-at-risk literature \citep{Adrian2019}, the evidence for negative skew is stronger still during the COVID-19 pandemic.} Panels (b) and (c) of Figure \ref{fig:diebold_plot} show the analogous plots for US inflation. Similar to panel (a), BPS-RT places more mass closer to the outturn and produces forecasts that are, in general, less disperse. Moreover, BPS-RT adjusts much more quickly to the increase in inflation  post-pandemic, both one-quarter- and one-year-ahead, attributing a higher probability to these outturns than BPS-RW. Consistent with the evidence in \cite{rossi2014evaluating} that combinations of predictive densities for US inflation appear to be approximately Gaussian, the inflation forecast densities from BPS-RT also tend to be symmetric (see Figure \ref{fig:skew} in the online appendix), although there is clear evidence of a spike in downside risks in 2011, a time when the Fed was engaged in quantitative easing to combat deflation threats.

\subsubsection{Interpretation: A Deeper Dive into the Mechanics of BPS-RT for US Inflation}

This section discusses how $\mathcal{D}$ can interpret the combined forecasts from BPS-RT. In so doing we continue to focus on the preferred BPS-RT specification in the US inflation application, not least because this is where we observe greater differences across the competing combination strategies. We first show how to quantify the degree of model set incompleteness, as a way of assessing how well the agents (the $J$ forecasting models) that BPS-RT is combining are actually able to forecast. Second, we assess the relative importance of individual weight modifiers in driving BPS-RT.

To measure model set incompleteness we compute an $R^2$-type measure. This estimates the proportion of the variation in $y_{t+h}$ that is explained by the $J$ agents. This measure is computed, for a specific period in the evaluation sample, as the ratio between the variation in the conditional mean in Eq. (\ref{eq:synthesis}) explained exclusively by the BPS-RT component --- which is the conditional mean in Eq. (\ref{eq:synthesis}) without the time-varying intercept $c_{t+h}$ --- and the overall variation of the target variable, $y_{t+h}$. $R^2$ values close to zero signify a high degree of model incompleteness, which means that the agents' forecasts are not informative about the target variable. Instead, the intercept and error term in the BPS synthesis function, Eq. (\ref{eq:synthesis}), explain a large portion of the total variation. In contrast, $R^2$  values close to one indicate that the agents' forecasts are informative and account for the majority of the variation, implying a complete model space.

\begin{figure}[!htbp]
\centering
\caption{Measuring model incompleteness: US inflation}\label{fig:mdlincmpl}
\begin{minipage}{0.49\textwidth}
\centering
(a) One-quarter-ahead ($h = 1)$  \\ 
\end{minipage}
\begin{minipage}{0.49\textwidth}
\centering
(b) One-year-ahead  ($h = 4)$ 
\end{minipage}
\begin{minipage}{0.49\textwidth}
\centering
\includegraphics[width=\textwidth,keepaspectratio]{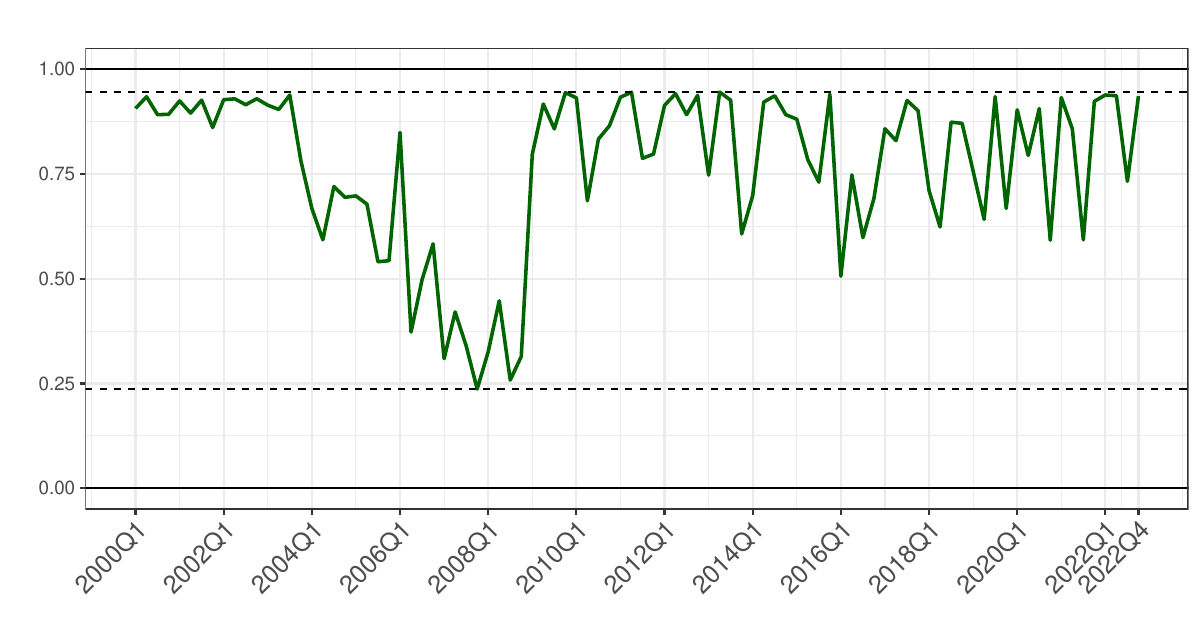}
\end{minipage}
\begin{minipage}{0.49\textwidth}
\centering
\includegraphics[width=\textwidth,keepaspectratio]{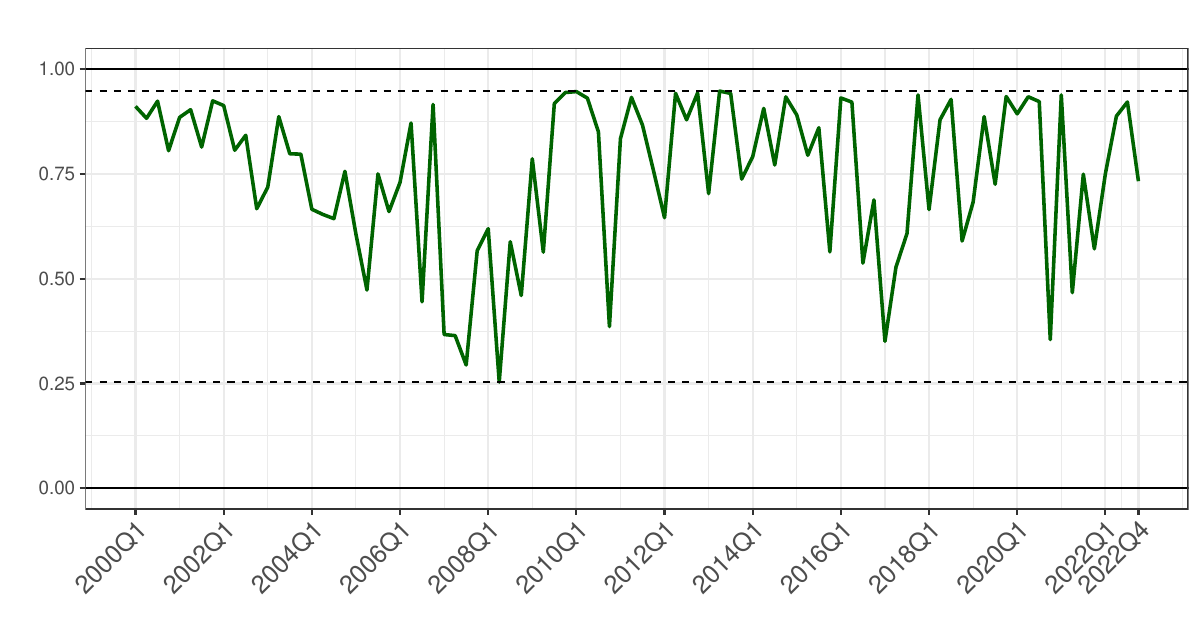}
\end{minipage}
\caption*{\scriptsize \textbf{Notes:} This figure shows the evolution of the model incompleteness measure over time. For each quarter in the evaluation sample this measure is computed for our preferred specification (homoskedastic \texttt{BPS-RT(TC): ALL} with a single tree) as the ratio between the variation explained exclusively by the BPS-RT part (i.e., the conditional mean without the time-varying intercept) and the total variation, which thus can be interpreted as an $R^2$ measure. The green solid lines represent the posterior median of this incompleteness $R^2$, which is bounded between zero and one. Values close to zero suggest that model incompleteness, as measured by the time-varying intercept and the error variance in Eq. (\ref{eq:synthesis}), plays an important role, while values close to one indicate that the BPS-RT part explains most of the variation.}
\end{figure}

Figure \ref{fig:mdlincmpl} plots this $R^2$-type estimate over the evaluation sample. Given that it is computed recursively, quarter-by-quarter, it experiences some volatility. But Figure \ref{fig:mdlincmpl}  still evidences meaningful temporal variations in the degree of model set incompleteness at both forecast horizons. We see higher model incompleteness for the one-year-ahead forecasts than for the one-quarter-ahead forecasts. This is not surprising, as producing longer-horizon forecasts is obviously more difficult. At both horizons, we see increases in model incompleteness during the period 2004-2008, a time of extreme oil price volatility as well as the global financial crisis and in the disinflation period after the 2015 oil price shock. 

Interestingly, there is no clear evidence of an increase in model incompleteness during the post-pandemic rise in inflation, reinforcing the message from Figure \ref{fig:diebold_plot} that BPS-RT was better able to anticipate the 2021 rise in US inflation.  

\newpage

\begin{figure}[!htbp]
\centering
\caption{Number of tree splits for BPS-RT ($S=1$) and relative importance for each weight modifier for US inflation: One-quarter-ahead forecasts} . \label{fig:adlsplits1}
\begin{minipage}{\textwidth}
\centering
(a) \textbf{Number of (total) tree splits:} Time-invariant and time-varying weights
\end{minipage}
\begin{minipage}{\textwidth}
\centering
\includegraphics[width=.9\textwidth,keepaspectratio]{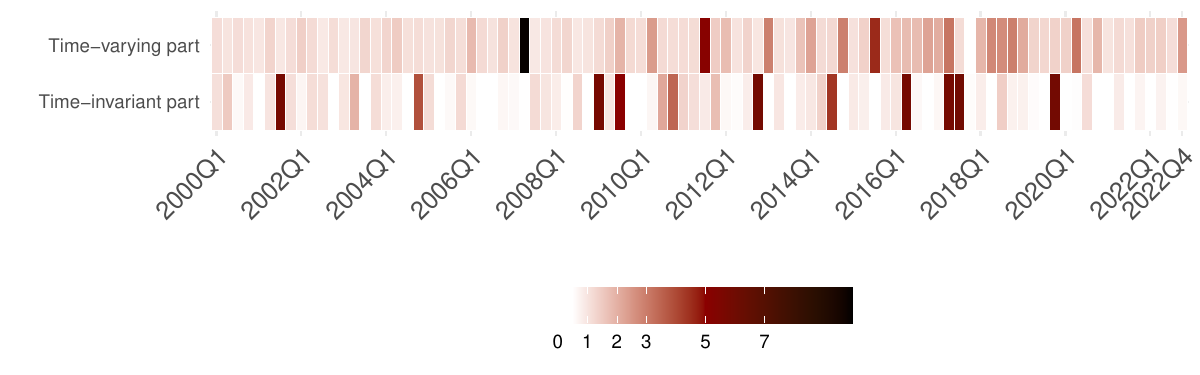}
\end{minipage}
\begin{minipage}{\textwidth}
\centering
(b) \textbf{Time-invariant weights:} Two weight modifiers and their relative importance
\end{minipage}
\begin{minipage}{\textwidth}
\centering
\includegraphics[width=.9\textwidth,keepaspectratio]{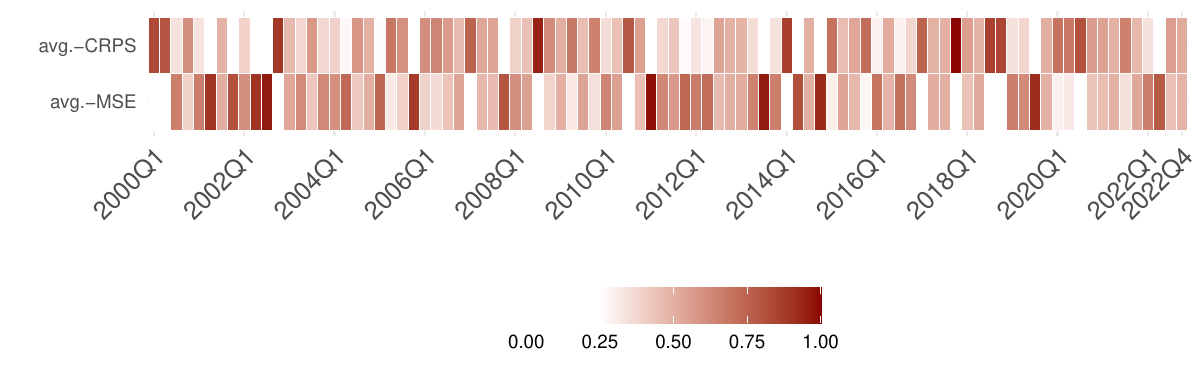}
\end{minipage}
\begin{minipage}{\textwidth}
\centering
(c) \textbf{Time-varying weights:} Ten weight modifiers and their relative importance
\end{minipage}
\begin{minipage}{\textwidth}
\centering
\includegraphics[width=.9\textwidth,keepaspectratio]{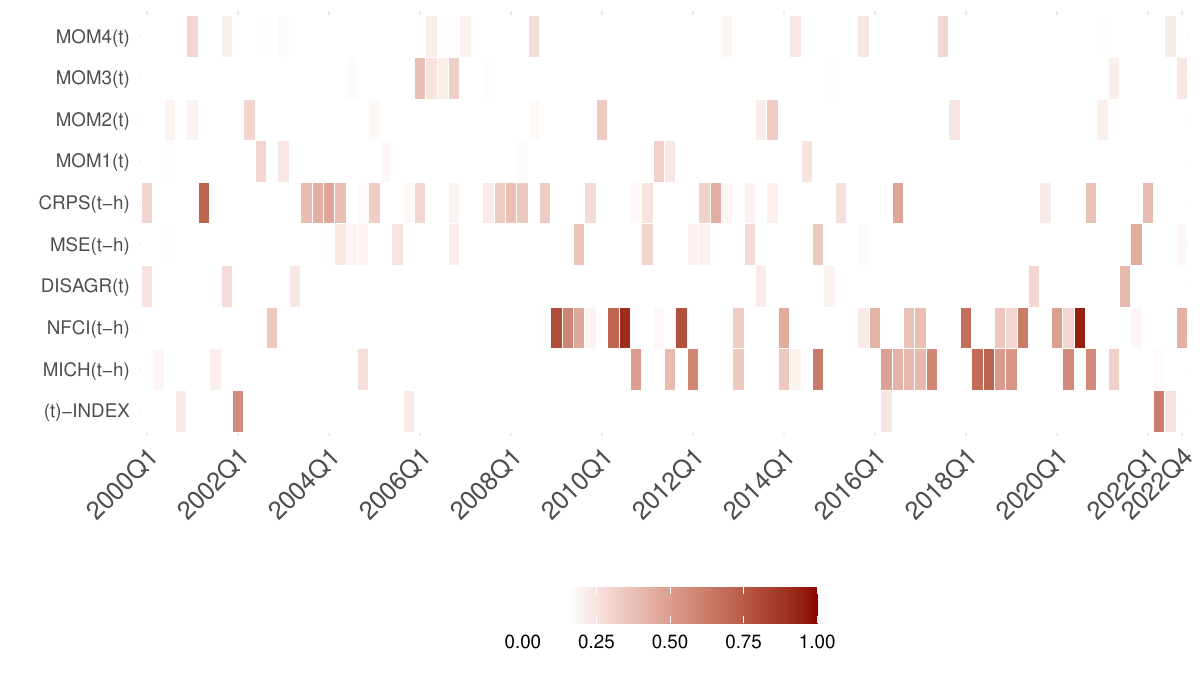}
\end{minipage}
\caption*{\scriptsize \textbf{Notes:} Panel (a) shows the evolution of the total number of tree splits over time, while panels (b) and (c) show the marginal importance of each weight modifier for each quarter in the evaluation sample. Relative importance is defined as the share of the total number of splits. For each quarter in the evaluation sample, we obtain the posterior mean for these measures for our preferred specification (homoskedastic \texttt{BPS-RT(TC): ALL} with a single tree). For the exogenous indicators and the MSE/CRPS scores, $(t-h)$ indicates that these measures are lagged by the forecast horizon $h$, while all other measures can be included contemporaneously.}
\end{figure}

\begin{figure}[!htbp]
\centering
\caption{\textbf{One-year-ahead horizon:} Number of (total) tree splits for our single-tree models and relative importance for each effect modifier. \label{fig:adlsplits4}}
\begin{minipage}{\textwidth}
\centering
(a) \textbf{Number of (total) tree splits:} Time-invariant versus time-varying part
\end{minipage}
\begin{minipage}{\textwidth}
\centering
\includegraphics[width=.9\textwidth,keepaspectratio]{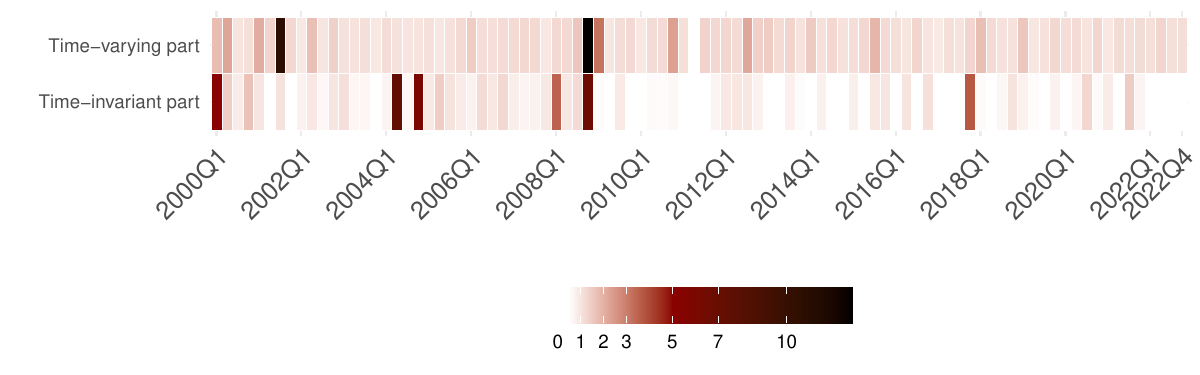}
\end{minipage}
\begin{minipage}{\textwidth}
\centering
(b) \textbf{Time-invariant part:} Two effect modifiers and their relative importance
\end{minipage}
\begin{minipage}{\textwidth}
\centering
\includegraphics[width=.9\textwidth,keepaspectratio]{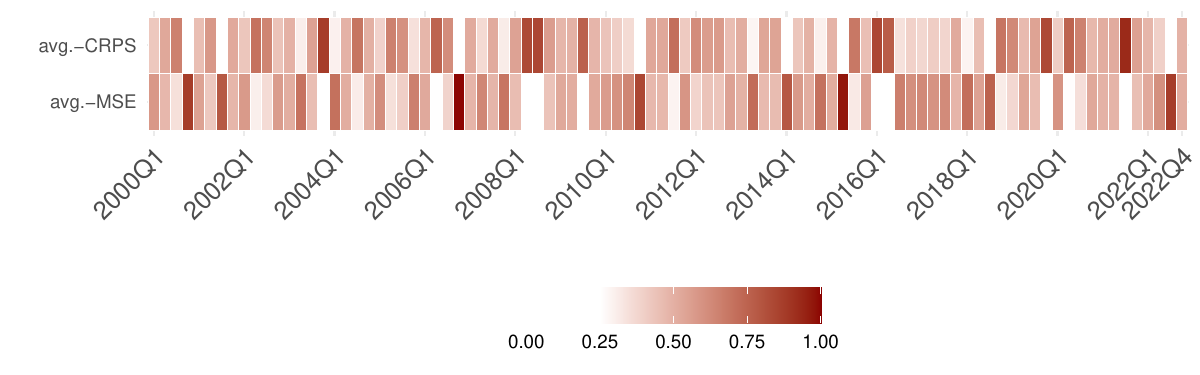}
\end{minipage}
\begin{minipage}{\textwidth}
\centering
(c) \textbf{Time-varying part:} Ten effect modifiers and their relative importance
\end{minipage}
\begin{minipage}{\textwidth}
\centering
\includegraphics[width=.9\textwidth,keepaspectratio]{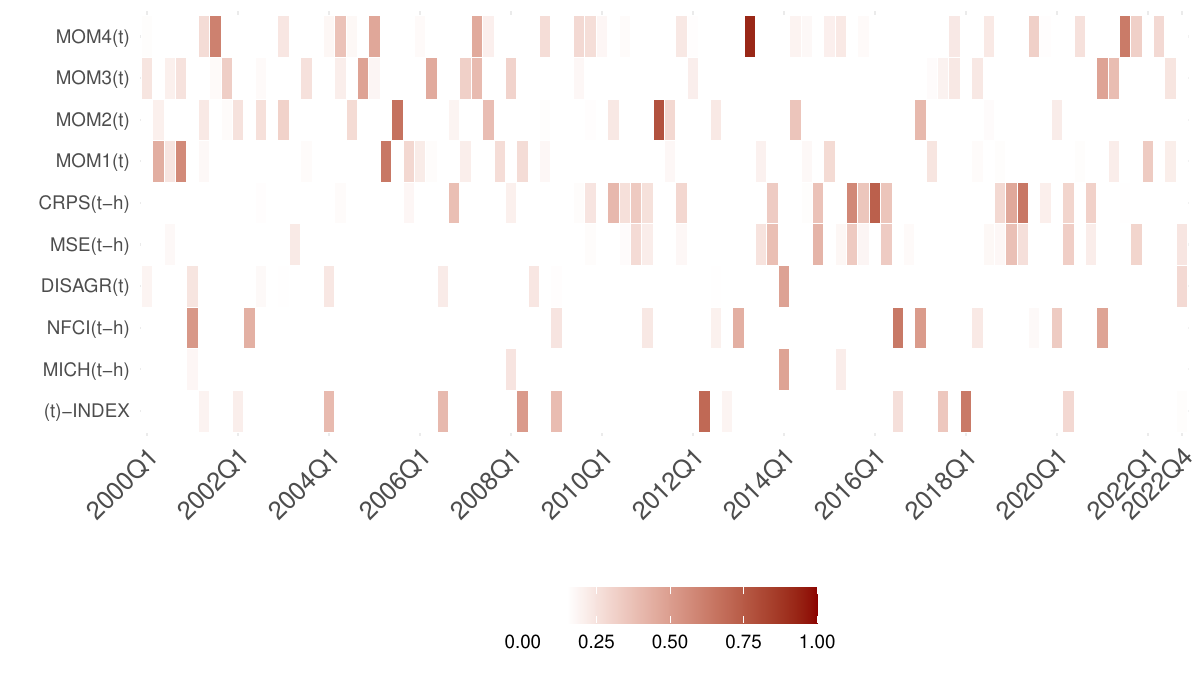}
\end{minipage}
\caption*{\scriptsize \textbf{Notes:} Panel (a) shows the evolution of the total number of tree splits over time, while panels (b) and (c) show the marginal importance of each weight modifier for each period in the evaluation sample. Relative importance is defined as the share of the total number of splits. For each period in the evaluation sample, we obtain the posterior mean for these measures for our preferred specification (homoskedastic \texttt{BPS-RT(TC): ALL} with a single tree). For the exogenous indicators and the MSE/CRPS scores, $(t-h)$ indicates that these measures are lagged by the forecast horizon $h$, while all other measures can be included contemporaneously.}
\end{figure}

We now turn to assessing the relative importance of the individual weight modifiers in driving the density forecasts from BPS-RT. We do so by looking first at the number of tree splits and then by calculating inclusion probabilities for each weight modifier. Inclusion probabilities are calculated as the number of splits associated with the respective weight modifier divided by the total number of splits. For space reasons, we focus our discussion on Figure \ref{fig:adlsplits1}, which examines the weight modifiers for forecasting US inflation one-quarter-ahead. Analogous results forecasting inflation one-year-ahead are reported in Figure \ref{fig:adlsplits4} and summarized below when the conclusions differ markedly from those discussed in greater detail for the one-quarter-ahead forecasts.

We start in panel (a) of Figure \ref{fig:adlsplits1} by plotting the evolution of the total number of tree splits over the evaluation sample. This panel indicates whether variability in the combination weights comes from the time-varying ($\beta_{jt+h}$) or constant component ($\gamma_j$) of BPS-RT. Panel (a) reveals that BPS-RT tends to select a relatively small number of tree splits, especially for the time-invariant weights. Typically for $\gamma_j$ we observe that the posterior mean of the number of tree splits lies between $0.52$ (lower quartile over the evaluation sample) and $1.15$ (upper quartile, with a few more exceptions in the upper tail), while the average over the evaluation sample is $1.28$. On the other hand, the posterior mean number of tree splits for the time-varying weights, $\beta_{jt+h}$, ranges from $1.08$ to $1.68$ (indicating the interquartile range) and has an average of $1.59$ over the evaluation sample. To place these numbers in the context of a single-tree split on, for example, $\gamma_j$ indicates that the combination weights tend to cluster around two distinct prior means. With this in mind, we interpret the results in panel (a) as showing that the combination weights often fall into a handful of clusters that are more likely to be determined by time-specific factors. However, the number of splits is modest, so the weights are relatively stable over time. This finding is consistent with the density forecast combination literature that finds that constant weight combinations can forecast well \citep[see, for example,][]{chernis2023BPS}.

Panels (b) and (c) of Figure \ref{fig:adlsplits1} then show the inclusion probabilities for each of the constant and time-varying weight modifiers. Panel (b) shows the inclusion probabilities for the weight modifiers (CRPS and MSE) used to model the time-invariant combination weights. Neither CRPS nor MSE is obviously more important. Both weight modifiers receive positive and often fairly similar probabilities of inclusion. This implies that BPS-RT does partition models on the basis of their historical forecast accuracy. 

Panel (c) of Figure \ref{fig:adlsplits1} shows the importance of both time-varying weight modifiers. The first thing to notice is that there is much more sparsity in terms of the weight modifiers BPS-RT selects. In the first half of the evaluation sample, we see that features of the individual density forecasts drive the posterior inclusion probabilities. Specifically, we see that the moments of the marginal densities and CRPS, lagged by the forecast horizon $h$, are selected. But in the second half of the evaluation sample, we see the largest proportion of tree splits attributed to the NFCI during and immediately after recessions. The Michigan survey expectations measure also receives more weight after the financial crisis. This is evidence that nonlinear features of BPS-RT are driven by  weight modifiers related to the business cycle. In other words, our BPS-RT model finds that the data support changing the combination weights abruptly with business cycle fluctuations. Finally, the time trend receives a higher weight in the post-COVID period of higher inflation seen in $2021$ and $2022$. This finding indicates that this inflationary episode ––– unprecedented within the sample ––– requires a substantial and rapid adjustment of the combination weights. These required weight dynamics cannot be fully captured by the business cycle weight modifiers. Instead, a time trend (or, more precisely, a time dummy) is ideal for modeling such a regime shift from low to high inflation during this exceptional period.

Finally we summarize the properties of the posterior median estimates of the combination weights that are plotted over the evaluation sample in Appendix \ref{app:B}. We draw out two conclusions for the combination weights estimated when forecasting US inflation one-quarter-ahead (see Figure \ref{fig:adlbeta1}). First, BPS-RT places more weight on those component models with SV, especially toward the end of the evaluation sample. This corresponds to the period when BPS-RT outperforms the BPS-RW benchmark (see Figure \ref{fig:crps-t-adlus}).

Second, among these SV models only a subset receives large, in absolute value, weights. This indicates that there is some pay off, in terms of forecast accuracy, to occasionally placing a significantly higher weight on a small subset of models. Interestingly, some models get large negative weights. This amounts to short-selling those models as a ``hedge'' against the models with higher weights. A roughly similar pattern is seen for the one-year-ahead forecast combination weights seen in Figure \ref{fig:adlbeta4}.\footnote{Figure \ref{fig:betasum} in the online appendix provides additional perspective on the temporal stability of the combination weights by plotting their sum over the evaluation sample. We see that when forecasting US inflation this sum becomes negative during the global financial crisis, indicating how BPS-RT is re-weighting most agents' densities in the face of temporal instabilities. The sum of the weights also spikes upward during the 2021-22 inflationary episode, again indicating how BPS-RT can quickly adapt to temporal change. }

 While this subsection has focused on the US inflation application, we end by returning briefly to the EA GDP growth forecasting application. Figure \ref{fig:eabeta1} in Appendix \ref{app:B}, shows that the combination weights on most individual forecasters from the ECB SPF are, as anticipated given our earlier results, closer to equal than in the inflation application, where there was greater sparsity in the weights. This said, we do still see higher weights on a couple of experts (forecasters 6 and 14). We take this contrasting evidence across the two applications as empirical proof that BPS-RT is sufficiently flexible to adjust to forecasting scenarios that exhibit different dependence structures between the agents' forecasts.

\section{Conclusion}
Within the general BPS framework of \cite{mcalinn2019dynamic}, this paper develops a method for nonparametric density forecast combination using regression trees: BPS-RT. While a handful of papers use nonparametric techniques to combine densities, ours is the first 
to use regression trees. In contrast to most applications of regression trees we model the coefficients, in our case the combination weights, instead of the variables using the regression trees. We show how this aids interpretation, since the combination model remains linear in the parameters. Additionally, regression trees use covariates, or weight modifiers, to drive changes in parameters, in contrast to conventional BPS applications where model parameters follow a random walk. Taken together, our approach is flexible but retains interpretability through linearity and the use of weight modifiers. We explain how BPS-RT can be used to understand the
role of model incompleteness, agent (forecast) clustering, and the time-varying importance of
the different weight modifiers. 

We test the performance of BPS-RT in two different applications -- combining model-based US inflation density forecasts and subjective histogram-based forecasts of euro area GDP growth. We find that, across both applications, BPS-RT forecasts well in terms of both relative and absolute accuracy. Interestingly, and in contrast to standard BART applications, we find that using a parsimonious single-tree specification outperforms models with more trees. Inspecting the best-performing specification, we observe that this superior performance is due to less disperse forecast densities and BPS-RT's ability to better accommodate the shocks associated with the global financial crisis (in the GDP application) and COVID-19 (in the inflation application). Our proposed measure of model set incompleteness suggests that BPS-RT is able to capture much of the post-COVID rise in inflation. Triggered by a rise in the relative importance of the time trend in determining tree splits, itself highlighting the unusual nature of this inflationary period, BPS-RT also shifts its combination weights toward component models with SV. This contrasts with the prior period of lower inflation, when the business cycle indicators were found to be more important weight modifiers. 

Future lines of research could involve investigating, in other forecasting applications and contexts, the usefulness of different sets of weight modifiers and the implications for weight structure. For instance, this could draw on the ability of BPS-RT, via its choice of weight modifiers, to capture general patterns of cross-sectional dependence between competing agents' probabilistic forecasts. Additional structure could be given to the clustering by, for example, letting the combination weight on a given individual agent's density forecast depend not only on characteristics of her own forecast (such as its mean or variance) but on characteristics of the other agents' forecasts.

\clearpage
\footnotesize{\setstretch{0.9}
\addcontentsline{toc}{section}{References}
\bibliography{lit}}\normalsize 
\clearpage

\begin{appendix}
\begin{center}
\textbf{\LARGE Online Appendix}
\end{center}
\setcounter{table}{0}
\setcounter{equation}{0}
\setcounter{figure}{0}
\renewcommand{\thetable}{A.\arabic{table}}
\renewcommand{\theequation}{A.\arabic{equation}}
\renewcommand{\thefigure}{A.\arabic{figure}}
\pagenumbering{arabic}
\doublespacing\normalsize

\section{Technical Appendix: Bayesian Inference}\label{app:A}
Before we start discussing the modeling choices, and the prior and the posterior sampler in detail, we introduce a bit of additional notation to simplify the exposition.

We can rewrite Eq. (\ref{eq:synthesis}) as a standard TVP regression:
\begin{equation}\label{eq:bpsapp}
y_t = c_t + \bm \gamma' \bm x_{t|t-h} + \bm \beta'_t \bm x_{t|t-h} + \sigma_t \nu_t, \quad \nu_t \sim \mathcal{N}(0, 1).
\end{equation}

The priors on $\bm \gamma$ and $\bm \beta_t$ can be written in the form of multivariate Gaussian distributions:
\begin{align*}
\bm \gamma &\sim \mathcal{N}(\mu^\gamma(\bm Z^\gamma), \underline{\bm V}^\gamma), \\
\bm \beta_t &\sim \mathcal{N}(\mu^\beta(\bm z_{t|t-h}^\beta), \underline{\bm V}^\beta).
\end{align*}

Here, $\mu^\gamma(\bm Z^\gamma)$ and  $\mu^\beta(\bm Z_{t|t-h}^\beta)$ are both prior mean functions of dimension $J$ and $\underline{\bm V}^n = \text{diag}(\tau^n_1, \dots, \tau^n_J)$ for $n \in \{\gamma, \beta\}$. The weight modifiers are stored, respectively, in a ($J\times K_\gamma$) matrix $\bm Z^\gamma$ with typical row $\bm z_j^\gamma$ and in a ($J T \times K_\beta$) matrix $\bm Z^\beta = (\bm z_1^\beta, \dots, \bm z_T^\beta)'$, with $\bm z_t^\beta = (\bm z^\beta_{1 t|t-h}, \dots,\bm z^\beta_{J t|t-h})$, $t = 1, \dots, T$.

Notice that the prior on stacked $\bm \beta =(\bm \beta'_1, \dots, \bm \beta'_T)$ can be written as a $JT-$dimensional Gaussian distribution:
\begin{equation*}
\bm \beta \sim \mathcal{N}(\mu^\beta(\bm Z^\beta), \bm I_T \otimes \underline{\bm V}^\beta), 
\end{equation*}
where the prior mean function $\mu^\beta(\bm Z^\beta)$ is now also of dimension $JT$.

\subsection{Additional Details about Our Modeling Choices and Priors}\label{app:prior}
In this sub-section we  provide additional details about our hierarchical prior setup used for $\bm \gamma$ and the time-varying part, $\bm \beta_t$, of the weights. In both cases, the prior mean functions, $\mu^\gamma(\bm Z^\gamma)$ and $\mu^\beta(\bm z_t^\beta)$, are approximated by tree functions (see Eq. \ref{eq:bart}), while the prior variances --- which define the degree of shrinkage toward these prior means --- are modeled with a horseshoe \citep[HS,][]{carvalho2010horseshoe} prior. In addition, we sketch the law of motion and modeling choices for the time-varying intercept and time-varying variances in Eq. (\ref{eq:bpsapp}), both of which capture the idea of model incompleteness.\\ 

\noindent \textbf{Tree functions to approximate the prior mean.} We closely follow here the suggestions of the Bayesian additive regression tree (BART) literature \citep{chipman1998bayesian, chipman2010bart} and use a similar prior setup for our tree structures $\mathcal{T}^{n}_s$ and terminal node parameters $\bm \phi^n_s$ for $n \in \{\beta, \gamma\}$. 
To generate the tree function, \cite{chipman1998bayesian} and \cite{chipman2010bart} suggest  using a stochastic process of the following form: 
\begin{enumerate}[leftmargin=*]
\item \textbf{Prior on the tree structure $\mathcal{T}^{n}_s$.} Impose a decreasing probability of growing more complex trees and that a terminal node is non-terminal. This probability is assumed to be
\begin{equation*}
\frac{c_0}{(1 + \vartheta)^{c_1}},
\end{equation*}
for a particular terminal node at depth $\vartheta$, with the hyperparameters $c_0 = 0.95$ and $c_1 = 2$ being two values that have been shown to be reasonable choices in much of the literature using Bayesian (additive) tree models. \cite{chipman2010bart} show that this choice works well even for single-tree models. Morever, for each splitting rule at each node, \cite{chipman2010bart} propose a prior that is  agnostic about the choice of the specific splitting variable and propose a natural default choice, which is to use a uniform prior on the splitting variables, treating each variable as equally likely to be used in a splitting rule. 

\item \textbf{Prior on the terminal node parameters $\bm \phi^n_s$ .} We use a Gaussian prior for the terminal node parameters. For a typical element in $\bm \phi^n_s$, that is
\begin{equation*}
\phi^n_{j,s} \sim \mathcal{N}(0, c_2/S),
\end{equation*}
where $c_2$ refers to a shrinkage parameter and $S$ to the number of trees. It is worth noting that –-– to avoid overfitting ––– the prior variances for these terminal parameters are scaled down by the number of trees and become tighter, so that each individual tree explains only a tiny fraction within the additive sum-of-tree function.

\end{enumerate}

\noindent \textbf{Shrinkage toward the prior mean through the horseshoe prior.}  The horseshoe prior  amounts to setting the scaling parameters as follows:
\begin{equation*}
\tau_j^n = \lambda^n \psi_j^n, \quad \lambda^n \sim \mathcal{C}^+(0, 1), \quad \psi_j^n \sim \mathcal{C}^+(0, 1), \text{ for } n \in \{\gamma, \beta\},
\end{equation*}
with $C^+(0, 1)$ denoting the half-Cauchy distribution. The key feature of this prior is that $\lambda^n$ serves as a global shrinkage parameter that pulls all weights toward the prior mean, whereas $\psi_j^n$ allows for agent-specific deviations from this common pattern. Another representation of this prior, which simplifies posterior sampling enormously, is based on introducing inverse Gamma distributed auxiliary variables \citep[see][]{makalic2015simple}:
\begin{align*}
    \lambda^n|\zeta^n  &\sim \mathcal{G}^{-1}(1/2, 1/\varphi^n), \quad \varphi^n \sim \mathcal{G}^{-1}(1/2, 1), \\
    \psi_j^n|\varpi_j^n &\sim \mathcal{G}^{-1}(1/2, 1/\varpi_j^n), \quad \varpi_j^n \sim \mathcal{G}^{-1}(1/2, 1).
\end{align*}
This representation is convenient since --- when combined with the likelihood --- it gives rise to a simple Gibbs sampling step that involves only inverse Gamma full conditionals (see Sub-section \ref{app:posterior} below).\\

\noindent \textbf{Controlling for model incompleteness.} A time-varying intercept $c_t$ and time-varying variances $\sigma_t^2$ both control for model incompleteness. The fact that both are potentially time-varying gives additional flexibility in the degree of model set incompleteness (as outlined in Sub-section 2.1.2).
The time-varying intercept follows a random walk (RW) law of motion with the state equation given by
\begin{equation*}
c_t = c_{t-1} + \eta_{c, t}, \quad \eta_{c, t} \sim \mathcal{N}(0,\sigma_c^2),
\end{equation*}
where $\sigma_c^2$ denotes the state innovation variance. To discipline $c_t$, we use a relatively tight Gamma prior on $\sigma_c^2$ and strongly push the state innovation variance toward a small positive value close to zero. 

The error variances $\sigma_t^2$ in Eq. (\ref{eq:bps}) can be time-varying or constant. The time-varying case is given by:
\begin{equation}\label{app:SV}
\varsigma_{t} = \mu_\varsigma + \rho_\varsigma (\varsigma_{t-1} - \mu_\varsigma) + \eta_{\varsigma, t}, \quad \eta_{\varsigma, t} \sim \mathcal{N}(0,\sigma_\varsigma^2),
\end{equation}
with $\mu_\varsigma$ denoting the unconditional mean, $\rho_\varsigma$ the persistence parameter, and $\sigma_\varsigma^2$ the state innovation variance of the log-volatility process. For SV we follow \cite{kastner2014ancillarity} and assume a Gaussian prior on $\mu_\varsigma \sim \mathcal{N}(0,10^2)$, a (transformed) Beta prior on $(\rho_\varsigma+1)/2\sim\mathcal{B}(5,1.5)$, and a Gamma prior on $\sigma_\varsigma^2\sim\mathcal{G}\left(0.5,0.5\right)$. Moreover, for the case of homoskedastic errors, we assume an inverse Gamma prior on $\sigma^2 \sim i\mathcal{G}(0.01,0.01)$.

\subsection{Posterior Simulation}\label{app:posterior}
The prior discussed in the previous section can be combined with the likelihood to derive the full posterior over all unknown quantities in our model. Since this joint density is untractable, we use Markov chain Monte Carlo (MCMC) methods to carry out posterior simulation. 
In what follows, we let $\bullet$ be generic notation that indicates that we condition on all other parameters/states of the model.

We start the discussion of our posterior sampler by first describing how we estimate the latent quantities that enter the synthesis function. This includes the static and dynamic weights, the error variances, the latent trend components, and the agent-specific forecasts.

\textbf{Sampling from $\{p(c_t|\bullet)\}_{t=1}^T$.} We sample the full history of the random walk intercept term conditional on all other unknowns in the model using a forward-filtering backward-sampling (FFBS) step \citep{carterkohn}. This is achieved by noting that
\begin{equation}
    \underbrace{y_t - \bm \gamma' \bm x_{t|t-h} - \bm \beta'_t \bm x_{t|t-h}}_{y^{\gamma, \beta}_t} = c_t + \sigma_t \nu_t,
\end{equation}
is a standard unobserved components model with heteroskedastic shocks.

\textbf{Sampling from $p(\bm \gamma | \bullet)$.} The time-invariant weights are sampled from a $J$-dimensional Gaussian full conditional posterior distribution:
\begin{equation}
    \bm \gamma|\bullet \sim \mathcal{N}(\overline{\bm \gamma}, \overline{\bm V}^\gamma),
\end{equation}
with moments given by:
\begin{align*}
    \overline{\bm V}^\gamma &= (\bm X' \bm \Sigma^{-1} \bm X + (\underline{\bm V}^\gamma)^{-1})^{-1}, \\
    \overline{\bm \gamma} &= \overline{\bm V}^\gamma \left(\bm X' \bm \Sigma^{-1} \bm y^{\beta, c} +  (\underline{\bm V}^\gamma)^{-1} \mu^\gamma(\bm Z^\gamma) \right),
\end{align*}
where $\bm X$ is a $T \times J$ matrix with $t^{th}$ row $\bm x'_{t|t-h}$, $\bm \Sigma = \text{diag}(\sigma_1^2, \dots, \sigma_T^2)$ and $\bm y^{\beta, c}$ is a $T$-dimensional vector with typical element $y_t - \bm \beta'_t\bm x_{t|t-h} - c_t$.

\textbf{Sampling $p(\bm \beta|\bullet)$.} The dynamic regression coefficients are simulated by writing the model in static form. The static form of the model reads:
\begin{equation}
    \bm y^{\gamma, c} = \bm W \bm \beta + \bm \nu, \quad \bm \nu \sim \mathcal{N}(\bm 0_T, \bm \Sigma),
\end{equation}
where $\bm y^{\gamma, c}$ is $T\times 1$ and has typical element $y_t - \bm \gamma' \bm x_{t|t-h} - c_t$ and $\bm W$ is a $T \times TJ$-dimensional block diagonal matrix with $\bm W = \text{bdiag}(\bm x'_{1|1-h}, \dots, \bm x'_{T|T-h})$.\footnote{Observations $-h, \dots, 0$ refer to a part of the sample that we use to initialize our models.} Under this static representation, the posterior of $\bm \beta$ takes a standard form and is multivariate Gaussian:
\begin{equation}
    \bm \beta|\bullet \sim \mathcal{N}(\overline{\bm \beta}, \overline{\bm V}^\beta),
\end{equation}
with posterior covariance matrix and mean vector given by, respectively:
\begin{align*}
    \overline{\bm V}^\beta &= \left(\bm W' \bm \Sigma^{-1} \bm W + (\bm I_T \otimes \underline{\bm V}^\beta)^{-1} \right)^{-1},\\
    \overline{\bm \beta}   &= \overline{\bm V}^\beta \left(\bm W' \bm \Sigma^{-1}\bm y^{\gamma, c} + (\bm I_T \otimes \underline{\bm V}^\beta)^{-1}\mu^\beta(\bm Z^\beta) \right).
\end{align*}

This distribution is high dimensional even for moderate values of $J$ and we thus use the efficient sampler  outlined in \cite{hauzenberger2022fast}.

\textbf{Sampling from $p(\sigma^2_1, \dots, \sigma^2_T|\bullet)$.} We sample the log-volatilities and associated state equation parameters using the algorithm  outlined in \cite{kastner2014ancillarity}. This step is implemented in the \texttt{R} package stochvol.

\textbf{Sampling from $\{p(\bm x_{t|t-h}|\bullet)\}_{t=1}^T$.} We draw from $\{p(\bm x_{t|t-h}|\bullet)\}_{t=1}^T$  on a $t$-by-$t$ basis. The time $t$ full conditional posterior of $\bm x_t$ is given by:
\begin{equation*}
    p(\bm x_{t|t-h}|\bullet) \propto \mathcal{N}(y_t|c_t + \bm \gamma' \bm x_{t|t-h} + \bm \beta'_t \bm x_{t|t-h}, \sigma_t^2) \prod_{j=1:J}\pi_{jt}(x_{j{t|t-h}}),
\end{equation*}
which, unless the agent densities $\pi_{jt}(x_{j{t|t-h}})$ are Gaussian, takes no well-known form. In our applications, the agent densities do not have analytical representations. For example, the ECB-SPF elicits histograms from survey respondents and in the US application the available forecasts are predictive draws based on model-specific Gibbs samplers. Accordingly, we sample $\bm x_t$ using an adaptive Metropolis Hastings step \citep[see, e.g.,][]{roberts2009examples}. This step proposes $\bm x^*_{t|t-h}$ from a mixture of Gaussian distributions:
\begin{equation}
    \bm x_{t|t-h}^*  \sim  (1 - \kappa) \mathcal{N}(\bm x_{t|t-h}, (2.38)^2 \hat{\bm Q}_{tm}/J) + \kappa \mathcal{N}(\bm x_{t|t-h}, (0.1)^2 \bm I_J/J)
\end{equation}
where $\kappa = 0.05$ is a small constant and $\hat{\bm Q}_{tm}/J$ is the empirical covariance matrix of the target distribution based on the first $m$ draws. Since this algorithm learns the proposal, it can quickly adjust to cases where the agent densities are non-Gaussian, feature multiple modes, or are heavy tailed. 

Next, we discuss the steps involved in sampling the parameters of the priors on the weights. 

\textbf{Sampling from $p(\mathcal{T}^n_1, \dots, \mathcal{T}^n_S, \bm \phi_1^n, \dots, \bm \phi_S^n|\bullet)$ for $n \in \{\gamma, \beta\}$.} We sample the tree structures and the terminal node parameters using the algorithm proposed in \cite{chipman2010bart}. This algorithm is applicable since, conditional on $\bm \gamma$ and $\bm \beta$, the corresponding priors can be interpreted as regression models. For instance, in the case of $\bm \beta$ notice that:
\begin{equation}
    \bm \beta_{jt} = \sum_{s=1}^S g(\bm z_{jt}^\beta|\mathcal{T}_s^\beta, \bm \phi_s^\beta) + \tau_j^\beta \nu_jt,\quad \nu_{jt} \sim \mathcal{N}(0, 1),
\end{equation}
which, in stacked form, can be written as:
\begin{equation}
    \bm \beta = \sum_{s=1}^S g(\bm Z^\beta|\mathcal{T}_s^\beta, \bm \phi_s^\beta) + \bm r, \quad \bm r \sim \mathcal{N}(\bm 0_{TJ} , \bm I_T \otimes \underline{\bm V}^\beta). \label{eq: BART_BETA}
\end{equation}

\autoref{eq: BART_BETA} is a standard BART regression with latent responses and heteroskedastic errors. For $\bm \gamma$, a similar regression representation can be derived. 

\textbf{Sampling from $p(\tau_1^n, \dots, \tau_J^n|\bullet)$ for $n \in \{\gamma, \beta\}$.} The scaling parameters are obtained using the algorithm described in \cite{makalic2015simple}.  This algorithm involves only inverse Gamma distributions and, for brevity, we do not discuss them in detail here. 

These steps form our MCMC algorithm. In all our empirical work we iteratively sample from the different full conditionals to obtain draws from the joint posterior of the coefficients and the latent states. Based on these draws, we back out the predictive distribution as described in the main text through Monte Carlo integration. That is, after obtaining a draw from the posterior, we use this draw to forecast $y_{t^*}$. This is done for every draw, leading to a posterior distribution over future values $y_{t^*}$. In all our empirical work, we repeat this $12,500$ times and discard the initial $2,500$ draws as a burn-in. We subsequently keep every second draw, yielding a total of $5,000$ draws from the joint posterior distribution.
 
\subsection{ADL Model Estimation}\label{app:adl}
Our US inflation forecasting exercise involves Bayesian estimation of ADL models involving different explanatory variables. Table \ref{tab:data} provides an overview of the $27$ variables each used as an exogenous predictor in Eq. (\ref{eq:adl}) and also highlights (in bold type face) the target variable.
\begin{table}[!htbp]
\centering
\begin{threeparttable}
\tiny
\caption{List of variables used for the autoregressive distributed lag (ADL) specifications.\label{tab:data}} 
\begin{tabular}{llc}
\toprule
\multicolumn{1}{l}{\textbf{Mnemonic}}&\multicolumn{1}{l}{\textbf{Description}}&\multicolumn{1}{l}{\textbf{Transformation}}\tabularnewline
\midrule
GDPC1&Real Gross Domestic Product& $100\times \Delta \log$\tabularnewline
PCECC96&Real Personal Consumption Expenditures& $100\times \Delta \log$\tabularnewline
FPIx&Real private fixed investment& $100\times \Delta \log$\tabularnewline
GCEC1&Real Government Consumption Expenditures and Gross Investment& $100\times \Delta \log$\tabularnewline
INDPRO& Total index Industrial Production Index & $100\times \Delta \log$ \tabularnewline
CUMFNS&Capacity Utilization:  Manufacturing (SIC) & none \tabularnewline
PAYEMS& Emp:Nonfarm All Employees: Total nonfarm & $100\times \Delta \log$ \tabularnewline
CE16OV&Civilian Employment &$100\times \Delta \log$ \tabularnewline
UNRATE&Civilian Unemployment Rate & $\Delta$ \tabularnewline
AWHMAN&Average Weekly Hours of Production and Nonsupervisory Employees:  Manufacturing Hours& none\tabularnewline
CES0600000007&Average Weekly Hours of Production and Nonsupervisory Employees:  Goods-Producing& $\Delta$ \tabularnewline
CLAIMSx&Initial Claims&$100\times \Delta \log$\tabularnewline
GDPCTPI&Gross Domestic Product: Chain-type Price Index&$100\times \Delta \log$ \tabularnewline
\bfseries CPIAUCSL& \bfseries Consumer Price Index for All Urban Consumers &$100\times \Delta \log$ \tabularnewline
PPIACO&Producer Price Index for All Commodities& $100\times \Delta \log$\tabularnewline
WPSID61&Producer Price Index by Commodity Intermediate Materials:  Supplies \& Components&$100\times \Delta \log$ \tabularnewline
WPSID62&Producer Price Index:  Crude Materials for Further Processing&$100\times \Delta \log$ \tabularnewline
COMPRNFB&Nonfarm Business Sector:  Real Compensation Per Hour &$100\times \Delta \log$ \tabularnewline
ULCNFB&Nonfarm Business Sector:  Unit Labor Cost &$100\times \Delta \log$ \tabularnewline
CES0600000008&Average Hourly Earnings of Production and Nonsupervisory Employees&$100\times \Delta \log$ \tabularnewline
FEDFUNDS&Effective Federal Funds Rate & $\Delta$ \tabularnewline
BAA10YM&Moodys Seasoned Baa Corporate Bond Yield Relative to Yield on 10-Year Treasury& none \tabularnewline
GS10TB3Mx&10-Year Treasury Constant Maturity Minus 3-Month Treasury Bill, secondary market& none \tabularnewline
CPF3MTB3Mx&3-Month Commercial Paper Minus 3-Month Treasury Bill, secondary market& none \tabularnewline
M2REAL&Real M2 Money Stock&$100\times \Delta \log$ \tabularnewline
BUSLOANSx&Real Commercial and Industrial Loans, All Commercial Banks&$100\times \Delta \log$ \tabularnewline
CONSUMERx&Real Consumer Loans at All Commercial Banks& $100\times \Delta \log$ \tabularnewline
S.P.500& S\&Ps Common Stock Price Index & $100\times \Delta \log$ \tabularnewline
\bottomrule
\end{tabular}
\begin{tablenotes}
\scriptsize
\item \textbf{Notes:} The variable in bold refers to the target inflation series.
\end{tablenotes}
\end{threeparttable}
\end{table}

To estimate these ADL specifications, we use standard Bayesian non-conjugate regression techniques with posteriors of standard form. The non-conjugate priors are weakly informative. We center both $\rho_{\pi}$ and $\alpha_{\pi}$ in Eq. (\ref{eq:adl}) on a prior mean of zero and assume a prior variance of $100$. For the case of homoskedastic errors, we assume an inverse Gamma prior on $\sigma_{\pi, t+h}^2 := \sigma_{\pi}^2 \sim i\mathcal{G}(0.01,0.01)$, while for SV we essentially use the setup sketched in Appendix \ref{app:prior} (see Eq. \ref{app:SV}). However, non-conjugacy (and SV in particular) leads to predictive densities for which there is no closed-form solution. Therefore we use MCMC methods and predictive simulation.

\setcounter{table}{0}
\setcounter{equation}{0}
\setcounter{figure}{0}
\renewcommand{\thetable}{B.\arabic{table}}
\renewcommand{\theequation}{B.\arabic{equation}}
\renewcommand{\thefigure}{B.\arabic{figure}}

\newpage
\clearpage
\section{Empirical Appendix: Additional Results}\label{app:B}
This empirical appendix contains supplementary results as referenced in the main paper. It is structured as follows. Section B.1 reports left and right tail CRPSs. Section B.2 presents the combination weights. Section B.3 presents the cumulative CRPS statistics. Section B.4 presents the fluctuation tests. Section B.5 presents the PITs tests. Section B.6 presents results showing how we can draw out the degree of shrinkage implied by BPS-RT. Section B.7 plots the predictive densities in both applications and examines their skewness. 

\subsection{Tail forecast accuracy}

\begin{figure}[!htbp]
\centering
\caption{Relative tail forecast accuracy: EA GDP growth.}\label{fig:easpfappend}
\includegraphics[width=0.95\textwidth,keepaspectratio]{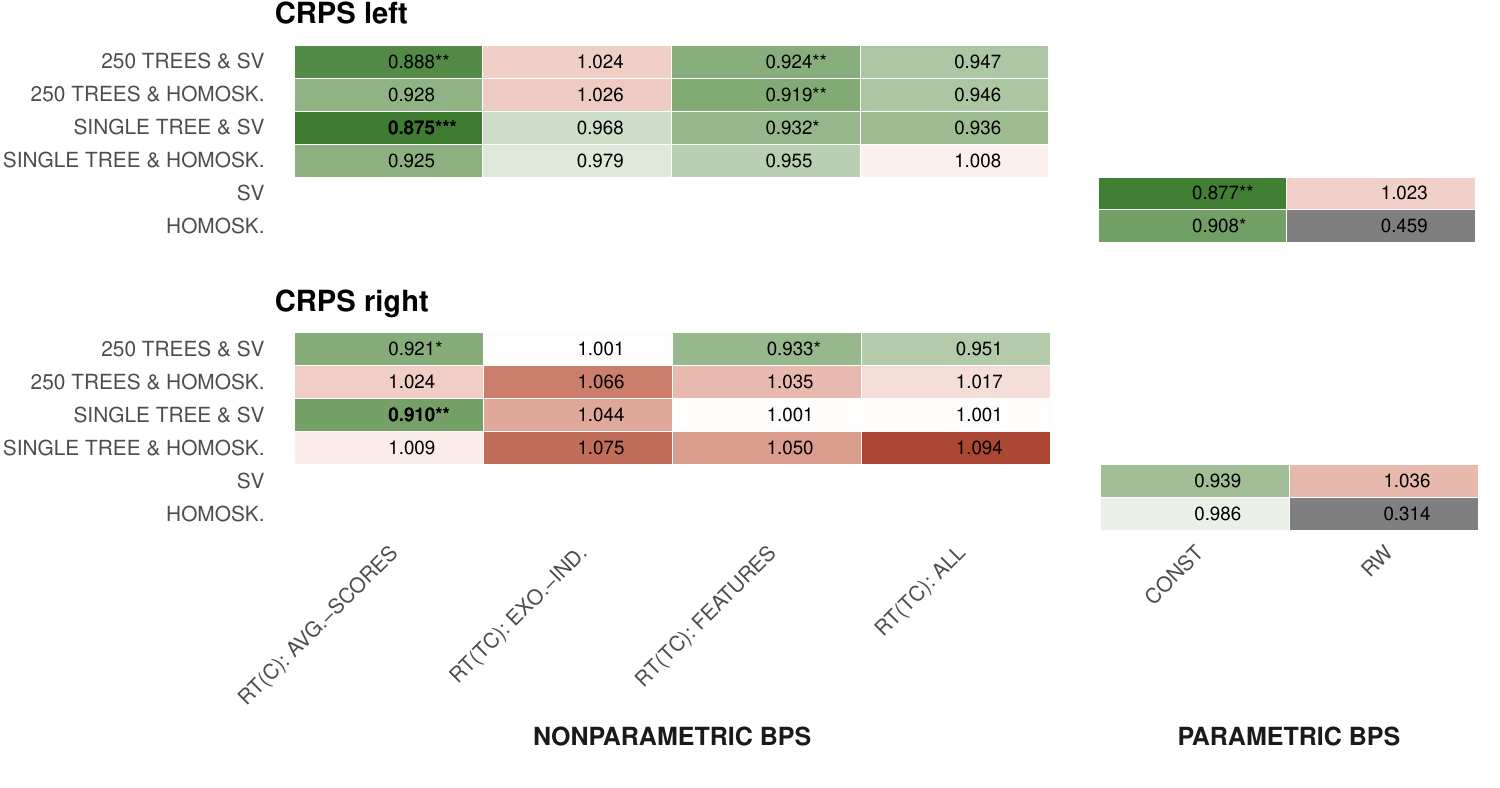}
\caption*{\scriptsize \textbf{Notes:} This figure shows two variants of quantile-weighted CRPS ratios, one focusing on the left tail and the other on the right tail. The gray shaded entries give the actual scores of our benchmark (BPS-RW with homoskedastic error variances). Green shaded entries refer to models that outperform the benchmark (with the forecast metric ratios below one), while red shaded entries denote models that are outperformed by the benchmark (with the forecast metric ratios greater than one). The best performing model specification by forecast metric is given in bold. Asterisks indicate statistical significance of the \cite{diebold1995dmtest} test, which assumes equal forecast performance for each model relative to the benchmark, at the $1$ ($^{***}$), $5$ ($^{**}$), and $10$ ($^{*}$) percent significance levels.}
\end{figure}

\begin{figure}[!htbp]
\centering
\caption{Relative tail forecast accuracy: US inflation.}\label{fig:usadlappend}
\begin{minipage}{0.5444\textwidth}
\centering
(a) One-quarter-ahead $(h = 1)$
\vspace*{2pt}
\end{minipage}
\begin{minipage}{\textwidth}
\centering
\includegraphics[width=0.95\textwidth,keepaspectratio]{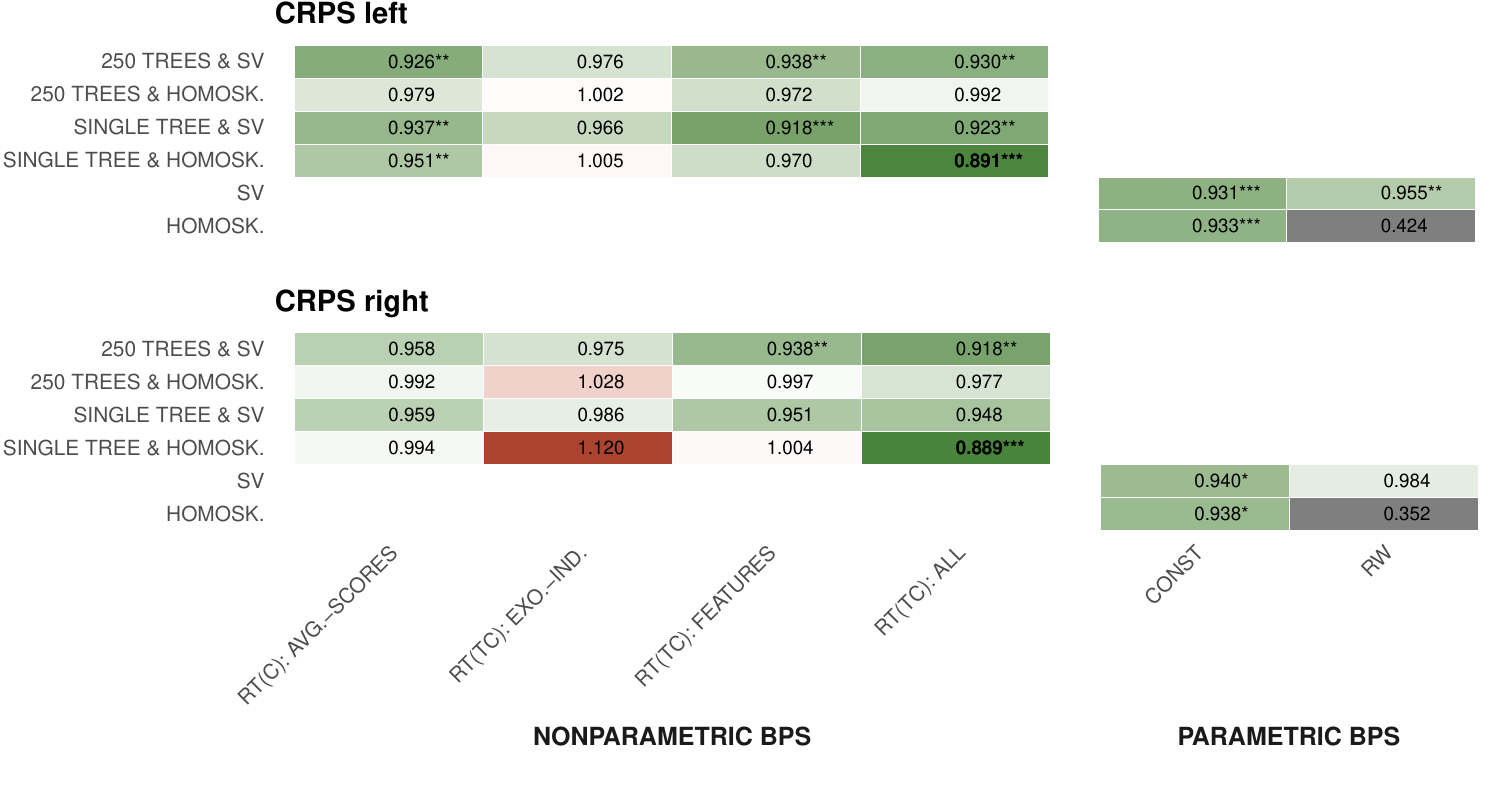}
\end{minipage}
\begin{minipage}{\textwidth}
\centering
\vspace*{10pt}
(b) One-year-ahead $(h = 4)$
\vspace*{2pt}
\end{minipage}
\begin{minipage}{\textwidth}
\centering
\includegraphics[width=0.95\textwidth,keepaspectratio]{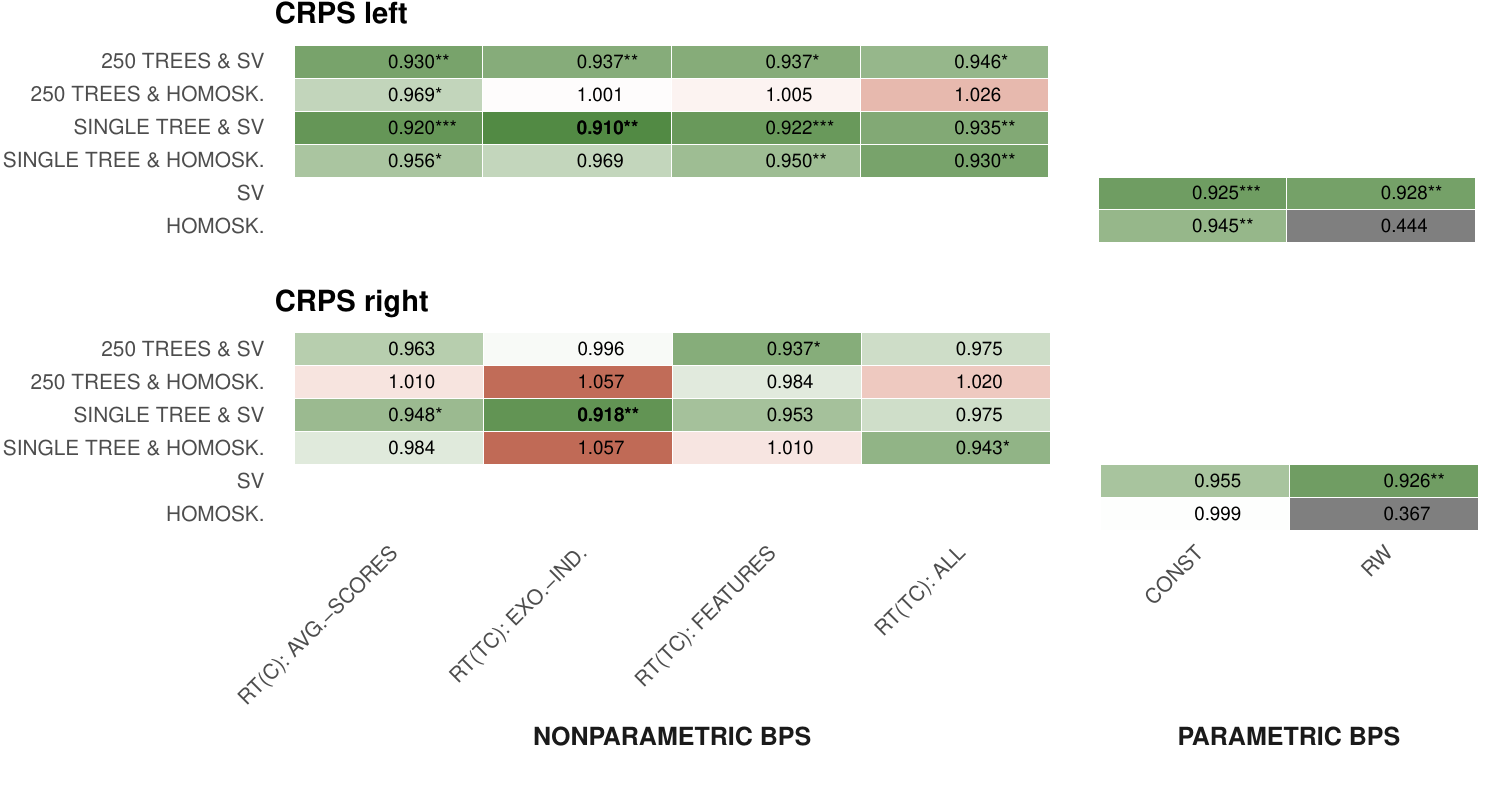}
\end{minipage}
\caption*{\scriptsize \textbf{Notes:} This figure shows two variants of quantile-weighted CRPS ratios, one focusing on the left tail and the other on the right tail. The gray shaded entries give the actual scores of our benchmark (BPS-RW with homoskedastic error variances). Green shaded entries refer to models that outperform the benchmark (with the forecast metric ratios below one), while red shaded entries denote models that are outperformed by the benchmark (with the forecast metric ratios greater than one). The best performing model specification by forecast metric is given in bold. Asterisks indicate statistical significance of the \cite{diebold1995dmtest} test, which assumes equal forecast performance for each model relative to the benchmark, at the $1$ ($^{***}$), $5$ ($^{**}$), and $10$ ($^{*}$) percent significance levels.}
\end{figure}

\clearpage

\subsection{Combination Weights}
\begin{figure}[!htbp]
\centering
\caption{Combination weights over the evaluation sample: EA GDP growth} \label{fig:eabeta1}
\includegraphics[width=\textwidth,keepaspectratio]{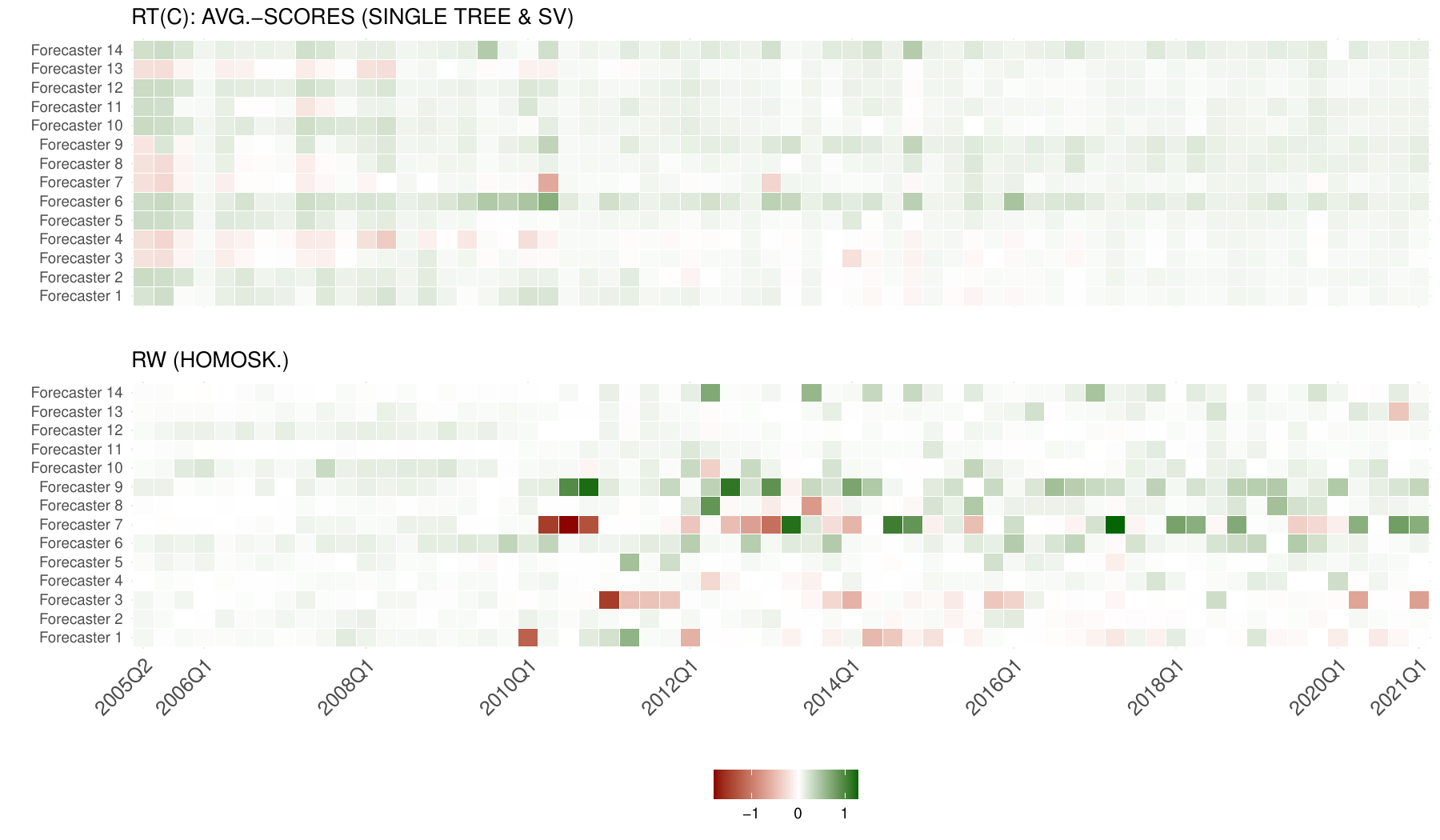}
\caption*{\scriptsize \textbf{Notes:} This figure shows the posterior median of the  combination weights, $(\gamma_{j} + \beta_{jt+h})$, for each of the $14$ SPF forecasters. Green (red) shaded cells indicate that calibration parameters are above (below) zero. The top panel corresponds to our preferred BPS-RT specification, while the bottom panel corresponds to the benchmark.}
\end{figure}

\begin{figure}[!htbp]
\centering
\caption{Combination weights over the evaluation sample: One-quarter-ahead US inflation\label{fig:adlbeta1}}
\includegraphics[width=0.8\textwidth,keepaspectratio]{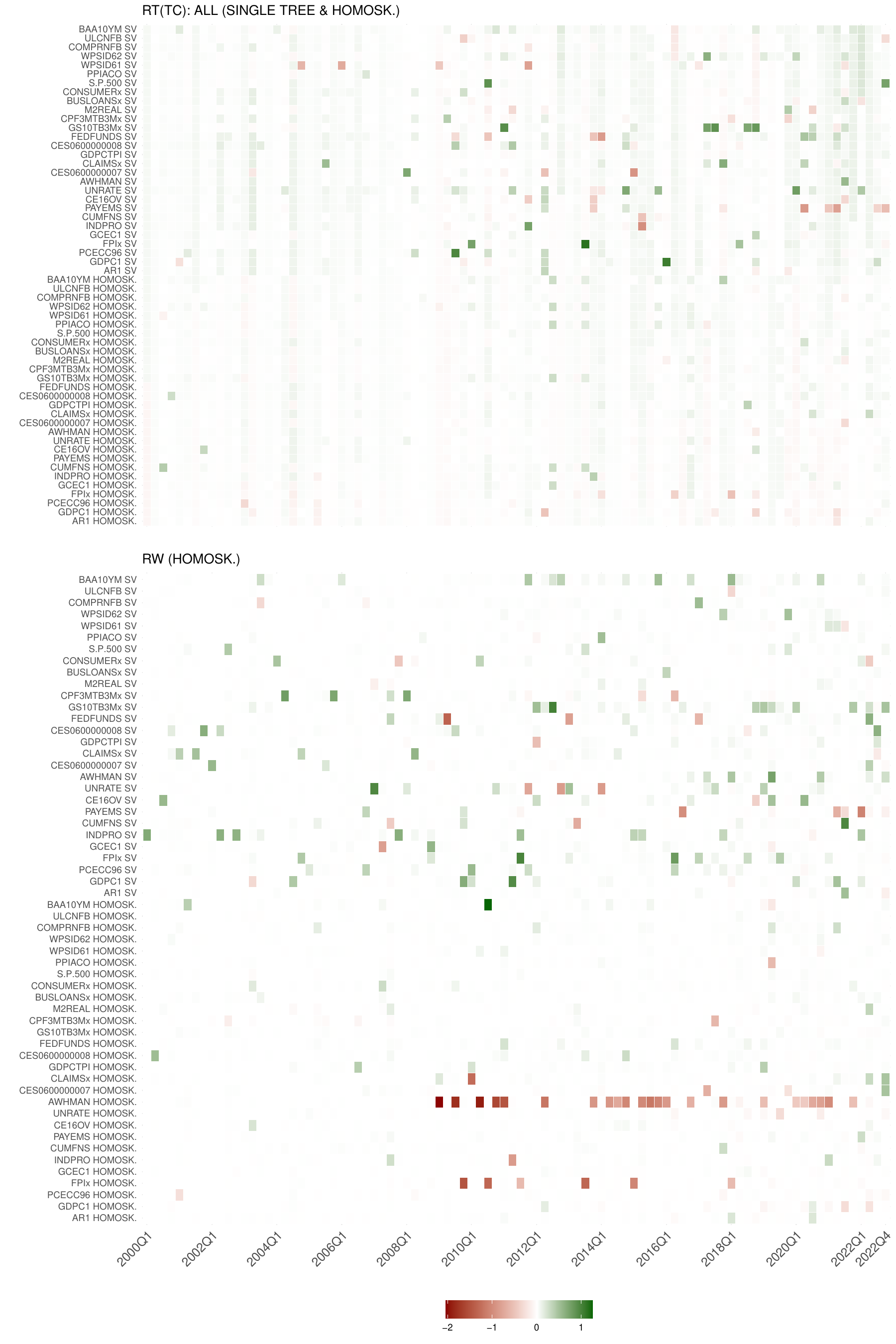}
\caption*{\scriptsize \textbf{Notes:} This figure shows the posterior median of the one-quarter-ahead combination weights, $(\gamma_{j} + \beta_{jt+h})$, for each of the $56$ ADL model variants. Green (red) shaded cells indicate that weights are above (below) zero. The top panel corresponds to our preferred BPS-RT specification, while the bottom panel corresponds to the benchmark.
}
\end{figure}

\begin{figure}[!htbp]
\centering
\caption{Combination weights over the evaluation sample: One-year-ahead US inflation $(h = 4)$ \label{fig:adlbeta4}}
\includegraphics[width=0.8\textwidth,keepaspectratio]{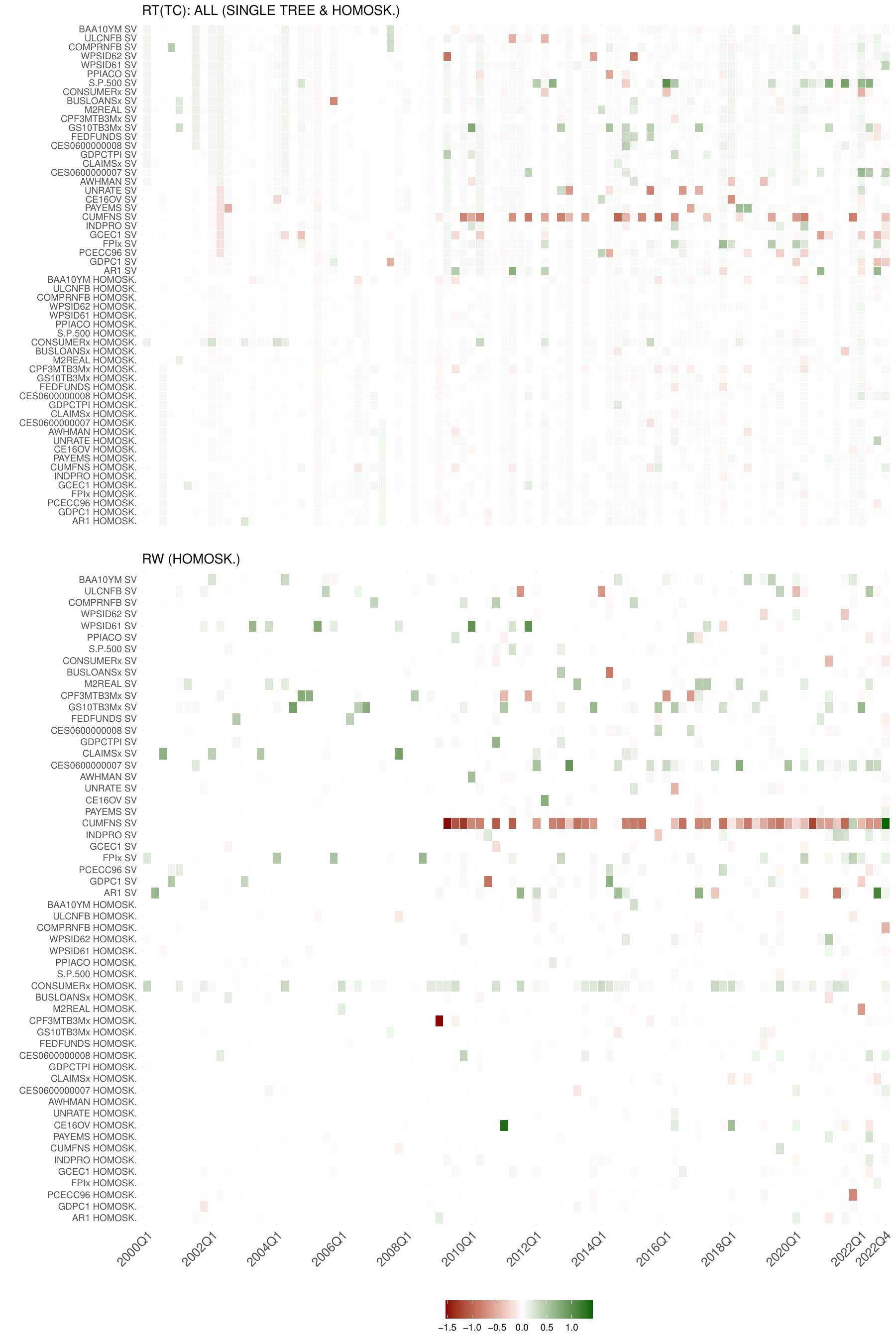}
\caption*{\scriptsize \textbf{Notes:} This figure shows the posterior median of the one-year-ahead combination weights, $(\gamma_{j} + \beta_{jt+h})$, of the best performing model parameters for each of the $56$ ADL model variants. Green (red) shaded cells indicate that weights are above (below) zero. The top panel corresponds to our preferred BPS-RT specification, while the bottom panel corresponds to the benchmark.}
\end{figure}

\begin{figure}[!htbp]
\centering
\caption{Sum of combination weights over the evaluation sample \label{fig:betasum}}
\begin{minipage}{\textwidth}
\centering
(a) EA GDP growth
\end{minipage}
\begin{minipage}{\textwidth}
\centering
\includegraphics[width=\textwidth,keepaspectratio]{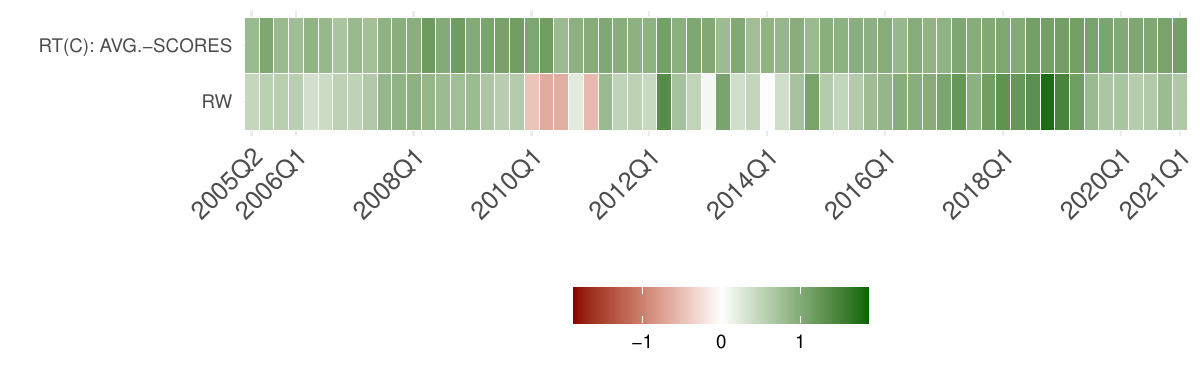}
\end{minipage}
\begin{minipage}{\textwidth}
\centering
(b) One-quarter-ahead US inflation ($h = 1$)
\end{minipage}
\begin{minipage}{\textwidth}
\centering
\includegraphics[width=\textwidth,keepaspectratio]{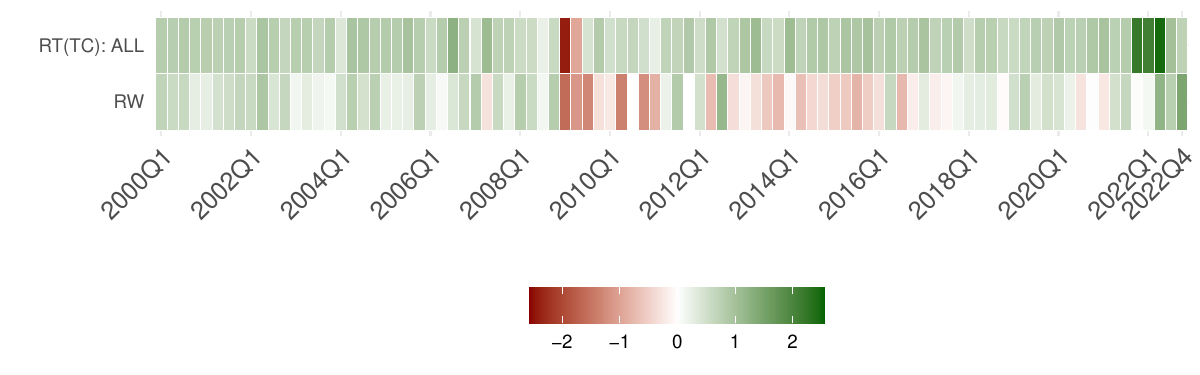}
\end{minipage}
\begin{minipage}{\textwidth}
\centering
(c) One-year-ahead US inflation ($h = 4$)
\end{minipage}
\begin{minipage}{\textwidth}
\centering
\includegraphics[width=\textwidth,keepaspectratio]{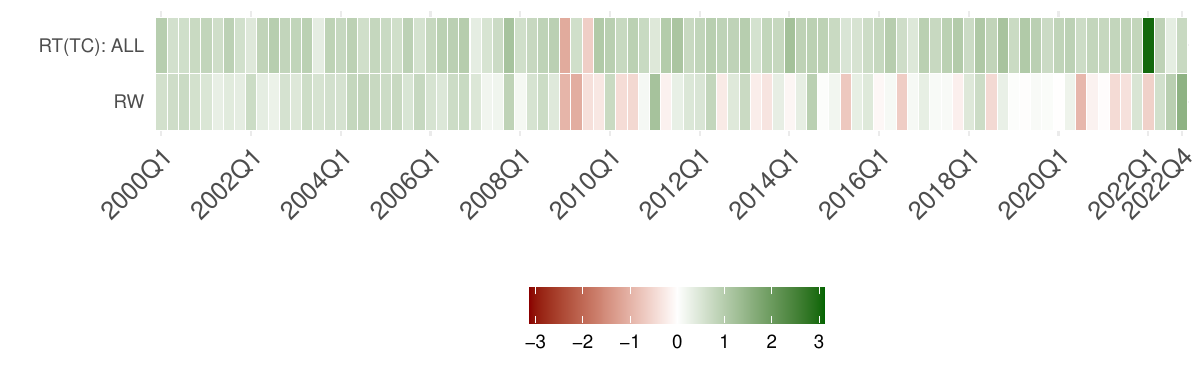}
\end{minipage}
\caption*{\scriptsize \textbf{Notes:} This figure shows the posterior median of the sum of the combination weights, $\sum_{j = 1}^{J}(\gamma_{j} + \beta_{jt+h})$, for the models shown in Figures \ref{fig:eabeta1}, \ref{fig:adlbeta1} and \ref{fig:adlbeta4}. 
Green (red) shaded cells indicate that the overall sum of weights is above (below) zero for a specific evaluation period.}
\end{figure}

\newpage
\subsection{Cumulative CRPS}

\begin{figure}[!htbp]
\centering
 \caption{Forecast performance of single-tree specifications with stochastic volatility: EA GDP growth \label{fig:crps-t-easpf}}
\includegraphics[width=0.8\textwidth,keepaspectratio]{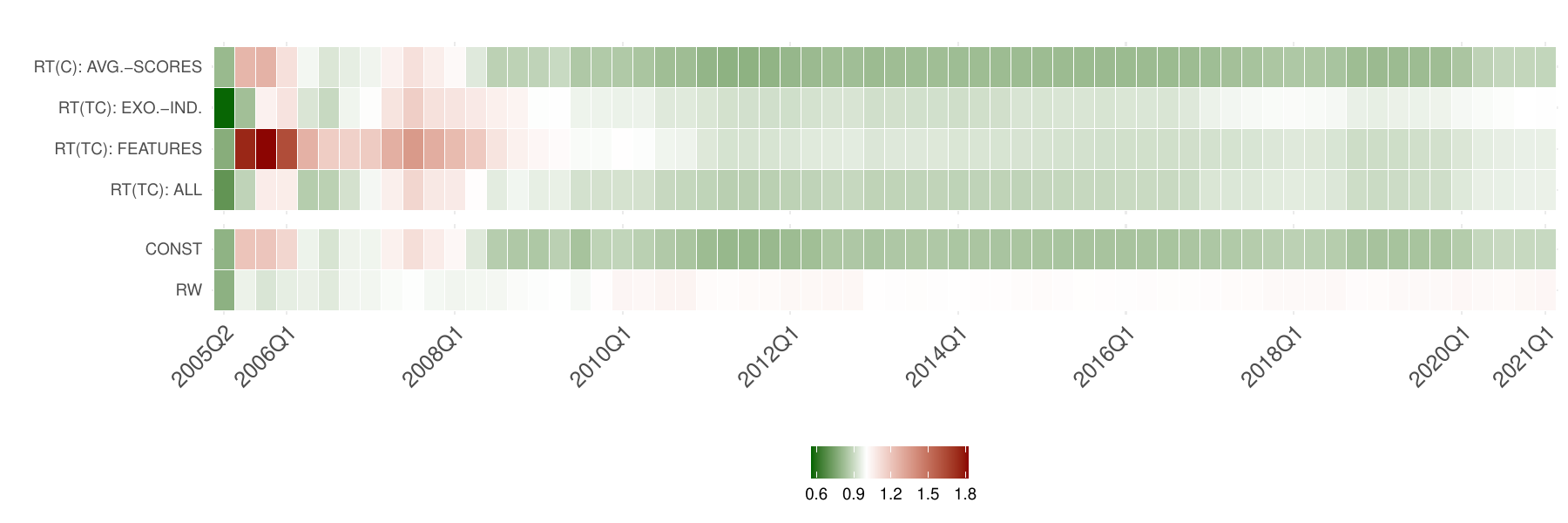}
\caption*{\scriptsize \textbf{Notes:} This figure shows relative cumulative continuous ranked probability scores (CRPSs) over the full evaluation sample, which ranges from $2005$Q$2$ to $2021$Q$1$. The benchmark model is a TVP regression with a random walk evolution of parameters (\texttt{BPS-RW}) and homoskedastic error variances. Green shaded entries indicate periods in which the respective model outperforms the benchmark (with the cumulative CRPS ratio below one), while red shaded entries denote periods in which the respective model is outperformed by the benchmark (with the cumulative CRPS ratio greater than one). We refrain from showing the forecast performance over time for all models, but focus on the class of models that contains the best performing specification in terms of CRPS, that is, all single-tree specifications with stochastic volatility.}
\end{figure}

\begin{figure}[!htbp]
\centering
\caption{Forecast performance of single-tree specifications with homoskedastic error variances: US inflation \label{fig:crps-t-adlus}}
\begin{minipage}{\textwidth}
\centering
(a) One-quarter-ahead ($h = 1$) 
\end{minipage}
\begin{minipage}{\textwidth}
\centering
\includegraphics[width=0.75\textwidth,keepaspectratio]{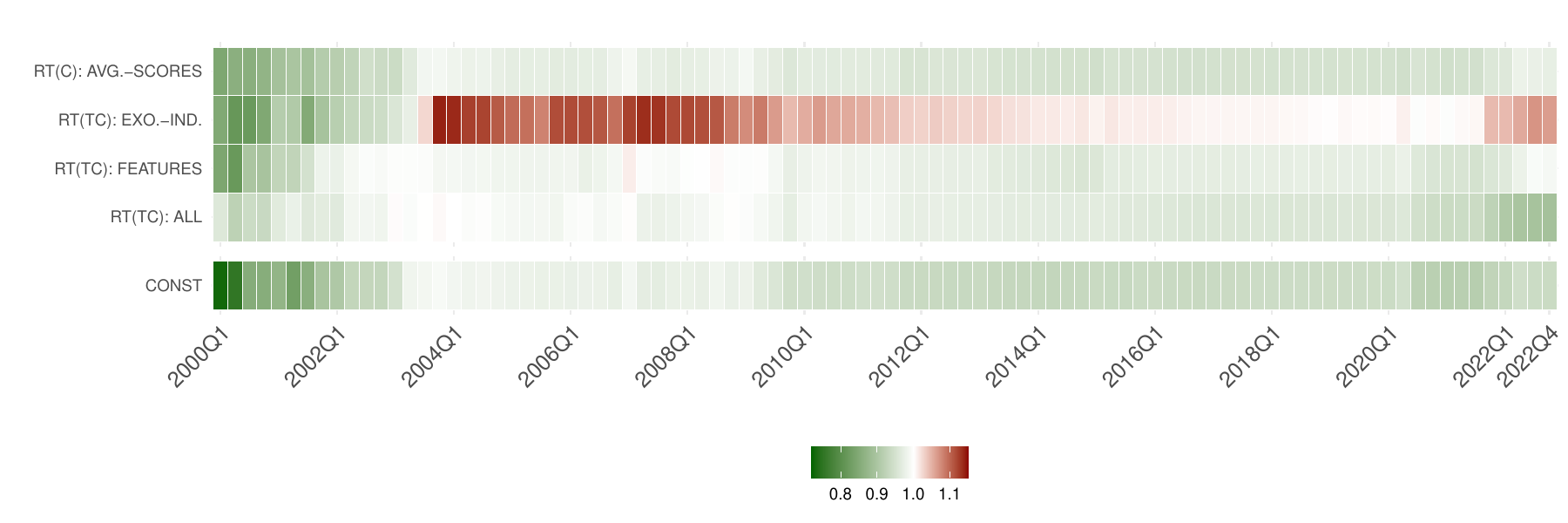}
\end{minipage}
\begin{minipage}{\textwidth}
\centering
\vspace*{10pt}
(b) One-year-ahead ($h = 4$) 
\vspace*{2pt}
\end{minipage}
\begin{minipage}{\textwidth}
\centering
\includegraphics[width=0.75\textwidth,keepaspectratio]{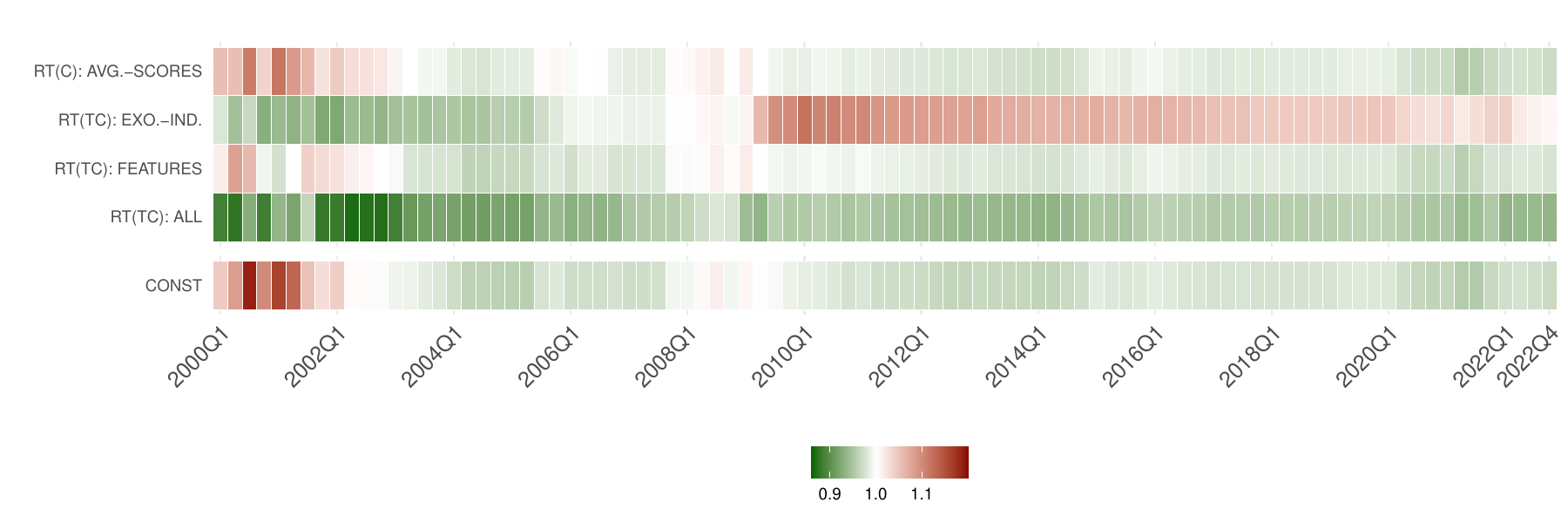}
\end{minipage}
\caption*{\scriptsize \textbf{Notes:} This figure shows average relative cumulative continuous ranked probability scores (CRPSs) over the evaluation sample, which ranges from $2000$Q$1$ to $2022$Q$4$. The benchmark model is a TVP regression with a random walk evolution of parameters (\texttt{BPS-RW}) and homoskedastic error variances. Green shaded entries indicate periods in which the respective model outperforms the benchmark (with the cumulative CRPS ratio below one), while red shaded entries denote periods in which the respective model is outperformed by the benchmark (with the cumulative CRPS ratio greater than one). We refrain from showing the forecast performance over time for all models, but focus on the class of models that forecast well in terms of CRPS, that is, all homoskedastic, single-tree specifications, the class containing the best performing specification for the one-quarter-ahead $(h = 1)$ horizon.}
\end{figure}

\newpage
\subsection{\cite{giacomini2010forecast} Fluctuation Test Statistic}

We focus on evaluating the BPS-RT specifications with the best overall forecast performance. As seen from Figures \ref{fig:easpfmain} and \ref{fig:usadlmain}, in the EA-GDP application this is the single-tree specification with SV using average scores as effect modifiers. For the US inflation application, this is the homoskedastic single-tree specification with the full set of weight modifiers.

\begin{figure}[!htbp]
\centering
\caption{Evolution of the \cite{giacomini2010forecast} fluctuation test statistic}\label{fig:fluctest}
\begin{minipage}{\textwidth}
\centering
(a) EA GDP growth
\end{minipage}
\begin{minipage}{\textwidth}
\centering
\includegraphics[width=0.49\textwidth,keepaspectratio]{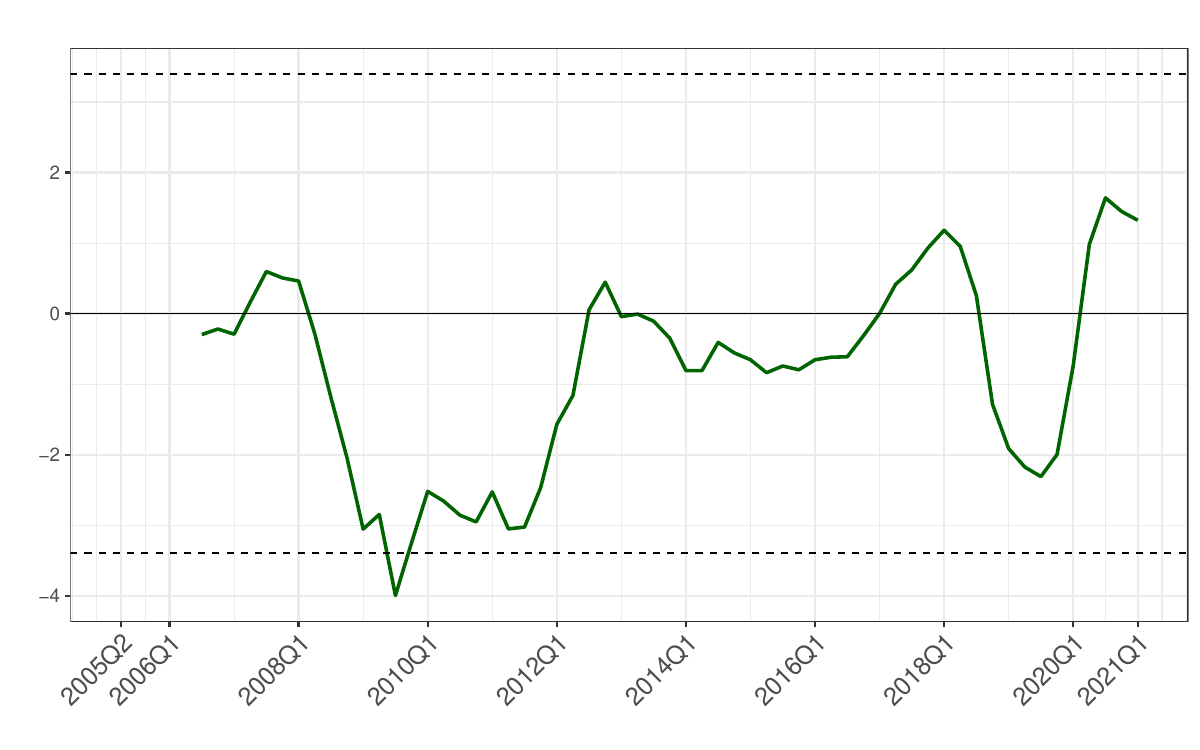}
\end{minipage}
\begin{minipage}{0.49\textwidth}
\centering
(b) One-quarter-ahead US inflation ($h = 1$)
\end{minipage}
\begin{minipage}{0.49\textwidth}
\centering
(c) One-year-ahead US inflation ($h = 4$)
\end{minipage}
\begin{minipage}{0.49\textwidth}
\centering
\includegraphics[width=\textwidth,keepaspectratio]{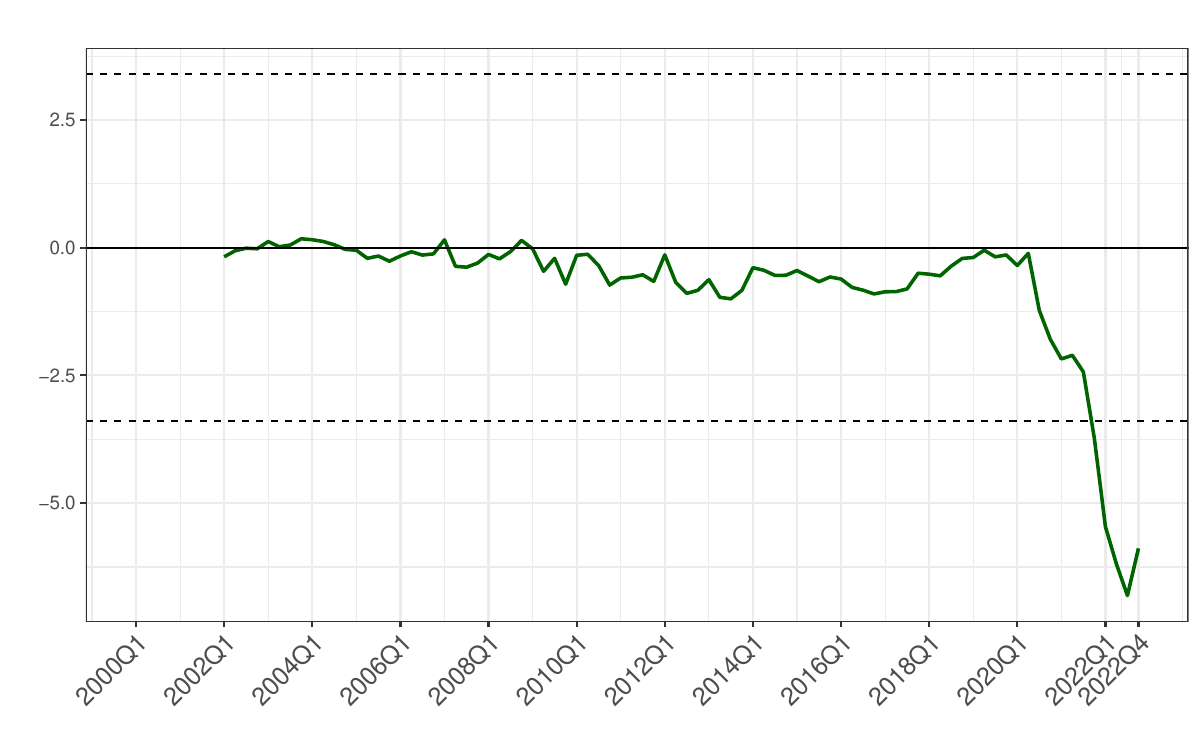}
\end{minipage}
\begin{minipage}{0.49\textwidth}
\centering
\includegraphics[width=\textwidth,keepaspectratio]{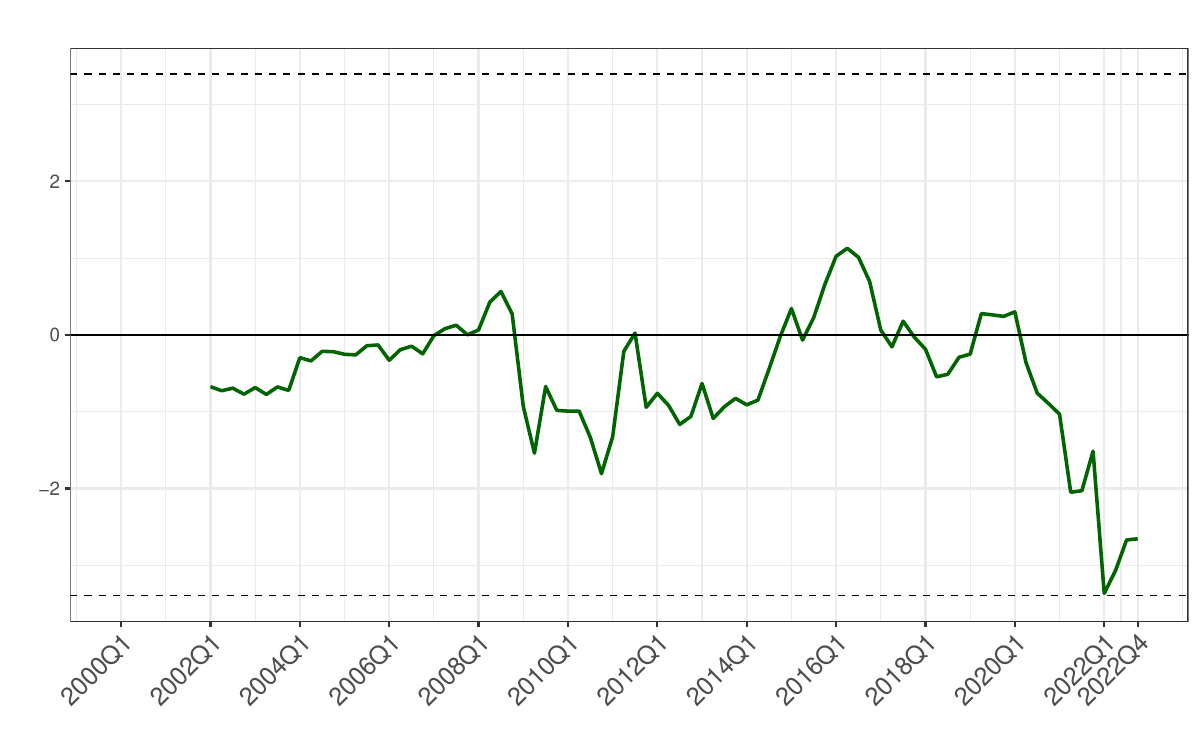}
\end{minipage}
\caption*{\scriptsize \textbf{Notes:} This figure shows the evolution of the \cite{giacomini2010forecast} fluctuation test statistic over time. The green solid line represents the test statistic, the black solid line marks the zero line, and the black dashed lines indicate the respective $95\%$ confidence bands. To compute this period-specific test statistic we use local relative continuous ranked probability scores (CRPSs) between the preferred BPS-RT specification and the benchmark (homoskedastic BPS-RW) over a rolling window comprising $10$\% of the evaluation sample. In panel (a), this implies that the rolling window is based on five observations (with the initial value of the test statistic available in $2006$Q$3$), while in panels (b) and (c), this implies the rolling window is based on eight observations (with the initial value of the test statistic available in $2002$Q$1$).}
\end{figure}

\clearpage

\subsection{Probability Integral Transforms (PITs)}
We focus on evaluating the BPS-RT specifications with the best overall forecast performance. As seen from Figures \ref{fig:easpfmain} and \ref{fig:usadlmain}, in the EA-GDP application this is the single tree specification with SV using average scores as effect modifiers. For the US inflation application, this is the homoskedastic single tree specification with the full set of weight modifiers.

\begin{figure}[!htbp]
\centering
\caption{Evaluating model calibration using probability integral transforms (PITs) \label{fig:pits}}
\begin{minipage}{\textwidth}
\centering
(a) EA GDP growth
\vspace*{2pt}
\end{minipage}
\begin{minipage}{\textwidth}
\centering
\includegraphics[width=0.55\textwidth,keepaspectratio]{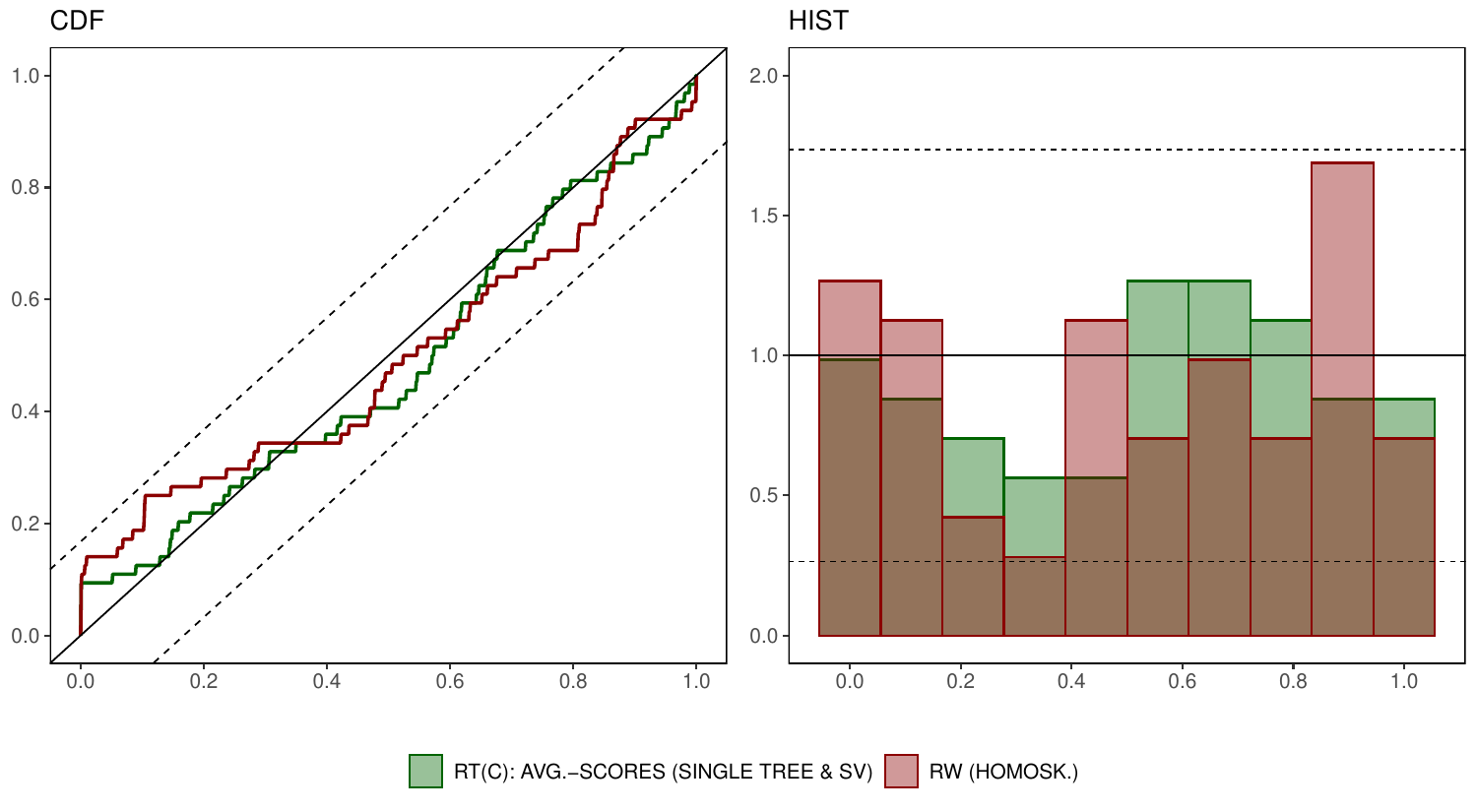}
\end{minipage}
\begin{minipage}{0.49\textwidth}
\centering
\vspace*{10pt}
(b) One-quarter-ahead US inflation ($h = 1$)
\vspace*{2pt}
\end{minipage}
\begin{minipage}{0.49\textwidth}
\centering
\vspace*{10pt}
(c) One-year-ahead US inflation ($h = 4$)
\vspace*{2pt}
\end{minipage}
\begin{minipage}{0.49\textwidth}
\centering
\includegraphics[width=1\textwidth,keepaspectratio]{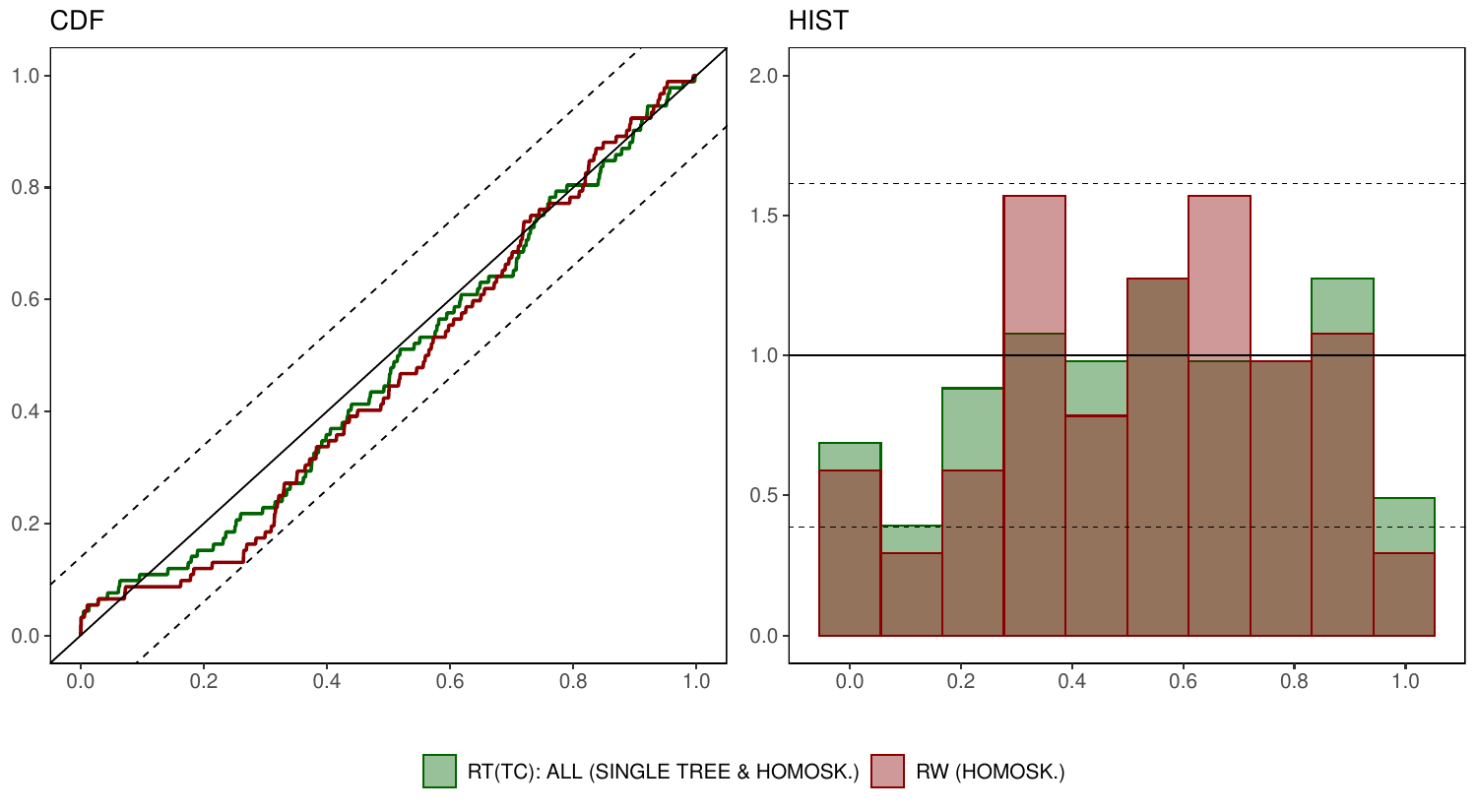} 
\end{minipage}
\begin{minipage}{0.49\textwidth}
\centering
\includegraphics[width=1\textwidth,keepaspectratio]{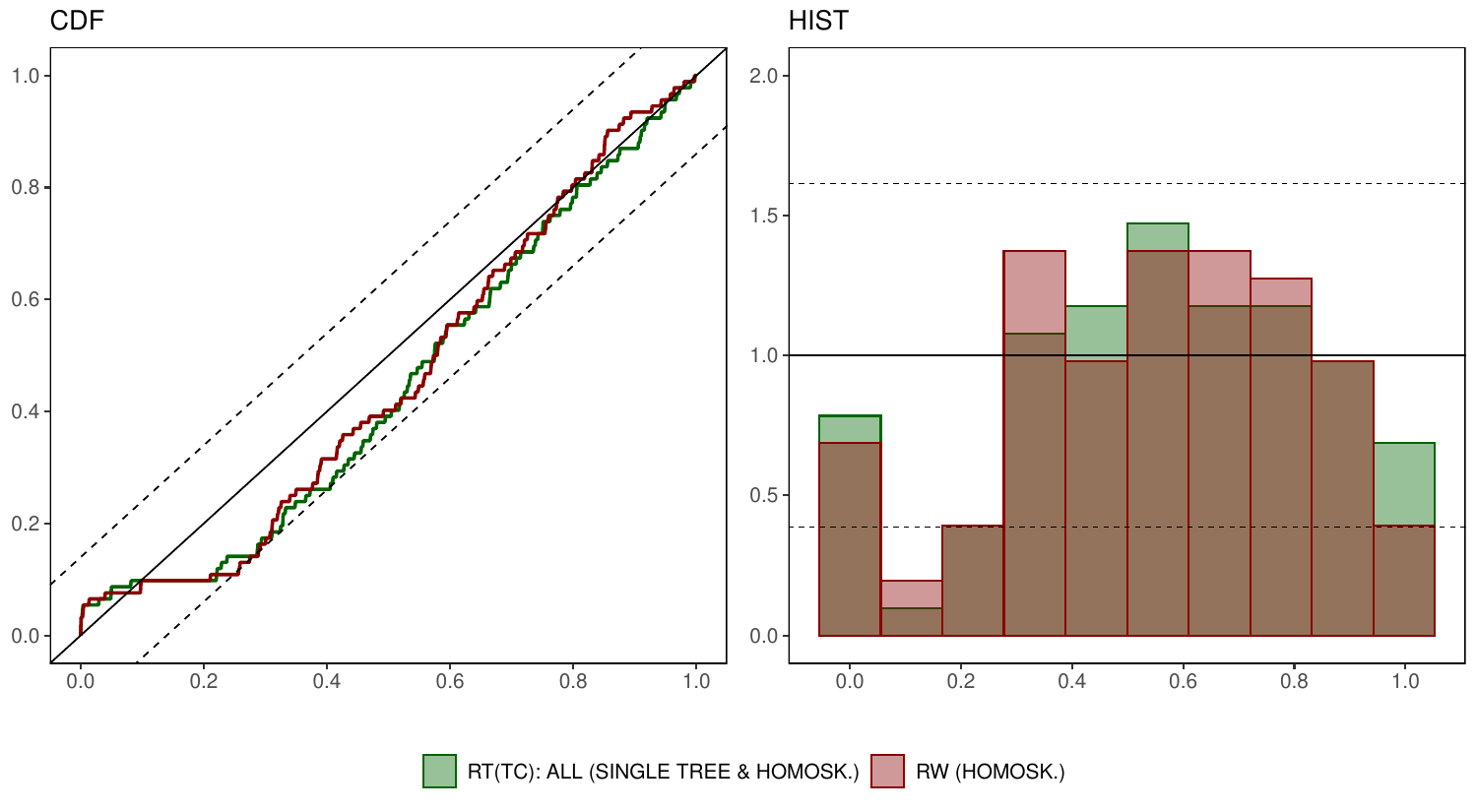}
\end{minipage}

\caption*{\scriptsize \textbf{Notes:} This figure shows the empirical cumulative density function of the PITs in the left panels and the histogram of the PITs in the right panels. A correctly specified model has PITs that are standard uniformly distributed. Such a specification is denoted by the black solid lines, with the black dashed lines denoting the respective $95\%$ confidence bands \cite[see][]{rossi2019alternative}. The preferred BPS-RT specification is shown in green, while the benchmark is indicated in red.} 
\end{figure}

\clearpage
\subsection{Measuring the Degree of Shrinkage}

\begin{figure}[!htbp]
\centering
\caption{Overall degree of shrinkage toward the prior mean for US inflation. \label{fig:usadlshrink}}
\begin{minipage}{\textwidth}
\centering
(a) One-quarter-ahead  ($h = 1)$  \\ 
\end{minipage}
\begin{minipage}{\textwidth}
\centering
\includegraphics[width=0.9\textwidth,keepaspectratio]{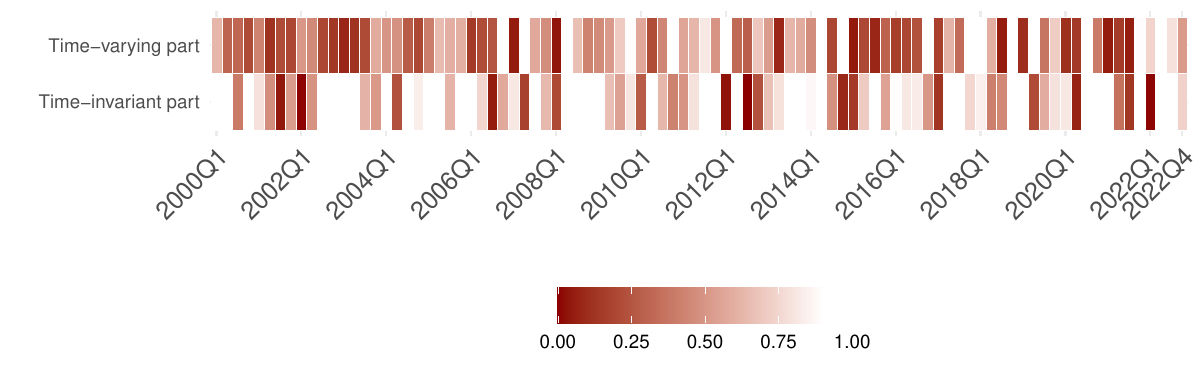}
\end{minipage}
\begin{minipage}{\textwidth}
\centering
(b) One-year-ahead ($h = 4)$ 
\end{minipage}
\begin{minipage}{\textwidth}
\centering
\includegraphics[width=0.9\textwidth,keepaspectratio]{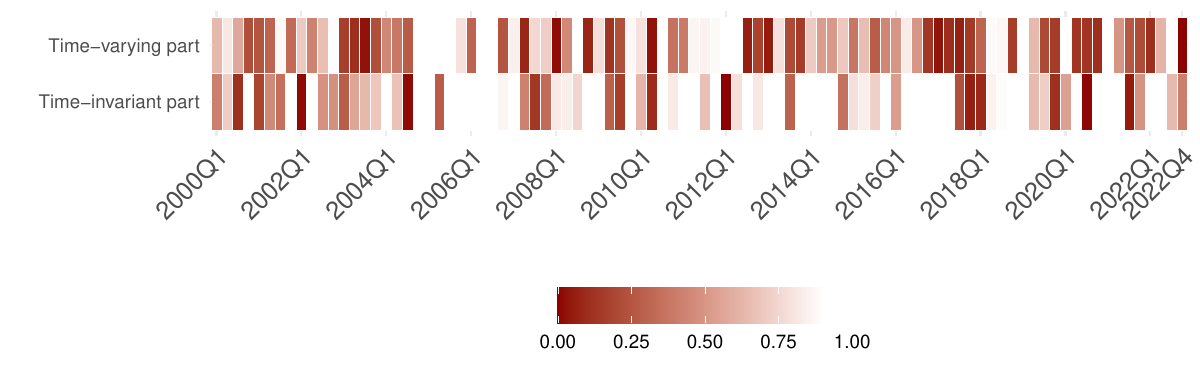}
\end{minipage}
\caption*{\scriptsize \textbf{Notes:} This figure shows the evolution of the degree of shrinkage measure over time. For each period in the evaluation sample, this measure is computed for our preferred specification (homoskedastic \texttt{BPS-RT(TC): ALL} with a single tree) as the ratio between the variation explained of the prior mean and the total variation of the respective coefficient part. This measure is bounded between zero and one. Values close to zero suggest that idiosyncratic deviations of coefficients (via the state innovation variances) dominate the prior mean (at least for some coefficients), while values close to one indicate that all coefficients are strongly pushed toward the prior mean in the respective period in the evaluation sample.}
\end{figure}

Figure \ref{fig:usadlshrink} shows a measure for overall shrinkage for both the time-invariant part ($\bm \gamma$) and the time-varying part ($\bm \beta_t$) of the weights. This measure effectively summarizes the overall variation explained by the prior mean as part of the overall variation in coefficients. It thus allows us to assess the relative importance of idiosyncratic innovations to the coefficients (i.e., innovations to the state equation) compared to the prior mean, and thus serves to quantify the overall degree of shrinkage by resembling something like a ``joint" $R^2$-type of measure, which is bounded between zero and one. In each of the state equations, the target variables are either the constant coefficients $\bm \gamma$ or the stacked time-varying coefficients $\bm \beta_t$ (for, $t = 1, \dots, T$). In such a hierarchical model, the prior mean can then be treated as the conditional mean (i.e., the fit), while the state innovations (i.e., the shocks) are mainly driven by the state innovation (or prior) variances. In the following, a low joint $R^2$ suggests that the state innovation variances play a significant role (at least for some of the coefficients), whereas a high joint $R^2$ suggests that the coefficients are heavily shrunk toward the prior mean. It is worth noting that in recessions the $R^2$ is typically essentially zero and thus the prior means are less informative in these periods and random innovations to the state equations provide/add more model flexibility, which is required/necessary in these highly volatile periods.

\subsection{Combined Predictive Densities and Their Skewness}

\begin{figure}[!htbp]
\centering
\caption{BPS-RT and BPS-RW predictive densities}\label{fig:preddens}
\begin{minipage}{\textwidth}
\centering
(a) EA GDP growth  
\vspace*{2pt}
\end{minipage}
\begin{minipage}{\textwidth}
\centering
\includegraphics[width=0.5\textwidth,keepaspectratio]{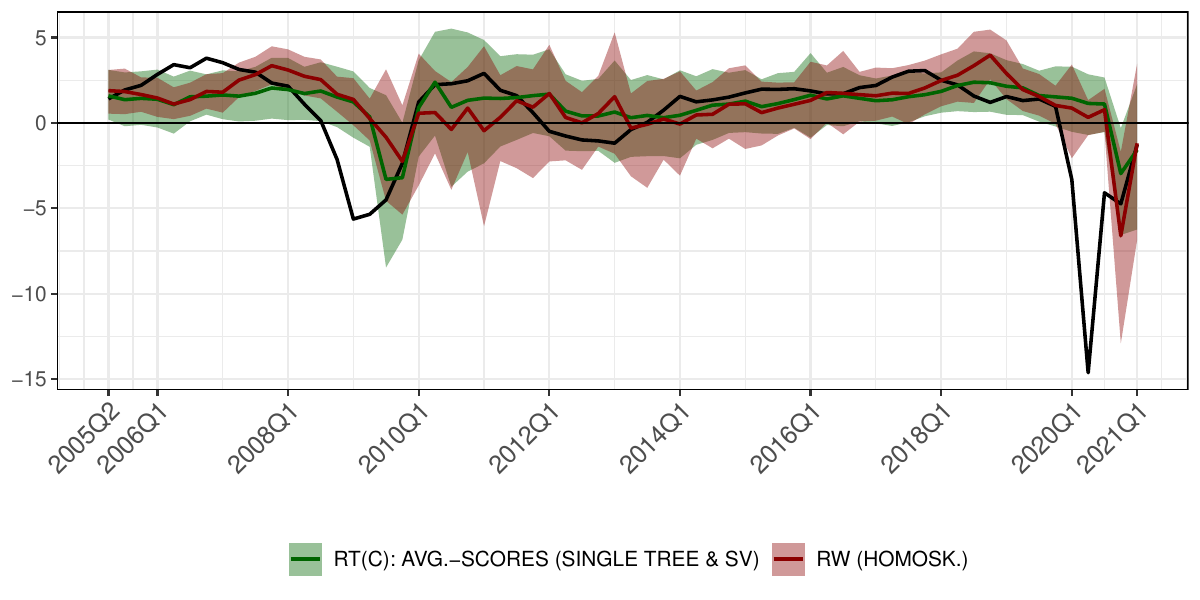}
\end{minipage}
\begin{minipage}{0.49\textwidth}
\centering
\vspace*{5pt}
(b) One-quarter-ahead US inflation ($h = 1$)
\vspace*{2pt}
\end{minipage}
\begin{minipage}{0.49\textwidth}
\centering
\vspace*{5pt}
(c) One-year-ahead US inflation ($h = 4$) 
\vspace*{2pt}
\end{minipage}
\begin{minipage}{0.49\textwidth}
\centering
\includegraphics[width=\textwidth,keepaspectratio]{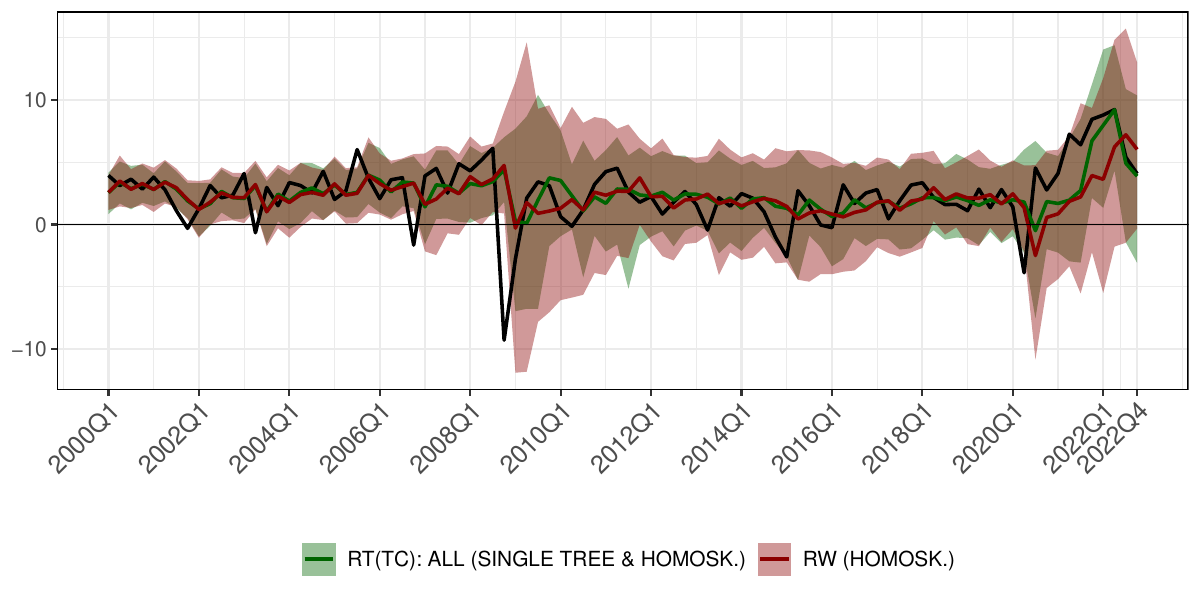}
\end{minipage}
\begin{minipage}{0.49\textwidth}
\centering
\includegraphics[width=\textwidth,keepaspectratio]{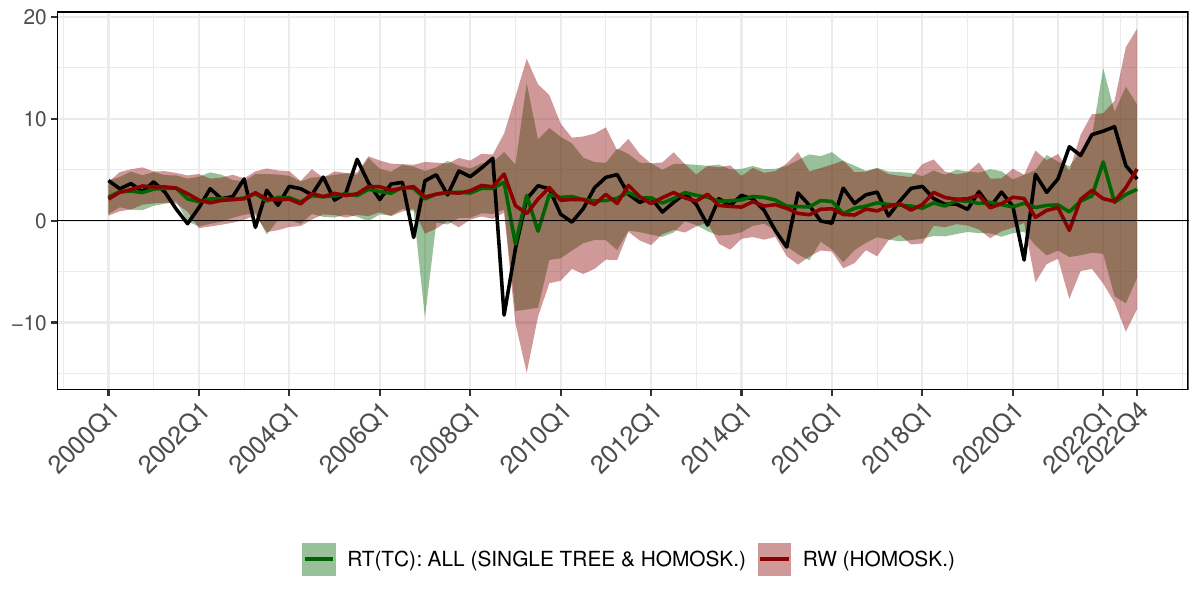}
\end{minipage}
\caption*{\scriptsize \textbf{Notes:} This figure displays the corresponding predictive densities. The colored shaded areas and the colored solid lines represent the $90\%$ confidence interval and the posterior median, respectively. The preferred BPS-RT specification is shown in green, while the benchmark is indicated in red. The black solid line in both panels refers to the respective realization.}
\end{figure}

\begin{figure}[!htbp]
\centering
\caption{Evolution of a quantile-based skewness measure for the BPS-RT predictive densities}\label{fig:skew}
\begin{minipage}{\textwidth}
\centering
(a) EA GDP growth
\end{minipage}
\begin{minipage}{\textwidth}
\centering
\includegraphics[width=0.49\textwidth,keepaspectratio]{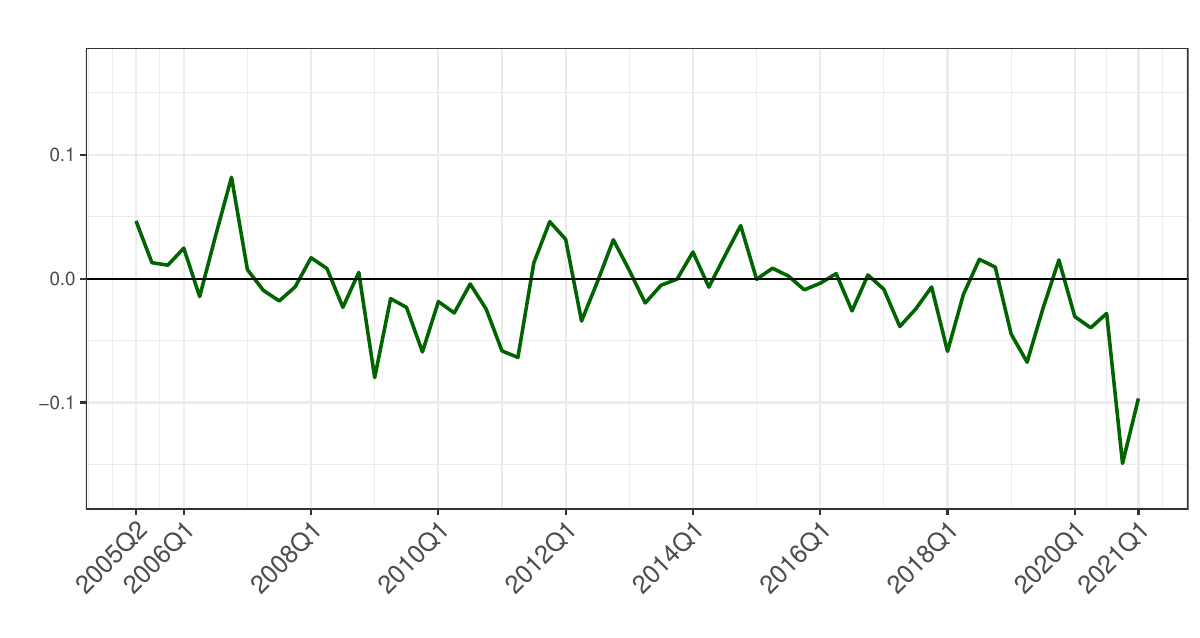}
\end{minipage}
\begin{minipage}{0.49\textwidth}
\centering
(b) One-quarter-ahead US inflation ($h = 1$)
\end{minipage}
\begin{minipage}{0.49\textwidth}
\centering
(c) One-year-ahead US inflation ($h = 4$)
\end{minipage}
\begin{minipage}{0.49\textwidth}
\centering
\includegraphics[width=\textwidth,keepaspectratio]{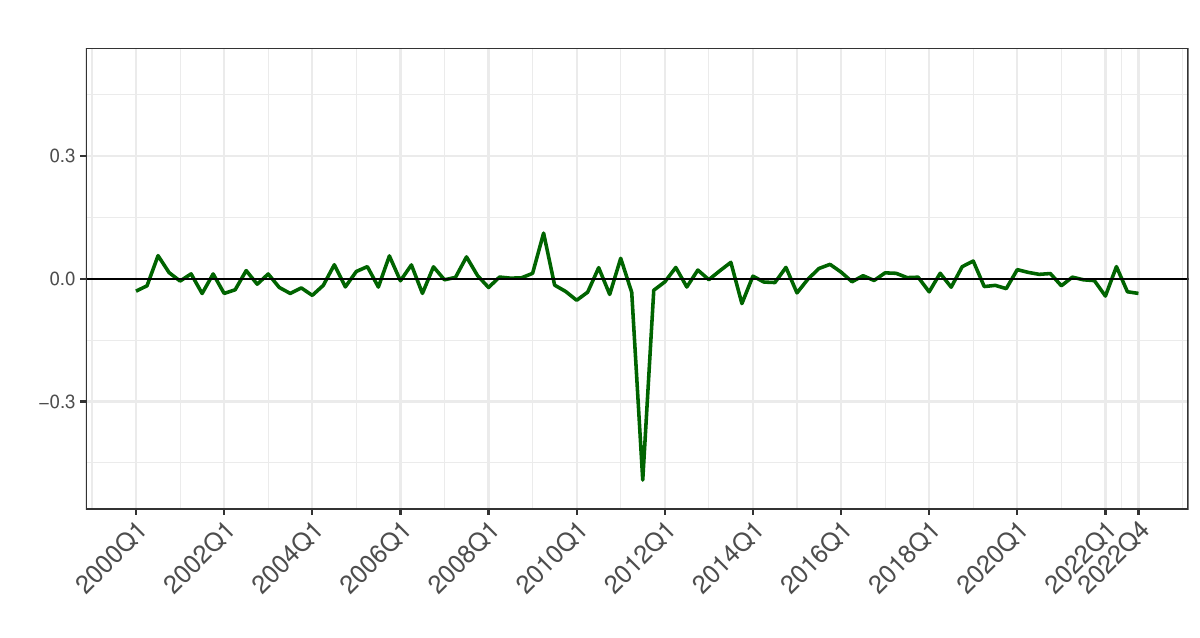}
\end{minipage}
\begin{minipage}{0.49\textwidth}
\centering
\includegraphics[width=\textwidth,keepaspectratio]{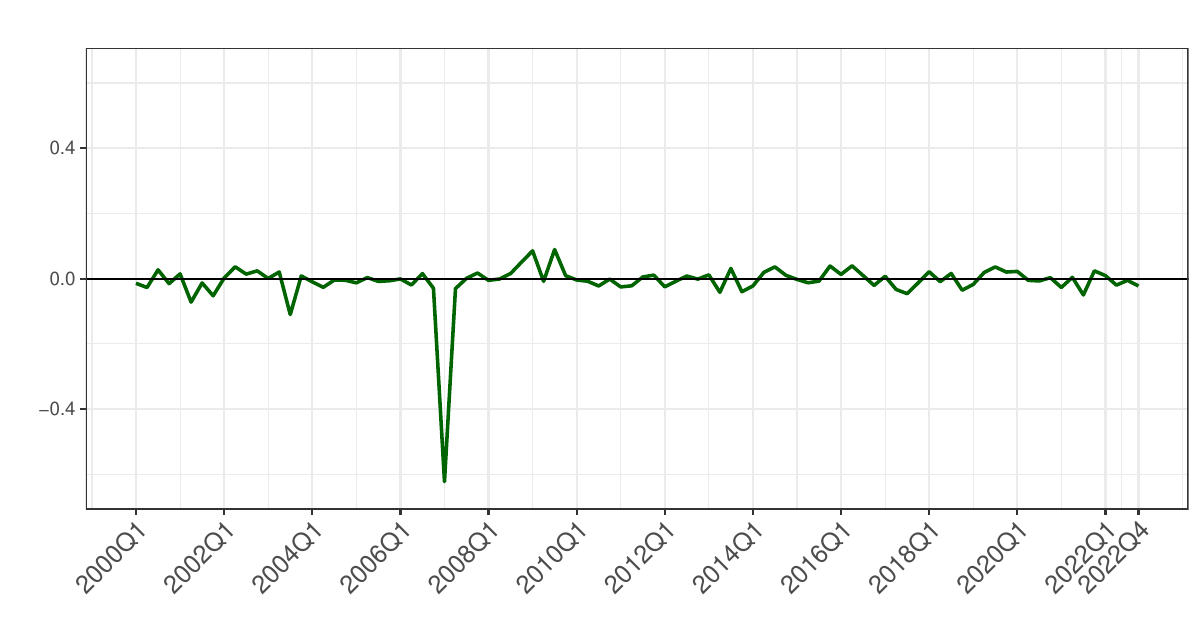}
\end{minipage}
\caption*{\scriptsize \textbf{Notes:} This figure shows the evolution of a quantile-based skewness measure for predictive densities in our preferred BPS-RT specification, as shown in Figure \ref{fig:preddens}. The quantile-based skewness measure is defined as $((q_{95\%} - q_{50\%}) - (q_{50\%} - q_{5\%}))/(q_{95\%} - q_{5\%})$, where $q_{5\%}$, $q_{50\%}$ and $q_{95\%}$ represent the $5\textsuperscript{th}$, $50\textsuperscript{th}$, and $95\textsuperscript{th}$ percentiles of the predictive densities, respectively. The green solid line represents the computed skewness measure, while the black solid line marks the zero line.}
\end{figure}

\end{appendix}

\end{document}